\documentclass[preprint,12pt]{elsarticle}

\usepackage{amssymb,amsmath}
\usepackage{textcomp}
\graphicspath{{./}{figures/}}
\usepackage{subfig}
\usepackage{placeins}
\usepackage{color}

\journal{Journal of Computational and Applied Mathematics}

\newcommand{\bfm}[1]{{\bf #1}}

\newcommand{\domain}{\Omega}

\newcommand{\Cauchy}{\boldsymbol{\sigma}}

\newcommand{\interface}{\Gamma}
\newcommand{\stresssurface}{\bfm{T}}

\newcommand{\surface}{\mathcal{S}}

\newcommand{\normale}{\bfm{n}}
\newcommand{\tangente}{\bfm{t}}
\newcommand{\curvature}{\kappa}
\newcommand{\position}{\bfm{R}}
\newcommand{\vS}{\bfm{S}}

\newcommand{\vV}{\bfm{V}}
\newcommand{\vP}{\bfm{P}}
\newcommand{\vT}{\bfm{T}}

\newcommand{\grad}{\operatorname{\nabla}}

\newcommand{\opdiv}{\operatorname{\nabla} \cdot}

\newcommand{\R}{\mathbb{R}}
\newcommand{\dsurface}{ds}
\newcommand{\dvolume}{dv}

\newcommand{\vh}{\bfm{h}}
\newcommand{\vb}{\bfm{b}}
\newcommand{\vw}{\bfm{w}}
\newcommand{\vv}{\bfm{v}}
\newcommand{\vx}{\bfm{x}}

\newcommand{\epp}{\dot{\boldsymbol{\varepsilon}}}

\newcommand{\identity}{\bfm{I}}
\newcommand{\tensionsuperficielle}{\gamma}

\newcommand{\ie}{\textit{i.e.} }
\newcommand{\maillage}{\mathcal{T}(\domain_h)}

\begin{document}

\begin{frontmatter}



\title{Finite element setting for fluid flow simulations with natural enforcement of capillary effects}


\author[EMSE]{J. Bruchon}
\ead{bruchon@emse.fr}
\author[SUNYATSEN]{Y. Liu\corref{cor1}}
\ead{liuyujie5@mail.sysu.edu.cn}
\author[EMSE]{N. Moulin}
\ead{nmoulin@emse.fr}

\address[EMSE]{SMS Centre \& LGF UMR CNRS 5307, Mines Saint-\'Etienne - Universit\'e de Lyon \\ 158 Cours Fauriel, CS 62362
F-42023 Saint- \'Etienne cedex 2}
\address[SUNYATSEN]{School of Data and Computational Science, Sun Yat-sen University \\ Guangzhou, 510275, P. R. China}

\cortext[cor1]{Corresponding author}

\begin{abstract}
Capillary phenomena are involved in many industrial processes, especially those dealing with composite manufacturing. However, their modelling is still challenging. Therefore, a finite element setting is proposed to better investigate this complex issue. The variational formulation of a liquid-air Stokes' system is established, while the solid substrate is described through boundary conditions. Expressing the weak form of Laplace's law over liquid-air, liquid-solid and air-solid interfaces, leads to a natural enforcement of the mechanical equilibrium over the wetting line, without imposing explicitly the contact-angle itself. The mechanical problem is discretised by using finite elements, linear both in velocity and pressure, stabilised with a variational multiscale method, including the possibility of enrichment of the pressure space. The moving interface is captured by a Level-Set methodology, combined with a mesh adaptation technique with respect to both pressure and level-set fields. Our methodology can simulate capillary-driven flows in 2D and 3D with the desired precision: droplet spreading, droplet coalescence, capillary rise. In each case, the equilibrium state expected in terms of velocity, pressure and contact angle is reached.
\end{abstract}


\end{frontmatter}


\section{Introduction}

Capillary phenomena are physical processes driven by the surface tension or surface energy of immiscible media~\cite{de_gennes_wetting:_1985}, \cite{blake}, \cite{snoeijer_moving_2013}, \cite{sui_efficient_2013}. The term ``surface tension'' is used when dealing with an interface between two fluids, while ``surface energy'' is employed when at least one of the domains in contact is a solid. For the sake of simplicity, these terms will be synonymously employed in this paper. However, there is still a fundamental difference between both of them: while surface tension refers to the stress state of the interface, surface energy refers to its energy density. In the situations investigated, three media, two fluids (liquid and gas) and a solid substrate, intersect at what is called a triple junction. This is a point, in two-dimensions, or a line, in three-dimensions, also called contact or wetting line. When in non-equilibrium, the force balance at the triple junction, causes the fluids to flow, and consequently induces a motion of the interface. This explains the spreading of a droplet on a solid substrate, as well as the rise of a liquid into a capillary tube, the two standard applications which will be discussed. Capillary phenomena have important implications in a wide range of industrial and scientific domains. For example, the fabrication of textured surfaces with superhydrophobic~\cite{kusumaatmaja_modeling_2007} properties, is still challenging for the automotive or aeronautics industry. In another context, the void formation observed in some bio-based composite materials, manufactured by liquid composite moulding, can be correlated to the value of the capillary number, that is the ratio between viscous and capillary forces~\cite{Park2011658,pucci_capillary_2015}.


Capillary phenomena will be described within a macroscopic continuum mechanics framework. In the literature dealing with computational aspects of capillary phenomena at continuum scale, the force balance at the triple junction is usually substituted for an angle condition to be enforced. Following the classification given by Sprittles and Shikhmurzaev~\cite{sprittles_finite_2012}, this enforcement is performed  either essentially or naturally through the variational formulation. When the contact-angle condition is seen as an essential boundary condition, some iterative scheme is usually employed to alternatively compute the contact interface velocity and modify the geometry of the interface until reaching a steady state. This strategy is adopted by Bellet~\cite{bellet_implementation_2001}, Liu~\cite{liu_towards_2016}, but also Spelt~\cite{Spelt2005389}. In this last reference, the geometry is modified in the reinitialisation step of the level-set interface-capturing method. The natural assignment of the contact-angle is based on the integration by part of the Laplace's law in the variational form of the mechanical problem. A boundary term, defined at the triple junction, appears when integrating by parts. Then, it can be replaced by the angle condition in the variational problem, as described by Sprittles in~\cite{sprittles_finite_2012,sprittles_finite_2013}. Usually, this condition connects the apparent angle, static angle, and the flow in the vicinity of the triple junction. Mechanically, as mentioned by Buscaglia and Ausas~\cite{buscaglia_variational_2011}, but also by different authors~\cite{ganesan_modelling_2008,Manservisi2009406,Afkhami20095370}, the angle condition considers a localised dissipation which could model the roughness or heterogeneity of surfaces, for example.

This paper offers an original finite element model of wetting problems, based on a variational formulation of the mechanical problem, in which the force balance is naturally imposed at the triple junction. The two fluids, assumed to be Newtonian, represent a liquid and a surrounding medium (typically air). To simplify the computational complexity, inertia effects are neglected. Therefore, the mechanical problem is reduced to the bifluid Stokes' equations, which describe a quasi-static system, allowing only the transient evolution of the liquid-gas interface. Consequently, the objective of this paper is mainly to study the ability of the proposed approach to predict successfully the final steady equilibrium state, and not to capture the transient dynamics with accuracy.

A crucial point of our approach is to fully consider all three interfaces meeting at the triple junction. Hence, working out the balance of forces acting on a surface element leads directly to the weak formulation of Laplace's law. Subsequently, considering this on liquid, gas and solid interfaces, provides a variational formulation of the mechanical problem, which can deal with the surface energy discontinuity across the triple junction, as well as express implicitly the force balance at the triple junction. Contrary to most of the previous cited articles, this force balance is not described in contact-angle terms,  but directly in terms of surface tension and surface energies. Consequently, the equilibrium value of the contact-angle is not imposed numerically, but results from the computation of this mechanical equilibrium at the triple junction. Computationally, the mechanical equations are discretised using a stabilised finite element (FE) technique. A level-set method, combined with an anisotropic mesh-adaptation strategy, is used to describe the moving interface with accuracy.

The rest of this paper is organised as follows. The mechanical problem, or bifluid Stokes' system, is presented and detailed in Section~\ref{sec:mechanical_equilibrium}, with a special focus on the balance of forces acting on the interfaces and at the triple junction. Section~\ref{sec:variational} establishes the variational formulation of this system. In these two sections, a tensor analysis setting, based on the introduction of the co- and contra-variant tangent bases, is used to mathematically describe surfaces and their geometry. Using tensor analysis describes surfaces embedded in $\R^3$ and curves embedded in $\R^2$ in a unified framework. The computational strategy is detailed in Section~\ref{sec:computation}. More precisely, the FE setting is given, including discrete problem stabilisation, pressure space enrichment, mesh adaptation strategy, and level-set method used to capture the interfaces. Finally, simulation results are shown in Section~\ref{sec:simulations}. First, accuracy of the FE framework is assessed through 2D-simulations of  droplet spreading. Second, the numerical developments are used to carry out 3D-simulations of droplet spreading and flows in a capillary tube as well.

\section{Mechanical problem}
\label{sec:mechanical_equilibrium}

\begin{figure}[htb!]
\begin{center}
\includegraphics[width=0.5\textwidth]{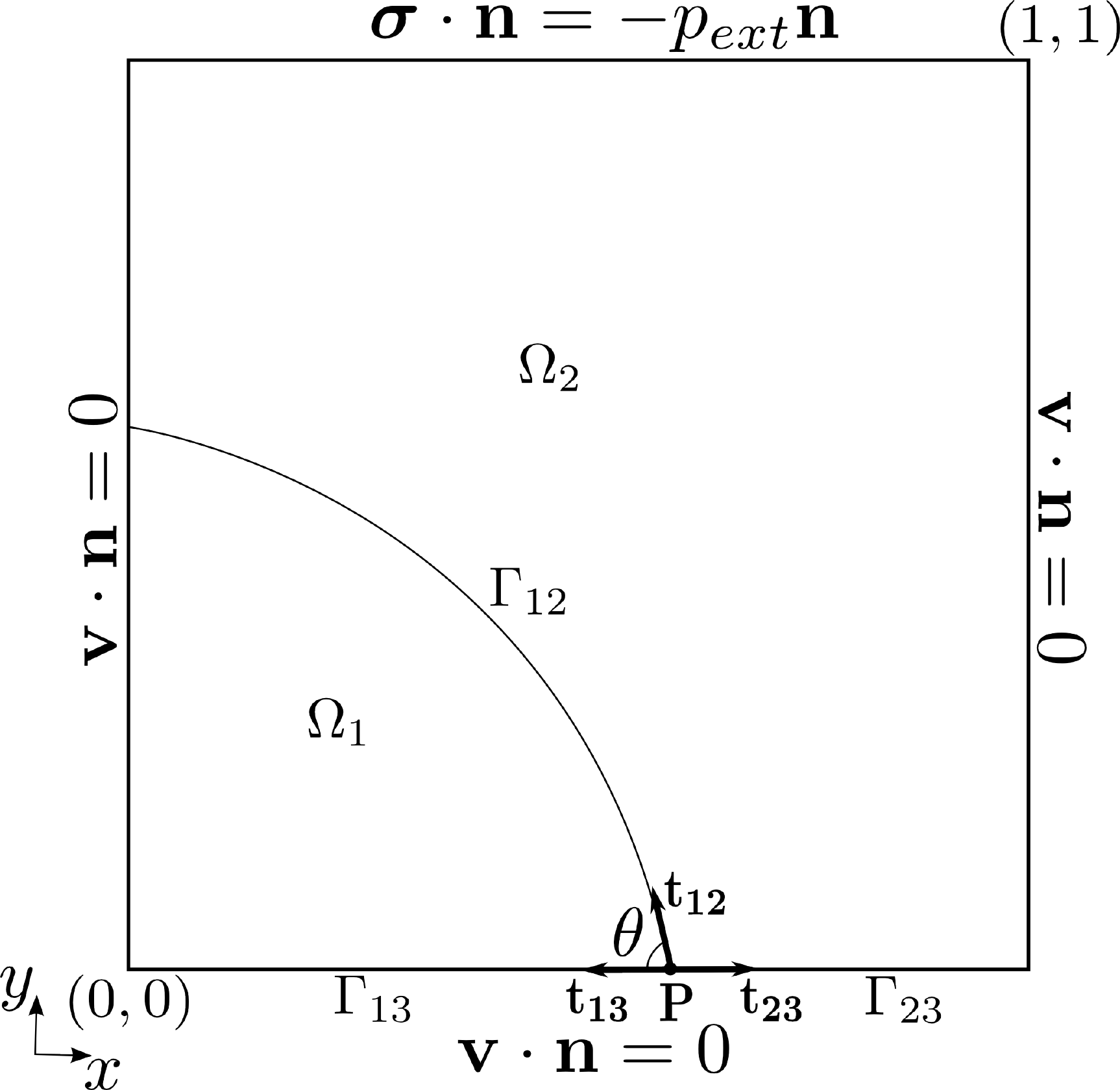}
\end{center}
\caption{Schematic of the computational domain $\domain = \domain_1 \cup \domain_2$. The boundary $\partial \domain$ is divided into two distinct parts, $\partial \domain = \Gamma_D \cup \Gamma_N$. In this 2D-case, $\Gamma_N = \{y=1\}$, and $\Gamma_D = \partial \domain \backslash \Gamma_N$.}
\label{fig:modele}
\end{figure}
Let $\domain$ be a bounded region of $\R^d$ ($d = 2,3$ is the spatial dimension), also referred to as the computational domain. This domain contains two immiscible parts: a liquid part, denoted $\domain_1$, immersed in a surrounding medium $\domain_2$, also referred to as a gaseous medium or the air. Furthermore, the liquid $\domain_1$ is lying on a rigid substrate, identified with a boundary of the computational domain. Such a situation is illustrated for simplicity in 2D (Figure~\ref{fig:modele}): $\domain$ is the unit square, while $\domain_1$ is half a liquid droplet spreading along the plane $\{y = 0\}$.

Both media $\domain_1$ and $\domain_2$ are assumed to behave as incompressible Newtonian fluids of viscosities $\eta_1$ and $\eta_2$, respectively, with $\eta_2 \ll \eta_1$. Neglecting inertial effects, momentum balance and incompressibility lead to Stokes' equations, expressed in terms of velocity $\vv$ and pressure $p$, and governing the flow into $\domain = \domain_1\cup \domain_2$:
\begin{equation}
\label{eq:momentum}
\opdiv \Cauchy = -\vb \;\; \Leftrightarrow \;\; \opdiv (2\eta\epp(\vv)) - \grad p = -\vb \mbox{ in } \domain,
\end{equation}
and
\begin{equation}
\label{eq:incompressibility}
\opdiv \vv = 0 \; \mbox{ in } \; \domain
\end{equation}
In momentum balance~\eqref{eq:momentum}, $\Cauchy$ is the Cauchy stress tensor\footnote{Here and in the following, all vectors and tensors are written in bold font.}, $\eta$ the global viscosity field, $\vb$ represents the body forces (\textit{e.g.} gravity), all these quantities being associated with the $i^{th}$ fluid in $\domain_i$. The strain rate tensor $\epp(\vv)$ is defined as the symmetric part of the velocity gradient, $\epp(\vv) = \frac{1}{2}(\grad \vv + (\grad \vv)^T)$.

This bifluid Stokes' system is closed when considering boundary conditions. As shown in Figure~\ref{fig:modele}, two types of conditions will be assumed in the simulations presented. Let us divide the boundary of the computational domain into two parts, $\Gamma_D$ and $\Gamma_N$: $\partial \domain = \Gamma_D \cup \Gamma_N$ and $\Gamma_D \cap \Gamma_N = \emptyset$. The stress vector is imposed over the boundary $\Gamma_N$ through a Neumann condition: $\Cauchy \cdot \normale = -p_{ext} \normale$, where $p_{ext}$ is a scalar (the external pressure), and $\normale$ the outward normal to $\Gamma_N$. Over $\Gamma_D$, the normal velocity is imposed to equal zero (Dirichlet condition).

However, this Stokes' system is physically irrelevant without taking into account additional mechanical equilibrium conditions. First, surface tension effects have to be considered at the liquid-gas interface denoted $\Gamma_{12}$ in Figure~\ref{fig:modele}, as well as surface energy effects at the liquid-solid and gas-solid interfaces, denoted $\Gamma_{13}$ and $\Gamma_{23}$, respectively, in the same figure. These effects are usually described by the Laplace's law: the jump in normal stress is proportional to the interface mean curvature. In subsection~\ref{sec:Laplace} this law is rederived in a form which does not involve explicitly the curvature. This expression of the surface tension/energy terms is not new, and was obtained by Buscaglia in~\cite{buscaglia_variational_2011} with energetic considerations, for example. Interestingly, our approach is based only on the mechanical equilibrium established over an interface. Furthermore, the derivation is performed with tensor analysis tools on surfaces. Even if they are not indispensable, the advantage of these tools is to treat curves (2D) and surfaces (3D) in a unified way. In particular, no projection or normal extension operators have to be explicitly introduced to describe a case in 3D. Not dealing explicitly with the curvature is computationally advantageous since this avoids manipulating second-order derivatives of the level-set function. Nevertheless, there is a deeper reason to use this form for interface mechanical equilibrium. Indeed, in Section~\ref{sec:variational}, with this expression, the mechanical equilibrium at the triple junction can be enforced as a Neumann condition in the variational form of the Stokes' system.  This mechanical equilibrium controls the value of the contact-angle $\theta$, and represents consequently, the second condition which has to be incorporated into the formulation of the mechanical problem.

Finally, before developing the points just mentioned, we have to complete the liquid-solid and gas-solid contact conditions by specifying a friction law. Indeed, the real motion of the contact line is complex, as mentioned by Sauer in~\cite{Sauer2016}, combining sliding and rolling, in a proportion which depends on physical parameters analysed experimentally in~\cite{thampi_liquid_2013}. Certainly, there is a body of work dealing with this subject. Among the boundary conditions proposed, the so-called Navier-slip condition has been more often used and is widely accepted~\cite{Ren2007,ganesan_modelling_2008}. For example, extending the work of Dussan and Davis~\cite{Dussan1974}, Buscaglia~\cite{buscaglia_variational_2011} considered a quasi-no-slip behaviour away from the contact line. A Navier condition is assumed in this reference, expressing the proportionality between the tangential force $\boldsymbol{\tau}$ and the velocity $\vv$: $\boldsymbol{\tau} = -f \vv$. The parameter $f$ is ``essentially $+\infty$ everywhere except in a very small vicinity'' (taken as the mesh size) of the contact line. In our work, we consider the usual Navier condition too, but with a parameter $f$ assumed to be constant over the solid substrate: the tangential force is proportional to the tangential velocity. Furthermore, we will see that treating the mechanical equilibrium at the triple junction as a Neumann condition, allows us to introduce ``naturally'' a dissipation term at this junction.

\subsection{Interface mechanical equilibrium}
\label{sec:Laplace}

\subsubsection{Mathematical preliminaries}
\label{sec:math}
\begin{figure}[htb!]
\begin{center}
\includegraphics[width=0.5\textwidth]{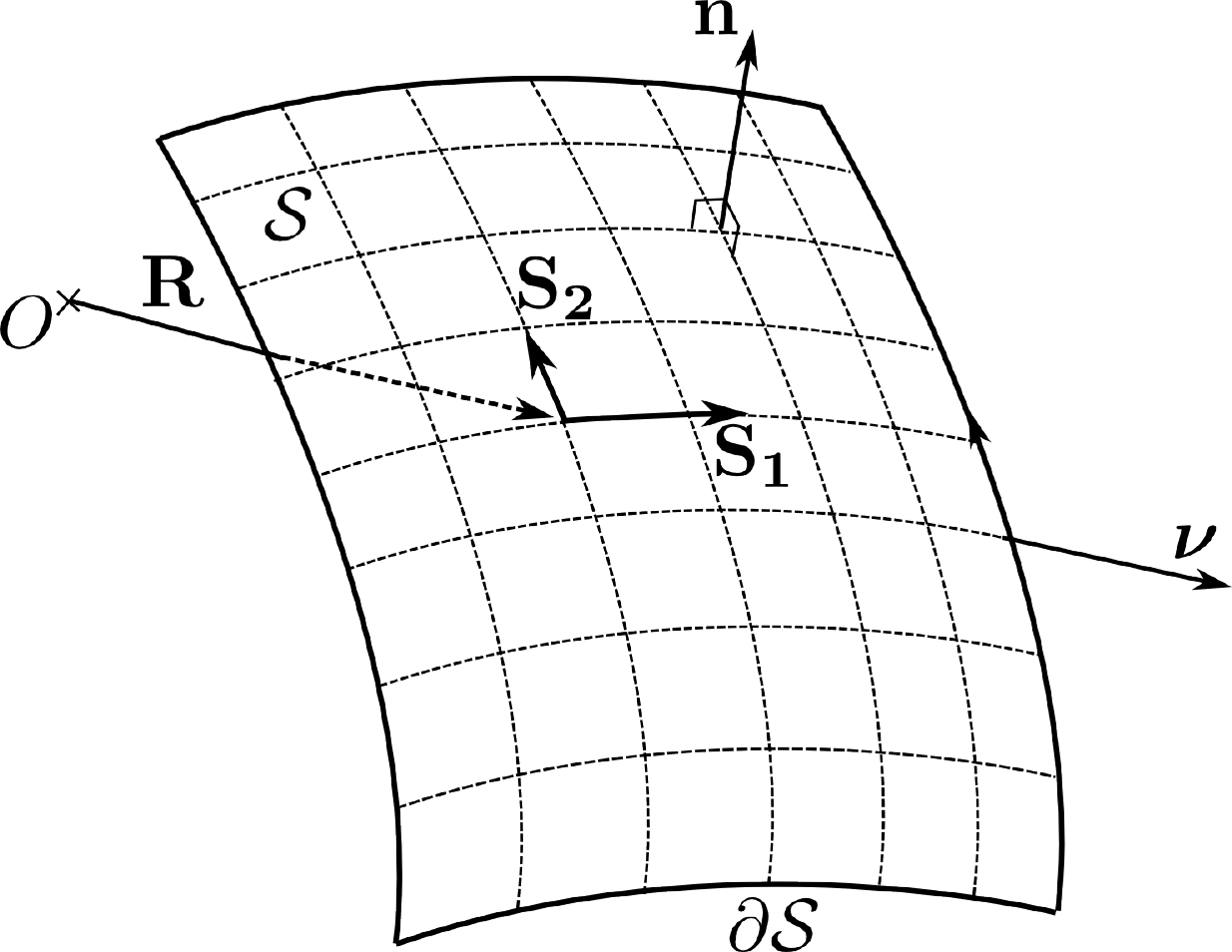}
\end{center}
\caption{Surface $\surface$ bounded by $\partial \surface$, represented with coordinate lines.}
\label{fig:surface}
\end{figure}

Let $\surface$ be any curved patch bounded by a contour $\partial \surface$. The position of any point of this surface can be represented by the position (or radius) vector $\position$, which can be viewed as a function of curvilinear surface coordinates $S^\alpha$ ($\alpha = 1,\cdots d-1$)\footnote{Here and in the following, letters from the Greek alphabet represent surface indices which assume values from 1 to $d-1$.}:
\begin{equation}
\label{eq:mapping}
\position = \position(S^\alpha)
\end{equation}
The vector $\position$ is expressed with respect to an arbitrary origin of the ambient Euclidean space $\R^d$. Note that, on the
left-hand side of Equation~\eqref{eq:mapping}, $\position$ represents the geometric position vector, called an invariant vector, that is an object which does not depend on the coordinate system. On the right-hand side, $\position$ stands for the vector-valued function that yields the position vector for every valid combination of coordinates. Consult the excellent tensor analysis textbook of Grinfeld~\cite{grinfeld_introduction_2013} for further details on the notions introduced and utilised here. Referring to Equation~\eqref{eq:mapping}, the definitions of the tangent vectors $\vS_\alpha = \partial \position/\partial S^\alpha$ (see Figure~\ref{fig:surface}), the surface metric tensor\footnote{The entries of the tensors are not in bold.} $S_{\alpha \beta} = \vS_\alpha \cdot \vS_\beta$, the dual tangent vectors $\vS^\alpha$, with $\vS_\alpha \cdot \vS^\beta = \delta^\beta_\alpha$, and the dual surface metric tensor $S^{\alpha \beta} = \vS^\alpha \cdot \vS^\beta$ follow. In these definitions, the delta symbol $\delta^\beta_\alpha$ is equal to 1 if $\alpha = \beta$ and to zero otherwise. Note that\footnote{Following the Einstein's convention adopted here, there is implicit summation over an index repeated twice, once as a subscript (covariant index) and once as a superscript (contravariant index).} $S_{\alpha \beta} S^{\beta\gamma} = \delta^\gamma_\alpha$, meaning that the two metrics are inverses of each other.
Furthermore, $\normale$, the vector normal to the surface $\surface$, is defined by $\normale \cdot \vS_\alpha = 0$ and $\normale \cdot \normale = 1$. Moreover, consider the unit vector $\boldsymbol{\nu}$ normal to the curve $\partial \surface$. This vector lies in the tangent plane of $\surface$, is pointing toward the exterior, perpendicular to the vector tangent to $\partial \surface$, and can be written as:
\begin{equation}
\label{eq:normale_tangente}
\boldsymbol{\nu} = \nu^\alpha \vS_\alpha = \nu_\alpha \vS^\alpha
\end{equation}
$\nu^\alpha$ and $\nu_\alpha$ are, respectively, the contravariant and covariant coordinates of vector $\boldsymbol{\nu}$. These coordinates satisfy the relation $\nu^\alpha = S^{\alpha\beta} \nu_\beta$, or $\nu_\alpha = S_{\alpha\beta} \nu^\beta$. Generally, these relations express the passage from the contravariant component $T^\alpha$ of a tensor to its covariant component $T_\alpha$, and vice versa.

The last object we have to introduce is the curvature tensor, denoted $B^\alpha_\beta$ and defined by
\begin{equation}
\label{eq:curvature_tensor}
\nabla_\beta \vS^\alpha = B^\alpha_\beta \normale
\end{equation}
where the differential operator $\nabla_\beta$ is the covariant derivative with respect to the coordinate $S^\beta$. This operator produces tensors out of tensors. Its formal definition is based on the Christoffel symbols, even if these are not explicitly needed in the following. The curvature tensor is symmetric, and consequently has two (if $d=3$) real eigenvalues $\kappa_1$ and $\kappa_2$, called principal curvatures. The trace of the curvature tensor, denoted $\curvature$, is obtained by contracting the indices $\alpha$ and $\beta$:
\begin{equation}
\label{eq:mean_curvature}
\kappa = B^\alpha_\alpha = \kappa_1 + \kappa_2
\end{equation}
The quantity $\curvature$ is the mean curvature. Note that the determinant of $B^\alpha_\beta$, equal to the product $\kappa_1 \kappa_2$, is the Gaussian curvature.

These mathematical preliminaries are concluded by giving the Gauss's (or divergence) theorem for surfaces, that is in a Riemannian form. Let $F$ be a sufficiently smooth enough scalar field, defined on $\surface$. Then, Gauss's theorem reads
\begin{equation}
\label{eq:Gauss_thm}
\int_\surface \grad_\alpha F \, \dsurface = \int_{\partial \surface} F \nu_\alpha \, dl
\end{equation}
where $\grad_\alpha F$ denotes the covariant derivative of $F$ with respect to the coordinate $\alpha$.

\subsubsection{Force balance}

Let us express the balance of forces acting on the interface $\interface_{ij}$ separating $\domain_i$ and $\domain_j$: $\interface_{ij} = \overline{\domain}_i \cap \overline{\domain}_j$, where $\overline{\domain}_i$ denotes the closure of the open domain $\domain_i$. From this point on, indices $i$ and $j$ can take the values 1 (liquid), 2 (gas) and 3 (solid), with $i \neq j$. They will never be considered as dummy indices, and consequently will never be involved in any implicit summation. Let $\surface$ be any curved patch embedded into $\interface_{ij}$ and bounded by the closed curve $\partial \surface$.

Similarly to the approach proposed by Fried and Gurtin~\cite{fried_unified_2004} in a 2D context, we consider the force and torque balances over the surface $\surface \subset \Gamma_{ij}$. In addition to the stresses $\Cauchy_i = \Cauchy_{|\domain_i}$ and $\Cauchy_j = \Cauchy_{|\domain_j}$, distributed over $\domain_i$ and $\domain_j$, an interface stress, characterised by a second-order tensor $\stresssurface$, with entries $T^{k\alpha}$, is assumed to be acting continuously along the interface $\surface$. Notice: the ambient index $k$ assumes values from 1 to $d$, while the surface index $\alpha$, from 1 to $d-1$. Hence, the interface stress vector acting along the curve $\partial \surface$ is given by $(T^{k\alpha} \nu_\alpha) \vx_k$, where $\vx_k$ is the $k^{th}$ vector of the standard basis of $\R^d$. The presence of ambient index $k$ in expression $T^{k\alpha}$ is ambiguous, one might confused and led  to a non-symmetric tensor and possibly out-of-plane components of the interface stress vector as well. However, the presence of an ambient index is only due to the fact that the interface stress vector has not been assumed tangent to the interface \textit{a priori}. This property is a consequence of the torque balance, as described in the following. Once it is proved, tensor $\bfm{T}$ could be rewritten in a symmetric form with two Greek indices using the entries $W^{\alpha \beta}$ of relation~\eqref{eq:cross_prod_eqzero_3}. Additionally, for simplicity, the friction force acting on a fluid-solid interface is not included in the developments given hereinafter. However, this will be expressed through a Navier-type boundary condition in the equations of Section~\ref{sec:Stokes_bifluid}.

Force and torque balances over $\surface$ are consequently expressed by
\begin{eqnarray}
\label{eq:balance_force}
\int_{\partial \surface} T^{k \alpha} \nu_\alpha \vx_k \, dl - \int_\surface \Cauchy_i ·\cdot \normale \dsurface +  \int_\surface \Cauchy_j \cdot \normale \dsurface = 0
\end{eqnarray}
\begin{eqnarray}
\label{eq:balance_torque}
\int_{\partial \surface} \position \times (T^{k \alpha} \nu_\alpha \vx_k) \, dl - \int_\surface \position \times (\Cauchy_i \cdot \normale) \, \dsurface +  \int_\surface \position \times (\Cauchy_j \cdot \normale) \dsurface = 0
\end{eqnarray}
with the convention that $\normale$ is pointing from $\domain_i$ to $\domain_j$.

Thanks to Gauss's theorem~\eqref{eq:Gauss_thm}, both previous 1D-line integrals can be turned into 2D-surface integrals:
\begin{equation}
\label{eq:Gauss_forces}
\int_{\partial \surface} T^{k \alpha} \vx_k \nu_\alpha \, dl = \int_\surface (\grad_\alpha T^{k \alpha}) \vx_k \, \dsurface
\end{equation}
and
\begin{equation}
\label{eq:Gauss_torque}
\int_{\partial \surface} \position \times (T^{k \alpha} \vx_k) \nu_\alpha \, dl = \int_\surface \grad_\alpha (\position \times T^{k\alpha} \vx_k) \, \dsurface
\end{equation}

Using the substitution provided by~\eqref{eq:Gauss_forces} in~\eqref{eq:balance_force} while satisfying for any surface $\surface \subset \Gamma_{ij}$, results in the local expression of the interface force balance
\begin{equation}
\label{eq:balance_force_local}
(\grad_\alpha T^{k \alpha}) \vx_k + [\Cauchy \cdot \normale]_{\Gamma_{ij}} = 0
\end{equation}
where $[\Cauchy \cdot \normale]_{\interface_{ij}}$ is the jump in the stress vector $\Cauchy \cdot \normale$ across $\interface_{ij}$
and, for any quantity $f$ and $\vx \in \surface$,
\begin{equation}
\label{eq:jump}
[f](\vx) = f^+(\vx) - f^-(\vx), \, \mbox{ with } \, f^\pm(\vx) = \lim\limits_{\begin{array}{l} \epsilon \to 0\\ \epsilon > 0 \end{array}} f(\vx \pm \epsilon \normale)
\end{equation}

Next, on the right-hand side of Equation~\eqref{eq:Gauss_torque}, the integrand can be developed by the product rule, as
\begin{equation}
\label{eq:torque_dev}
\grad_\alpha (\position \times T^{k\alpha} \vx_k) = \vS_\alpha \times \vx_k T^{k\alpha} + \position \times \nabla_\alpha(T^{k\alpha} \vx_k)
\end{equation}
when taking into account the fact that the surface covariant derivative of the invariant $\position$ is nothing but its partial derivative, which in turn is, by definition, a tangent vector: $\nabla_\alpha \position = \partial \position/\partial S^\alpha = \vS_\alpha$.

Relations~\eqref{eq:Gauss_torque} and~\eqref{eq:torque_dev} rewrite the torque balance equation~\eqref{eq:balance_torque} as
\begin{equation}
\int_\surface \vS_\alpha \times \vx_k T^{k\alpha} \, \dsurface + \int_\surface \position \times \left((\grad_\alpha T^{k \alpha}) \vx_k + [\Cauchy \cdot \normale]_{\Gamma_{ij}} \right)\dsurface = 0
\end{equation}

From local force balance equation~\eqref{eq:balance_force_local}, it can be concluded that this relation is satisfied for any $\surface \subset \Gamma_{ij}$ if tensor $T^{k\alpha}$ fulfils
\begin{equation}
\label{eq:cross_prod_eqzero}
\vS_\alpha \times (\vx_k T^{k\alpha}) = 0
\end{equation}

Let us split the vectors $\vx_k T^{k\alpha}$ into their tangential and normal components:
$$ \vx_k T^{k\alpha} = W^{\alpha \beta} \vS_\beta + W^\alpha \normale $$
Then, Equation~\eqref{eq:cross_prod_eqzero} becomes
\begin{equation}
\label{eq:cross_prod_eqzero_2}
\vS_\alpha \times (W^{\alpha \beta} \vS_\beta + W^\alpha \normale) = 0
\end{equation}
Since $\vS_\alpha \times \vS_\beta$ is proportional to the normal vector $\normale$ if $\alpha \neq \beta$, and $\vS_\alpha \times \normale$ lies in the tangent plane, relation~\eqref{eq:cross_prod_eqzero_2} implies $W^\alpha  = 0$. Actually, by expanding the cross product, it can be shown that the cross components $W^{12}$ and $W^{21}$ must be equal. In other words, relation~\eqref{eq:cross_prod_eqzero_2} leads to
\begin{equation}
\label{eq:cross_prod_eqzero_3}
T^{k1} \vx_k = W^{11} \vS_1 + W^{12} \vS_2 \;\; \mbox{ and } \;\; T^{k2} \vx_k = W^{12} \vS_1 + W^{22} \vS_2
\end{equation}

In this work, we assume that the surface is isotropic, that is $W^{12} = 0$ and $W_{11} = W_{22} = \tensionsuperficielle_{ij}$. This scalar is the surface tension or surface energy at the interface between media $i$ and $j$. Consequently, we can write
\begin{equation}
\label{eq:tension_superficielle}
T^{k\alpha} \vx_k = \tensionsuperficielle_{ij} \vS^\alpha
\end{equation}

Finally, considering this last expression, the definitions of the curvature tensor~\eqref{eq:curvature_tensor} and the mean curvature~\eqref{eq:mean_curvature}, we can rewrite local interface force balance~\eqref{eq:balance_force_local} as
\begin{eqnarray}
\label{eq:laplace}
[\Cauchy \cdot \normale]_{\interface_{ij}} = -\grad_\alpha(\tensionsuperficielle_{ij} \vS^\alpha) = -\left(\tensionsuperficielle_{ij} \curvature \normale + \frac{\partial \tensionsuperficielle_{ij}}{\partial S^\alpha} \vS^\alpha\right)  \;\;\; \mbox{ on } \interface_{ij}
\end{eqnarray}

Force balance~\eqref{eq:laplace} is nothing but the usual condition expressing the discontinuity of the stress vector across an interface. The normal component of the stress vector jump, depends on the mean curvature $\curvature$ and is associated with the Laplace's law. The tangential part of this jump, depends on the variation of the parameter $\tensionsuperficielle_{ij}$, giving rise to the Marangoni's effect. However, the real key point of expression~\eqref{eq:laplace} is the fact that the stress jump can be directly related to the surface derivatives of $\tensionsuperficielle_{ij} \vS^\alpha$, without expanding these derivatives. Indeed, this relation, which does not involve explicitly the curvature but contains intrinsically both Laplace and Marangoni components, will give rise to what we will call the weak formulation of the Laplace's law, discussed in Section~\ref{sec:variational}.

\subsection{Triple junction mechanical equilibrium}
\label{sec:triple_junction}
Force balance~\eqref{eq:laplace} holds over the interfaces $\interface_{ij}$, except at the triple junction. According to the contact theory for surface tension driven systems developed by Sauer in~\cite{sauer_contact_2016}, the static equilibrium condition over a triple line is expressed as
\begin{equation}
\label{eq:force_triangle}
\tensionsuperficielle_{12} \tangente_{12} + \tensionsuperficielle_{13} \tangente_{13} + \tensionsuperficielle_{23} \tangente_{23} + \epsilon \Cauchy_3 \cdot \normale = 0
\end{equation}
where, for simplicity, equilibrium relation~\eqref{eq:force_triangle} is first presented in the 2D-version of Figure~\ref{fig:modele}:  $\tangente_{ij}$ is the vector tangent to interface $\Gamma_{ij}$ and plays the role of $\vS_\alpha$. $\Cauchy_3$ denotes the Cauchy stress tensor of the solid body $\Omega_3$, $\normale$ the unit normal pointing outward $\Omega_3$, and $\epsilon$ the distance spanned by the wetting ridge. Furthermore, note that, as $\epsilon \to 0$ we have $\Cauchy_3 \cdot \normale \to \infty$, since, in order to balance the normal component of $\tensionsuperficielle_{12} \tangente_{12}$,  the product $\epsilon \Cauchy_3 \cdot \normale$ must remain finite. The point load $\epsilon\Cauchy_3 \cdot \normale$ will act on $\Omega_3$ but has no influence on the problem if $\Omega_3$ is rigid.

This vector relation can also be described in term of the contact-angle $\theta$ made, in Figure~\ref{fig:modele}, by the droplet with the substrate. Since the substrate is assumed to be a rigid body, we take $\epsilon = 0$ and consider the projection of above relation onto the substrate tangent plane:
\begin{equation}
\label{eq:angle}
\cos \theta = \frac{\tensionsuperficielle_{13} - \tensionsuperficielle_{23}}{\tensionsuperficielle_{12}}
\end{equation}

\noindent This type of Young's relation is often used in the literature to enforce force balance at triple junction by imposing the contact-angle, called in this case static angle. On the contrary, the variational formulation of the mechanical problem presented in Section~\ref{sec:variational}, takes directly into account force balance~\eqref{eq:force_triangle}.

More precisely, in general situations ($d = 2, 3$), the force triangle~\eqref{eq:force_triangle} can be reformulated using the interface stress tensor $T^{k\alpha}$ introduced in Section~\ref{sec:Laplace}. As we are dealing with a rigid substrate, $\epsilon = 0$, and
\begin{equation}
\label{eq:triple_junction0}
\sum_{i, j>i} (T^{k\alpha} \nu_\alpha)_{|\Gamma_{ij}} \vx_k = 0 \;\; \mbox{ on } \bigcap_{i, j>i} \overline{\interface}_{ij} =:\partial \Gamma
\end{equation}
where $\sum_{i,j>i}$ stands for the summation over the $d$ interfaces meeting at the contact line $\partial \Gamma$, that is $\sum_{i,j>i} \equiv \sum_{i=1}^{d-1} \sum_{j=i+1}^d$. The terms involved in this sum are subsequently evaluated for each interface along the contact line.

Taking into account expression~\eqref{eq:tension_superficielle} of $T^{k\alpha}\vx_k$, the following expression of the mechanical equilibrium at the triple junction is considered:
\begin{equation}
\label{eq:triple_junction}
\sum_{i, j>i} \tensionsuperficielle_{ij} (\vS^\alpha \nu_\alpha)_{|\Gamma_{ij}} = \vh \;\; \mbox{ on } \partial \Gamma
\end{equation}
The additional term $\vh$ represents a possible dissipation along the wetting line~\cite{buscaglia_variational_2011}. The existence of a dynamic contact angle due to viscous dissipation near the triple junction, was originally postulated by Qu\'er\'e in \cite{quere_inertial_1997}. This idea has been taken up by numerous authors, especially T.D. Blake \textit{et al}~\cite{martic_molecular_2002,coninck_wetting_2008}. An informative analysis of dissipative term $\vh$ and values found by molecular dynamics can be found in the work of Ren, E \textit{et al}~\cite{ren_continuum_2010,ren_contact_2011}. Here, we just want to point out how such a dissipation can naturally be included into the variational formulation of a wetting problem. However, the purpose of this paper is not to investigate the modelling of wetting dynamics. Consequently, in most of the simulations presented in the following, we will consider $\vh = 0$.

\subsection{Full bifluid Stokes' system}
\label{sec:Stokes_bifluid}
The mechanical problem consists in the bifluid Stokes' system~\eqref{eq:momentum}-\eqref{eq:incompressibility} with the different boundary conditions specified which must be enforced on $\Gamma_D$ and $\Gamma_N$, as well as with the mechanical equilibrium conditions~\eqref{eq:laplace} and~\eqref{eq:triple_junction}, considered for each interface and at the triple junction respectively.

This can be summarised as: find the velocity and pressure fields, $\vv$ and $p$, satisfying
\begin{equation}
\label{eq:Stokes_momentum}
\opdiv(2\eta \epp(\vv)) - \grad p = -\vb \; \mbox{ in } \; \domain
\end{equation}
\begin{equation}
\label{eq:Stokes_incompressible}
\opdiv \vv = 0 \; \mbox{ in } \; \domain
\end{equation}
\begin{equation}
\label{eq:Stokes_sigma_n_jump}
[\Cauchy \cdot \normale]_{\interface_{12}} = -\grad_\alpha(\tensionsuperficielle_{12} \vS^\alpha) \; \mbox{ on } \; \Gamma_{12}
\end{equation}
\begin{equation}
\label{eq:Stokes_sigma_n2_fs}
\Cauchy \cdot \normale = \grad_\alpha(\tensionsuperficielle_{13} \vS^\alpha)  - f(\vv \cdot \vS^\alpha)\vS_\alpha \; \mbox{ on } \; \Gamma_{13}
\end{equation}
\begin{equation}
\label{eq:Stokes_sigma_n2_ls}
\Cauchy \cdot \normale = \grad_\alpha(\tensionsuperficielle_{23} \vS^\alpha)  - f(\vv \cdot \vS^\alpha)\vS_\alpha \; \mbox{ on } \; \Gamma_{23}
\end{equation}
\begin{equation}
\label{eq:Stokes_sigma_n1}
\Cauchy \cdot \normale = -p_{ext} \normale \; \mbox{ on } \; \Gamma_N
\end{equation}
\begin{equation}
\label{eq:Stokes_GammaD2}
(\Cauchy \cdot \normale)\cdot \vS_\alpha = 0 \; \mbox{ on } \; \Gamma_D \backslash (\Gamma_{13}\cup\Gamma_{23})
\end{equation}
\begin{equation}
\label{eq:triple_junction2}
\sum_{i,j>i} \tensionsuperficielle_{ij} (\vS^\alpha \nu_\alpha)_{|\Gamma_{ij}} = \vh \; \mbox{ on } \; \partial \Gamma
\end{equation}
\begin{equation}
\label{eq:Stokes_GammaD}
\vv \cdot \normale = 0 \; \mbox{ on } \; \Gamma_D
\end{equation}
Note that condition~\eqref{eq:Stokes_GammaD2} holds for $\alpha = 1, 2$, meaning that the shear stress vanishes on $\Gamma_D \backslash (\Gamma_{13}\cup\Gamma_{23})$. Furthermore, an important point has to be outlined. Conditions~\eqref{eq:Stokes_sigma_n2_fs} and~\eqref{eq:Stokes_sigma_n2_ls} combined with~\eqref{eq:Stokes_GammaD} do not impose the normal stress $\normale \cdot \Cauchy \cdot \normale$ over the liquid-solid and gas-solid interfaces. This quantity is an unknown of the problem. However, this combination of conditions expresses the tangential part of the stress vector $\Cauchy \cdot \normale$ with respect to the variation of the surface energy holding between liquid-solid and gas-solid interfaces. Additionally, still in~\eqref{eq:Stokes_sigma_n2_fs} and~\eqref{eq:Stokes_sigma_n2_ls}, $f$ is the  friction-coefficient associated to the Navier boundary condition, while $(\vv \cdot \vS^\alpha)\vS_\alpha$ is the tangential velocity. Considering the no-penetration condition~\eqref{eq:Stokes_GammaD}, velocity and tangential velocity are equal at the solid-fluid interfaces.

In the two dimensional illustration (Figure~\ref{fig:modele}), $\Gamma_N \equiv \{y = 1\}$, while $\Gamma_D$ is the complement, $\Gamma_D \equiv \partial \domain \backslash \Gamma_N$. Furthermore, still in 2D, $\alpha = 1$, $\grad_\alpha \equiv \partial/\partial s$ where $s$ is the arc-length, $\nu_\alpha = \pm 1$, and $\vS^\alpha$ is the unit vector tangent to the interface considered (denoted $\tangente_{12}$, $\tangente_{13}$, and $\tangente_{23}$ in the figure).

\section{Variational formulation}
\label{sec:variational}
In order to approximate the solution $(\vv,p)$ of the previous Stokes' system~\eqref{eq:Stokes_momentum}-\eqref{eq:Stokes_GammaD} using finite elements, the variational form of this set of equations is first derived. The challenging step is to formulate the variational form of momentum equation, since it has to incorporate all the natural boundary and interface conditions.

\subsection{Momentum equation variational form}

The primal variables of Stokes' equations in their variational form, are the velocity and pressure fields, $\vv$ and $p$, which belong to the functional spaces $H^1(\domain)^d$  and $L^2(\domain)$ respectively. For the sake of simplicity, the weak form of the momentum equation is first established without expressing the stress tensor $\Cauchy$ with respect to the velocity and pressure.

Let $\vw \in H^1(\domain)^d$  be any test function satisfying $\vw \cdot \normale = 0$ on $\Gamma_D$. Considering momentum equation~\eqref{eq:momentum} in $\domain_i$, $i = 1,2$, integrating the dot product of this equation by $\vw$, and using the divergence theorem, provides
\begin{equation}
\label{eq:momentum_weak1}
\int_{\domain_i} \Cauchy : \grad \vw \, \dvolume = \int_{\partial \domain_i} (\Cauchy \cdot \normale) \cdot \vw \, \dsurface + \int_{\domain_i} \vb \cdot \vw \, \dvolume
\end{equation}
Summing up the contributions of the liquid and gaseous parts, $\domain_1$ and $\domain_2$, leads to
\begin{eqnarray}
\label{eq:momentum_weak2}
\int_\domain \Cauchy : \grad \vw \, \dvolume &=& \int_\domain \vb \cdot \vw \, \dvolume \nonumber -\int_{\interface_{12}} [\Cauchy \cdot \normale]_{|\interface_{12}} \cdot \vw \, \dsurface \\ & & + \int_{\Gamma_N} (\Cauchy \cdot \normale) \cdot \vw \, \dsurface + \int_{\Gamma_D} (\Cauchy \cdot \normale) \cdot \vw \, \dsurface
\end{eqnarray}
where the minus sign in front of the interface integral, results from the definition~\eqref{eq:jump} of the jump operator, because in this integral, $\normale$ is pointing from $\domain_1$ to $\domain_2$.

Dividing the test function $\vw$ into normal and tangential components, $\vw = (\vw \cdot \normale) \normale + (\vw \cdot \vS^\alpha) \vS_\alpha$, since $\vw \cdot \normale = 0$ over $\Gamma_D$, we have
\begin{equation}
\label{eq:momentum_weak3}
\int_{\Gamma_D} (\Cauchy \cdot \normale) \cdot \vw \dsurface = \int_{\Gamma_D} (\Cauchy \cdot \normale) \cdot \vS_\alpha (\vw \cdot \vS^\alpha) \, \dsurface
\end{equation}

Therefore, taking into account conditions~\eqref{eq:Stokes_GammaD2}, \eqref{eq:Stokes_sigma_n2_fs} and~\eqref{eq:Stokes_sigma_n2_ls}:
\begin{equation}
\label{eq:momentum_weak4}
\begin{array}{l}
\displaystyle \int_{\Gamma_D} (\Cauchy \cdot \normale) \cdot \vw \, \dsurface = -\int_{\Gamma_{13} \cup \Gamma_{23}} f \vv \cdot \vw \, \dsurface +  \\
\hspace{1cm} \displaystyle \int_{\Gamma_{13}} (\grad_\beta(\tensionsuperficielle_{13} \vS^\beta)) \cdot \vS_\alpha (\vw \cdot \vS^\alpha) \dsurface + \int_{\Gamma_{23}} (\grad_\beta(\tensionsuperficielle_{23} \vS^\beta)) \cdot \vS_\alpha (\vw \cdot \vS^\alpha) \, \dsurface
\end{array}
\end{equation}
The first integral of the right-hand side, expressing the Navier condition, will remain unchanged in the next developments and consequently will be denoted just by $A_f$. Note that, following expression~\eqref{eq:curvature_tensor}, on the right-hand side of~\eqref{eq:momentum_weak4}, the derivatives of tangent vectors $\vS^\beta$ are proportional to the normal vector, and consequently perpendicular to $\vS_\alpha$. Moreover, by definition of the covariant and contravariant bases, recall that $\vS^\beta \cdot \vS_\alpha = \delta^\beta_\alpha$. Ergo, Equation~\eqref{eq:momentum_weak4} can be simplified in
\begin{equation}
\label{eq:momentum_weak5}
\int_{\Gamma_D} (\Cauchy \cdot \normale) \cdot \vw \dsurface = \sum_{i=1}^2 \int_{\Gamma_{i3}} (\grad_\alpha\tensionsuperficielle_{i3}) (\vw \cdot \vS^\alpha) \, \dsurface - A_f
\end{equation}

Taking into account that $(\grad_\alpha\tensionsuperficielle_{i3}) (\vw \cdot \vS^\alpha) = \grad_\alpha(\tensionsuperficielle_{i3} \, \vw \cdot \vS^\alpha) - \tensionsuperficielle_{i3} \grad_\alpha (\vw \cdot \vS^\alpha)$, and that $\grad_\alpha (\vw \cdot \vS^\alpha) = \vS^\alpha \cdot \grad_\alpha \vw$, since $\vw \cdot \normale = 0$ on $\Gamma_D$ and $\grad_\alpha \vS^\alpha = \kappa \normale$ by virtue of Equation~\eqref{eq:curvature_tensor}, Gauss's theorem~\eqref{eq:Gauss_thm} leads to
\begin{equation}
\label{eq:momentum_weak6}
\int_{\Gamma_D} (\Cauchy \cdot \normale) \cdot \vw \dsurface = \sum_{i=1}^2 \left(\int_{\partial \Gamma_{i3}} \tensionsuperficielle_{i3} (\vw \cdot \vS^\alpha) \nu_\alpha \, dl -  \int_{\Gamma_{i3}} \tensionsuperficielle_{i3} \vS^\alpha \cdot \grad_\alpha \vw \, \dsurface \right) - A_f
\end{equation}

Similarly, we can rewrite the interface integral involved in expression~\eqref{eq:momentum_weak2}. The jump condition~\eqref{eq:Stokes_sigma_n_jump}, decomposition of $\vw$ into normal and tangential components, and Gauss's theorem~\eqref{eq:Gauss_thm}, imply
\begin{equation}
\label{eq:momentum_weak7}
-\int_{\interface_{12}} [\Cauchy \cdot \normale]_{|\interface_{12}} \cdot \vw \, \dsurface =
\int_{\partial \interface_{12}} \tensionsuperficielle_{12} (\vS^\alpha \cdot \vw) \nu_\alpha \, dl -
\int_{\interface_{12}} \tensionsuperficielle_{12} \vS^\alpha \cdot \grad_\alpha\vw \, \dsurface
\end{equation}

Hence, using expressions~\eqref{eq:momentum_weak6} and~\eqref{eq:momentum_weak7}, including Neumann condition~\eqref{eq:Stokes_sigma_n1}, relation~\eqref{eq:momentum_weak2} can be expressed as
\begin{equation}
\label{eq:momentum_weak8}
\begin{array}{l}
\displaystyle \int_\domain \Cauchy : \grad \vw \, \dvolume = \int_\domain \vb \cdot \vw \, \dvolume - \int_{\Gamma_N} p_{ext} \vw \cdot \normale \, \dsurface - \int_{\Gamma_{13} \cup \Gamma_{23}} f \vv \cdot \vw \, \dsurface \\
\hspace{1cm} + \displaystyle \sum_{i,j>i} \left(\int_{\partial \interface_{ij}} \tensionsuperficielle_{ij} (\vS^\alpha \cdot \vw) \nu_\alpha \, dl - \int_{\interface_{ij}} \tensionsuperficielle_{ij} \vS^\alpha \cdot \grad_\alpha\vw \, \dsurface \right)
\end{array}
\end{equation}

To conclude, first note that $\vS^\alpha  \cdot \grad_\alpha \vw$ is nothing but the surface divergence of the test function $\vw$, which can be rewritten $(\identity - \normale \otimes \normale) : \grad \vw$ in dyadic notation, where $\identity$ is the identity tensor and $\identity - \normale \otimes \normale$ is the projection operator onto the tangent plane of normal $\normale$~\cite{buscaglia_variational_2011, sprittles_finite_2013, pino_munoz_finite_2013} (further details are given in~\ref{appendix:A}). Second, triple junction condition~\eqref{eq:triple_junction2} can be included in~\eqref{eq:momentum_weak8} as a Neumann condition.

Consequently, the variational formulation corresponding to the momentum equation~\eqref{eq:momentum} with conditions~\eqref{eq:Stokes_sigma_n_jump}-\eqref{eq:Stokes_GammaD} reads:
\begin{equation}
\label{eq:weak_form2}
\begin{array}{l}
\displaystyle \int_\domain \Cauchy : \grad \vw \, \dvolume = \int_\domain \vb \cdot \vw \, \dvolume - \int_{\Gamma_{13} \cup \Gamma_{23}} f \vv \cdot \vw \, \dsurface + \int_{\partial \Gamma} \vh  \cdot \vw \, dl  \\ \hspace{2cm} \displaystyle - \sum_{i, j>i} \int_{\interface_{ij}} \tensionsuperficielle_{ij} (\identity - \normale \otimes \normale) : \grad \vw \, \dsurface  - \int_{\Gamma_N} p_{ext} \vw \cdot \normale \, \dsurface
\end{array}
\end{equation}
for any test function $\vw \in H^1(\domain)^d$ with $\vw \cdot \normale = 0$ on $\Gamma_D$. Note that this formulation can still be derived (in even a more direct way) when the Dirichlet condition $\vv \cdot \normale = 0$ is not considered over the rigid substrate.


\subsection{Velocity-pressure mixed variational formulation}

Finally, the mixed variational form of the Stokes' system~\eqref{eq:Stokes_momentum}-\eqref{eq:Stokes_GammaD}, expressed in terms of velocity and pressure, is obtained by substituting $\Cauchy$ for the fluid constitutive law in weak momentum  equation~\eqref{eq:weak_form2}, and completing the system with the weak form of the incompressibility constraint.

Defining $\tensionsuperficielle$ as a global surface tension parameter, equal to $\tensionsuperficielle_{ij}$ on $\Gamma_{ij}$, the mixed form is

Find $(\vv, p) \in H^1(\domain)^d \times L^2(\domain)$, with $\vv \cdot \normale = 0$ on $\Gamma_D$, such that
\begin{equation}
\label{eq:weak_form3}
\begin{array}{l}
\displaystyle \int_\domain 2\eta \epp(\vv) : \epp(\vw) \, \dvolume + \int_{\Gamma_{13} \cup \Gamma_{23}} f \vv \cdot \vw \, \dsurface - \int_{\partial \Gamma} \vh(\vv) \cdot \vw \, dl  - \int_\domain p \opdiv \vw \, \dvolume = \\ \displaystyle \hspace{1cm} \int_\domain \vb \cdot \vw \, \dvolume - \int_{\Gamma_N} p_{ext} \vw \cdot \normale \, \dsurface - \int_{\Gamma_{12}\cup\Gamma_{13}\cup\Gamma_{23}} \tensionsuperficielle (\identity - \normale \otimes \normale) : \grad \vw \, \dsurface  \\
\vspace{0.1cm}
\displaystyle \int_\domain q \opdiv \vv \, \dvolume = 0
\end{array}
\end{equation}
for any two test functions $(\vw,q) \in H^1(\domain)^d \times L^2(\domain)$, with $\vw \cdot \normale = 0$ on $\Gamma_D$.

The dissipation term $\vh$ is usually related to the velocity, and is consequently on the left-hand side of Equation~\eqref{eq:weak_form3}. When we will take into account this term, a simple relation of proportionality will be assumed,
\begin{equation}
\label{eq:dissiative_term}
\vh = -\xi \vv
\end{equation}
where $\xi$ is the proportionality constant.

Finally, it has to be emphasised that in the 2D-case shown in Figure~\ref{fig:modele}, the triple junction is a point, $\partial \Gamma \equiv \{P\}$, and the integral over the triple junction for this reason is calculated by evaluating the integrand at this point: $\int_{\partial \Gamma} \vh  \cdot \vw \, dl = \vh(P) \cdot \vw(P)$.

\section{Computational strategy}
\label{sec:computation}

The finite element discretisation of mechanical problem~\eqref{eq:weak_form3} is first formulated. Next, a front capturing method is presented, and the time-stepping strategy given.

The whole computational domain $\domain$ is discretised by an unstructured mesh $\maillage$ made up of elements $K$, triangles in $2D$, or tetrahedrons in $3D$, whose characteristic size is denoted by $h_K$. Analogously, denoting by $t_f > 0$ the time when the simulation stops, the time interval $[0,t_f]$ is discretised by a set of points $\{t_n\}_{n=0,\cdots,N}$ satisfying $t_0 = 0$, $t_N = t_f$ and $t_{n+1} > t_n$. In our simulations, these points are assumed to be uniformly distributed. The time step is then defined as $\Delta t = t_{n+1} - t_n$.

\subsection{Finite element discretisation}

\subsubsection{Choice of the FE spaces}

Both velocity and pressure fields are approximated by continuous piecewise-linear functions, $\vv_h$ and $p_h$ ($P_1$-approximation). However, such linear - linear elements do not fulfil the inf-sup condition of the Brezzi - Babu\u{s}ka theory~\cite{brezzi_mixed_1991} and hence the corresponding discrete system is unstable. The FE formulation is subsequently stabilised by using the Variational Multiscale Method (VMS) framework, formalised by Hughes~\cite{hughes_multiscale_1995,hughes_variational_1998}, and more precisely the Algebraic SubGrid Scale (ASGS) method introduced and analysed by Codina~\cite{codina_stabilized_2001}, which can circumvent the inf-sup condition. VMS methods decompose the unknown fields of a problem, into a computable component, typically the FE solution, and an uncomputable component, belonging to the fine or subgrid scale. For the Stokes' problem, this decomposition is applied only to the velocity: $$ \vv = \vv_h + \tilde{\vv} $$ Using ASGS, the subgrid component $\tilde{\vv}$ is approximated by a quantity proportional to the residue obtained by replacing $\vv$ and $p$ by $\vv_h$ and $p_h$ in Equation~\eqref{eq:Stokes_momentum}. Since $\vv_h$ is piecewise linear, its second-order derivatives vanish, and the residue is just $-\vb_h + \grad p_h$. Hence, on a mesh element $K$, the ASGS method furnishes the following expression of the fine-scale velocity,
\begin{equation}
\label{eq:fine_scale}
\tilde{\vv}_{|K} = -\tau_K(-\vb_h + \grad {p_h}_{|K}), \;\; \mbox{ for any } K \in \maillage
\end{equation}
where $\tau_K$ is the proportionality constant such that $\tau_K^{-1}$ ``approximates'' the effect of the Laplace operator.

\noindent Finally, substituting $\vv$ for $\vv_h + \tilde{\vv}$ in the second equation of~\eqref{eq:weak_form3}, integrating by parts the subgrid term and neglecting the resulting boundary integral~\cite{codina_stabilized_2001,abouorm_comparaison_2014}, provides the additional term required to stabilise the FE formulation,
\begin{equation}
\label{eq:stabilisation}
\sum_K \int_K \tau_K (-\vb + \grad p_h) \cdot \grad q_h \, dK
\end{equation}
where the symbol $\sum_K$ stands for the summation over all the mesh elements $K$, while $q_h$ are the discrete test functions associated to the pressure (in practice, the piecewise linear hat functions). The simulations presented here have been carried out with a stabilisation parameter $\tau_K$ equal to
\begin{equation}
\label{eq:parametre_stab}
\tau_K = \beta_K h_K^2
\end{equation}
where $h_K$ is the characteristic size of element $K$. The parameter $\beta_K$ is equal to $1/40\eta_1$, with $\eta_1$ the liquid viscosity, if element $K$ is crossed by the interface $\Gamma_{12}$, and otherwise to $1/2\eta$, with $\eta$ equal to the liquid or gas viscosity. The pressure enrichment technique used in this work and described in subsection~\ref{sub:enrichment}, requires that elemental matrices of the form~\eqref{eq:stabilisation} must be invertible in the elements crossed by $\Gamma_{12}$. That is why $\beta_K$ can not be set to zero in such elements, but instead is considered as small as possible.

\subsubsection{Semi-implicit discretisation of the surface tension term}
\label{sec:semi_implicit}

According to Brackbill~\cite{Brackbill_1992}, ``the explicit treatment of surface tension is stable when the time step resolves the propagation of capillary waves'', while an ``implicit treatment of surface tension would remove this constraint''. In our context, capillary waves are spurious waves of short wavelengths (see Figure~\ref{fig:oscillations}), initiated by small local motions of the moving interface $\Gamma_{12}$ due to discretisation errors as explained in subsection~\ref{sub:enrichment}. However, Denner and van Wachen have deeply revisited the subject in a recent paper~\cite{Denner201524}. They claim that if the capillary time-step constraint ``is violated, even a numerically unconditionally stable differencing scheme may lead to inaccurate results and to unsustainable numerical errors''. Using a Volume Of Fluid methodology with a Continuum Surface Force approach, they compare an explicit treatment of surface tension (\textit{i.e.} the surface force remains constant during the resolution of fluid equations) with a fully implicit treatment, \textit{i.e.} fluid equations and interface advection equation are solved through a unique algebraic system. Among other results, their numerical simulations show that when time-step restriction is not satisfied, initial perturbations are unphysically amplified regardless of the surface tension treatment.

However, satisfying such a rigid restriction, implies unacceptable small time steps, especially when considering the very small mesh size provided by the adaptation technique used in this work. This difficulty can be overcome by adopting what is called a semi-implicit treatment of the surface tension term $\int_{\Gamma_{12}} \tensionsuperficielle (\identity - \normale \otimes \normale):\grad\vw \, \dsurface$ , following the work of B{\"a}nsch~\cite{bansch_finite_2001,bansch_finite_2005}. This widely used treatment, detailed for example in~\cite{buscaglia_variational_2011}, computes the projection matrix $(\identity-\normale\otimes\normale)$ of the surface tension term at time $t_n$, using a prediction of the interface at time $t_{n+1}$. That leads to an additional term on the left-hand side of Equation~\eqref{eq:weak_form3},
\begin{equation}
\label{eq:implicite}
\Delta t \int_{\Gamma_{12}} \gamma \left(\grad \vv \cdot (\identity - \normale\otimes\normale)\right) : \grad \vw \, \dsurface,
\end{equation}
corresponding to the surface Laplacian of the displacement $\Delta t \vv$, see~\ref{appendix:A}. In fact, according to the work of Denner and van Wachen and their conclusions exposed in~\cite{Denner_conf14} on this semi-implicit technique, the additional interfacial shear stress~\eqref{eq:implicite}  that dissipates surface energy, is most effective at shortest wavelength and consequently avoids accumulation of small numerical errors. Hence, the capillary time-step constraint can be overcome not because considering a semi-implicitation technique, but because adding a numerical dissipative term.

\subsubsection{Discrete mixed variational formulation}

Considering the previous sections, and denoting by $P_1(\domain_h)$ the set of scalar functions, continuous on $\domain_h$ and linear on each mesh element $K$, the discrete form of the Stokes' problem~\eqref{eq:weak_form3} reads:
Find $(\vv_h, p_h) \in P_1(\domain_h)^d \times P_1(\domain_h)$, with $\vv_h \cdot \normale = 0$ on $\Gamma_D$, such that
\begin{equation}
\label{eq:weak_discrete}
\begin{array}{l}
\displaystyle \int_{\domain_h} 2\eta_h \epp(\vv_h) : \epp(\vw_h) \, \dvolume + \Delta t \int_{{\Gamma_h}_{12}} \gamma_h \left(\grad \vv_h \cdot (\identity - \normale_h\otimes\normale_h)\right) : \grad \vw_h \, \dsurface \\ \displaystyle + \int_{{\Gamma_h}_{13} \cup {\Gamma_h}_{23}} f \vv_h \cdot \vw_h \, \dsurface - \int_{\partial \Gamma_h} \vh(\vv_h)  \cdot \vw_h \, dl  - \int_{\domain_h} p_h \opdiv \vw_h \, \dvolume = \\ \displaystyle \hspace{1cm} \int_{\domain_h} \vb_h \cdot \vw_h \, \dvolume - \int_{{\Gamma_h}_N} p_{ext} \vw_h \cdot \normale_h \, \dsurface \\ \displaystyle \hspace{3.5cm} - \int_{{\Gamma_h}_{12}\cup{\Gamma_h}_{13}\cup{\Gamma_h}_{23}} \tensionsuperficielle_h (\identity - \normale_h \otimes \normale_h) : \grad \vw_h \, \dsurface  \\  \\
\displaystyle \int_{\domain_h} q_h \opdiv \vv_h \, \dvolume + \sum_K \int_K \tau_K \grad p_h \cdot \grad q_h \, dK = \sum_K \int_K \tau_K \vb_h \cdot \grad q_h \, dK
\end{array}
\end{equation}
for any couple of test functions $(\vw_h,q_h) \in P_1(\domain_h)^d \times P_1(\domain_h)$, with $\vw_h \cdot \normale = 0$ on $\Gamma_D$.

Some remarks can be made concerning this discrete formulation:
\begin{itemize}
\item Interface $\Gamma_{12}$ is approximated by ${\Gamma_h}_{12}$, a set of continuous segments (in 2D) or planes (in 3D), as detailed in~\cite{pino_munoz_finite_2013}.
\item Discrete viscosity $\eta_h$ is equal to the liquid or air viscosity. When an element is cut by the interface ${\Gamma_h}_{12}$, this element is  divided into sub-elements accordingly. The parameter $\eta_h$ is then evaluated on each sub-element.

\item Surface or line integrals are evaluated in a standard manner by applying Gaussian quadrature rules, over ${\Gamma_h}_{12}$, ${\Gamma_h}_{13}$, ${\Gamma_h}_{23}$ and $\partial \Gamma_h$ (a set of segments or a point). Furthermore, the discrete normal $\normale_h$ is taken piecewise constant.

\item The discrete surface tension parameter $\tensionsuperficielle_h$ is equal to $\gamma_{ij}$ on ${\Gamma_h}_{ij}$. The triple junction lies on a boundary of the computational domain, as shown in Figure~\ref{fig:modele}. Consequently, a face cut by the triple junction, is divided into sub-faces (segments or triangles) accordingly. The parameter $\tensionsuperficielle_h$ is then evaluated on each sub-face.
\end{itemize}

\subsection{Interface capturing method}
\label{sec:interface}
The liquid-gas moving interface $\Gamma_{12} \equiv \Gamma_{12}(t)$ is described via a Level-Set methodology~\cite{sethian_level_1999}. Let $\alpha(\vx,t) : \domain\times \R^+ \mapsto \R$ be the level-set function, that is a continuous function positive on one side of $\Gamma_{12}$ and negative on the other. The surface $\Gamma_{12}$ is thus identified with the zero-isovalue of $\alpha$: $$ \Gamma_{12}(t) \equiv \{\vx \in \domain \, ; \, \alpha(\vx,t) = 0\} $$

At time $t=0$, $\alpha(\vx,0)$ is set equal to the signed distance to a given initial interface $\Gamma_{12}(0)$.  Assuming the velocity field $\vv$ known at every time $t$, the change over time in the function $\alpha$ is then provided by the usual transport equation
\begin{equation}
\label{eq:transport}
\frac{\partial \alpha}{\partial t} + \vv \cdot \grad \alpha = 0
\end{equation}

Computationally, $\alpha$ is approximated by $\alpha_h$, a continuous piecewise-linear function. Transport equation~\eqref{eq:transport} is time-discretised using an implicit Eulerian scheme, written as
\begin{equation}
\label{eq:transport2}
\frac{1}{\Delta t} \alpha_h(\vx,t_{n+1}) + \vv_h(\vx,t_n) \cdot \grad \alpha_h (\vx,t_{n+1}) = \frac{1}{\Delta t} \alpha_h(\vx,t_n)
\end{equation}

Since the Galerkin discretisation method is unstable when applied to first-order hyperbolic equation~\eqref{eq:transport}, a standard Streamline Upwind Petrov-Galerkin method is used to obtain a stable FE formulation. In order to preserve the distance property of $\alpha_h$, that is $\|\grad \alpha_h\| = 1$, throughout the simulation, a reinitialisation step is performed at each time increment. More precisely, the methodology~\cite{shakoor_efficient_2015} by Shakoor and coauthors, rebuilds node-to-node the signed distance to ${\Gamma_h}_{12}$, in a given width around the interface.
%

\subsection{Pressure enrichment}
\label{sub:enrichment}
A characteristic and critical issue arising when simulating a bifluid system with surface tension forces, is what is called spurious or parasitic currents~\cite{Gross200740,gross_finite_2007,Zahedi_2013,Magnini20166811}. Such currents are non-physical velocities, located in the neighbourhood of the interface, with an intensity inversely proportional to the capillary number. They are due to the discrete form of the surface tension term and to the approximation of the discontinuous pressure with a continuous field. If not controlled, they disturb the interface evolution, leading possibly to a significant loss of mass, and even to the interface degradation.

In order to capture the pressure discontinuities, and consequently reduce these parasitic currents, the discrete pressure space is enriched following the method proposed by Ausas and coauthors in reference~\cite{ausas_new_2012}. In an element $K \in \maillage$ cut by the liquid-air interface, this approach writes the discrete pressure as
\begin{equation}
\label{eq:pressure_enrichment}
p_h(\vx) = \sum_{i=1}^{d+1} p_i N_i(\vx) + C_1 M_1(\vx) + C_2 M_2(\vx), \;\; \forall \vx  \in K
\end{equation}
where the $N_i$ are the usual $d+1$ linear or ``hat'' shape functions, $p_i$ the associated pressure nodal values, while $M_1$ and $M_2$ are the two additional enrichment functions, local to the element, and $C_1$, $C_2$ the associated degrees of freedom. However, since these additional degrees are local to each element, they can be eliminated at the elementary level before the final assembly. This condensation provides a ``velocity-velocity'' matrix term, to be added to the left-hand side of the first equation in discrete Stokes' system~\eqref{eq:weak_discrete}. Consequently, the pressure $p_h$ remains piecewise linear. All the details of this method are given in~\cite{ausas_new_2012}. Only the expressions of $M_1$ and $M_2$ are given:
\begin{equation}
M_1(\vx) = (1-S(\vx)) \chi_+(\vx) \; \mbox{ and } \; M_2(\vx) = S(\vx) (1-\chi_+(\vx)),
\end{equation}
where $\chi_+$ is equal to 1 on one side of the interface cutting element $K$, and to zero on the other side. Furthermore,
\begin{equation}
\label{eq:S}
S(\vx) = \sum_{i\in\mathcal{S}_+} N_i(\vx)
\end{equation}
Denoting by $\bfm{s}_i$, $i=1,\cdots,d+1$, the vertices of element $K$, the set of indices $\mathcal{S}_+$ is such that $\alpha_h(\bfm{s}_i) > 0$ for any $i \in \mathcal{S}_+$.

Finally, note: when the body forces $\vb$ are not neglected, as in subsection~\ref{sec:capillary_rise}, they generally present a discontinuity across the liquid-gas interface. This implies that the pressure gradient is discontinuous too. When taking into account only a pressure piecewise linear approximation, without enrichment~\eqref{eq:pressure_enrichment}, the corresponding gradient can be discontinuous only from-element-to-element, but not inside a mesh element, leading again to spurious velocities as analysed in reference~\cite{Gerbeau1997}. As for the pressure discontinuity, several methods have been developed in the literature for this purpose: using a XFEM technique in~\cite{Gross200740}, building an auxiliary function to account for the discontinuity in~\cite{Discacciati2013}, enriching the pressure shape functions on the elements cut by the interface in~\cite{Owen2005}. Nevertheless, in this work and for the values of $\vb$ used in subsection~\ref{sec:capillary_rise}, numerical simulations show that no enrichment other than~\eqref{eq:pressure_enrichment} is needed.

\subsection{Anisotropic mesh adaptation strategy}

Mesh adaptation can be used, especially in 3D-cases, in order to enhance the approximation of the pressure, velocity and level-set fields, while controlling the number of elements. The technique we applied is detailed by Mesri and coauthors in~\cite{mesri_2008}, based on an \textit{a posteriori} error analysis. Without developing this aspect in detail too much, this analysis depends on the reconstruction of the Hessian of the field with respect to which the mesh is then adapted. Moreover, this takes into account both a target error and a target of number of nodes. Further considerations on this subject can also be found in the paper of Coupez~\cite{coupez_metric_2011}, or in the work proposed by Hachem~\cite{Hachem2016238} showing an application to two-phase compressible-incompressible flows.

In our simulations, the mesh is adapted with respect to two variables: the pressure and a smoothed delta function $\delta_\varepsilon$ defined by
\begin{equation}
\label{eq:delta}
\delta_\varepsilon(\vx) = \left\{\begin{array}{ll} \displaystyle
                                 \frac{1}{2\varepsilon}\left(1 + \cos\left(\frac{\alpha_h(\vx)\pi}{\varepsilon}\right)\right) & \displaystyle \, \mbox{ if } \, |\alpha_h(\vx)| \leq \varepsilon \\ 0 & \, \mbox{ otherwise }
                                \end{array} \right.
\end{equation}
where $\varepsilon$ is the width of the zone in which the mesh is adapted around the interface. The error analysis conducted with these two variables provides two metric tensors specified at each node of the current mesh. The metric intersection procedure, defined in~\cite{alauzet_3d_2007} is subsequently used to obtain the final metric tensor. This metric is defined by the largest ellipsoid included into the intersection of the ellipsoids associated with the two metric tensors.

This strategy typically results in the anisotropic mesh of Figure~\ref{fig:droplet_maillage}, with elements stretched in the direction perpendicular to the pressure gradient. Therefore, the mesh is well-adapted to capture the variation of the pressure across the interface. However, this does not mean that mesh adaptation alone can prevent parasitic currents, since pressure remains approximated with a continuous field. Additionally, the description of the liquid-air interface is improved as well, since the mesh is refined in its vicinity. To conclude, notice: only the level-set function must be projected onto the new adapted mesh. This is achieved by linear interpolation. In fact, due to the reinitialisation step, only the zero-isosurface must be preserved through this projection. This is ensured by the linearity of the level-set function in the neighbourhood of the interface.

\subsection{Time-stepping strategy}

The time-stepping strategy that carries out simulations is summarised as follows.

At time $t = 0$, compute the signed distance function $\alpha_h(\vx,t=0)$ with respect to a given liquid-air interface ${\Gamma_h}_{12}$. Next, repeat until reaching a given final time $t = t_f$, the steps:
\begin{itemize}
\item At time $t = t_n$, knowing $\alpha_h(\vx,t_n)$, compute $\vv_h(\vx,t_n)$ and $p_h(\vx,t_n)$ by solving Stokes' system~\eqref{eq:weak_form3}.
\item Knowing $\alpha_h(\vx,t_n)$ and $\vv_h(\vx,t_n)$, compute $\alpha_h(\vx,t_{n+1})$ by solving transport equation~\eqref{eq:transport}.

\item Reinitialisation: rebuild the signed distance with respect to ${\Gamma_h}_{12}$.

\item If necessary, mesh adaptation step.

\item $t \leftarrow t_{n+1}$
\end{itemize}

It is important to mention that the numerical tools used to solve system~\eqref{eq:weak_discrete} have been implemented in the FE library CIMLIB~\cite{digonnet_cimlib_2007,mesri_advanced_2009}. This highly parallel code has been developing for more than a decade. It is used in a wide range of scientific domains, ranging from numerical microstructure analysis, to the simulation of manufacturing or forming processes, or, again, fluid - solid interaction, both in academic and industrial contexts. CIMLIB software offers a framework with facilities in mesh adaptation and front capturing methods.

\section{Numerical simulations of capillary-driven flows}
\label{sec:simulations}
\subsection{Simulation units}

Stokes' equations do not determine uniquely the units of the different physical quantities. For example, let us introduce  characteristic length $\overline{x} = 10^{-3}$m, characteristic pressure $\overline{p} = 1$Pa, and assume that fluid viscosity is expressed in Pa.s. Looking at Table~\ref{tab:pamaters}, which provides the numerical values of the simulation parameters, it is then possible to define the characteristic time $\overline{t} = \eta_1/\overline{p} = 3\times10^{-2}$s, velocity $\overline{v} = \overline{x}/\overline{t}$ m.s$^{-1}$, and surface tension $\overline{\gamma} = \overline{p} \times \overline{x} = 10^{-3}$ N.m$^{-1}$.  Ergo, from now on, whatever the physical quantity $q$ considered (especially the values of surface tensions shown in Table~\ref{tab:pamaters}), it is always the dimensionless part $q^*$ which is given. The physical value of $q$ is then equal to $q^* \overline{q}$.

\subsection{Numerical tests: droplet spreading 2D-simulations}
\label{sec:test}
The ability to simulate wetting problems by using the previous numerical developments, is now assessed. The spreading of a droplet is investigated in a 2D-configuration, on different structured and unstructured meshes, with and without pressure enrichment. For each case, pressure, mass-loss, parasitic currents, contact angle, etc., are evaluated.

\begin{table}[!hb]
\centering \begin{tabular}{|c|c|c|c|c|c|c|c|}
\hline
$\tensionsuperficielle_{12}$ & $\tensionsuperficielle_{13}$ & $\tensionsuperficielle_{23}$ & $\xi$ & f & $\eta_1$ & $\eta_2$ & $p_{ext}$  \\
\hline
1.0 & 0.5 & 0.0 or 1.0 & 0 & 0.1 & $3.0\times 10^{-2}$ & $3.0\times 10^{-5}$ & 0\\
\hline
\end{tabular}
\caption{Parameters used in the simulations: surface tension/energy parameters, $\xi$-friction parameter, Navier-condition friction parameter, liquid viscosity, ``surrounding medium'' viscosity and external pressure. $\gamma_{23}=0$ leads to a static angle equal to 120\textdegree, while this angle is of 60\textdegree~for $\gamma_{23} = 1$.}
\label{tab:pamaters}
\end{table}

\begin{figure}[!hb]
  \begin{center}
    \subfloat[t=0]{
      \includegraphics[width=0.5\textwidth]{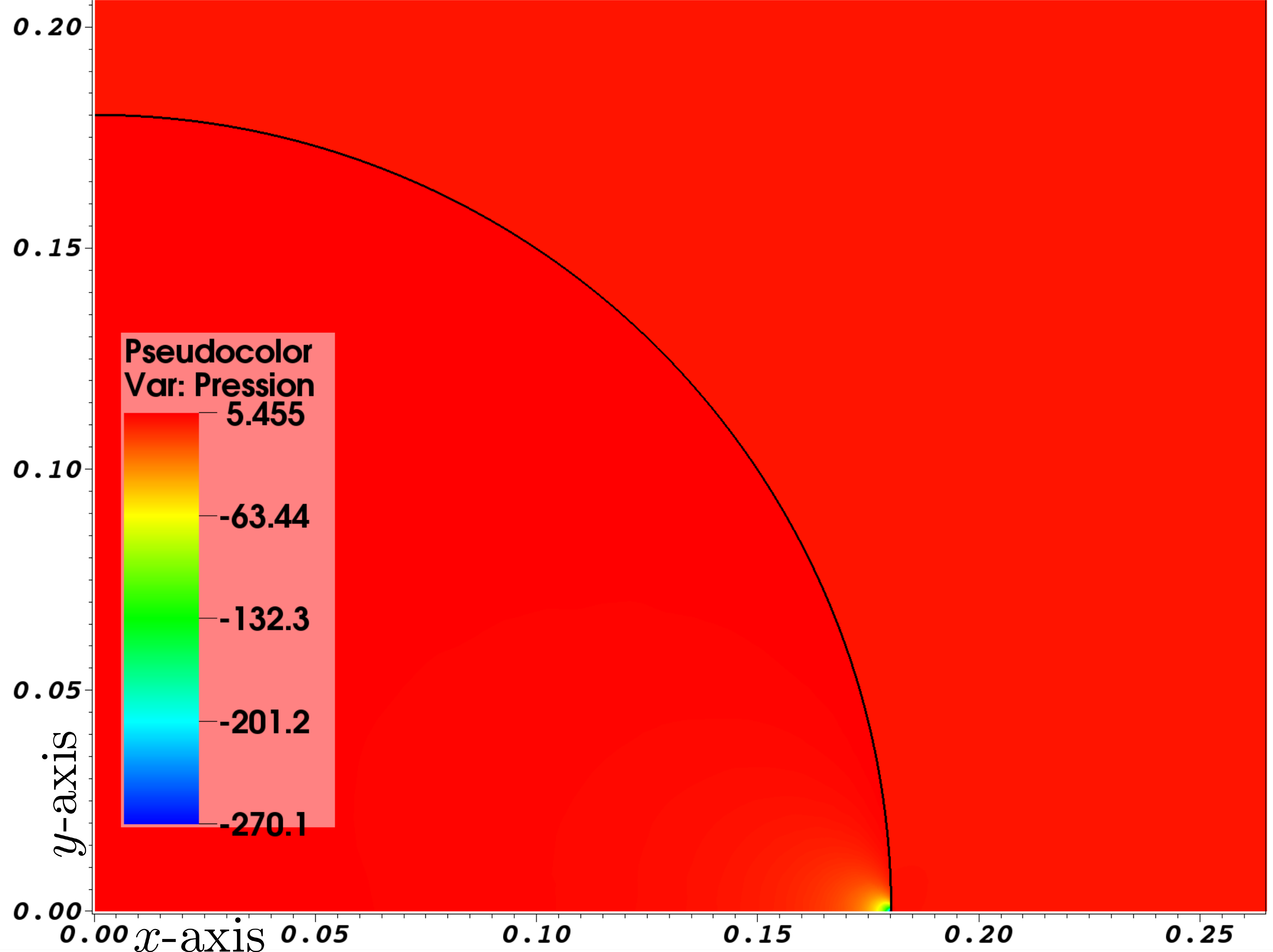}
      \label{sub:droplet_0} }
    \subfloat[$t = 1.0 \times 10^{-2}$]{
      \includegraphics[width=0.5\textwidth]{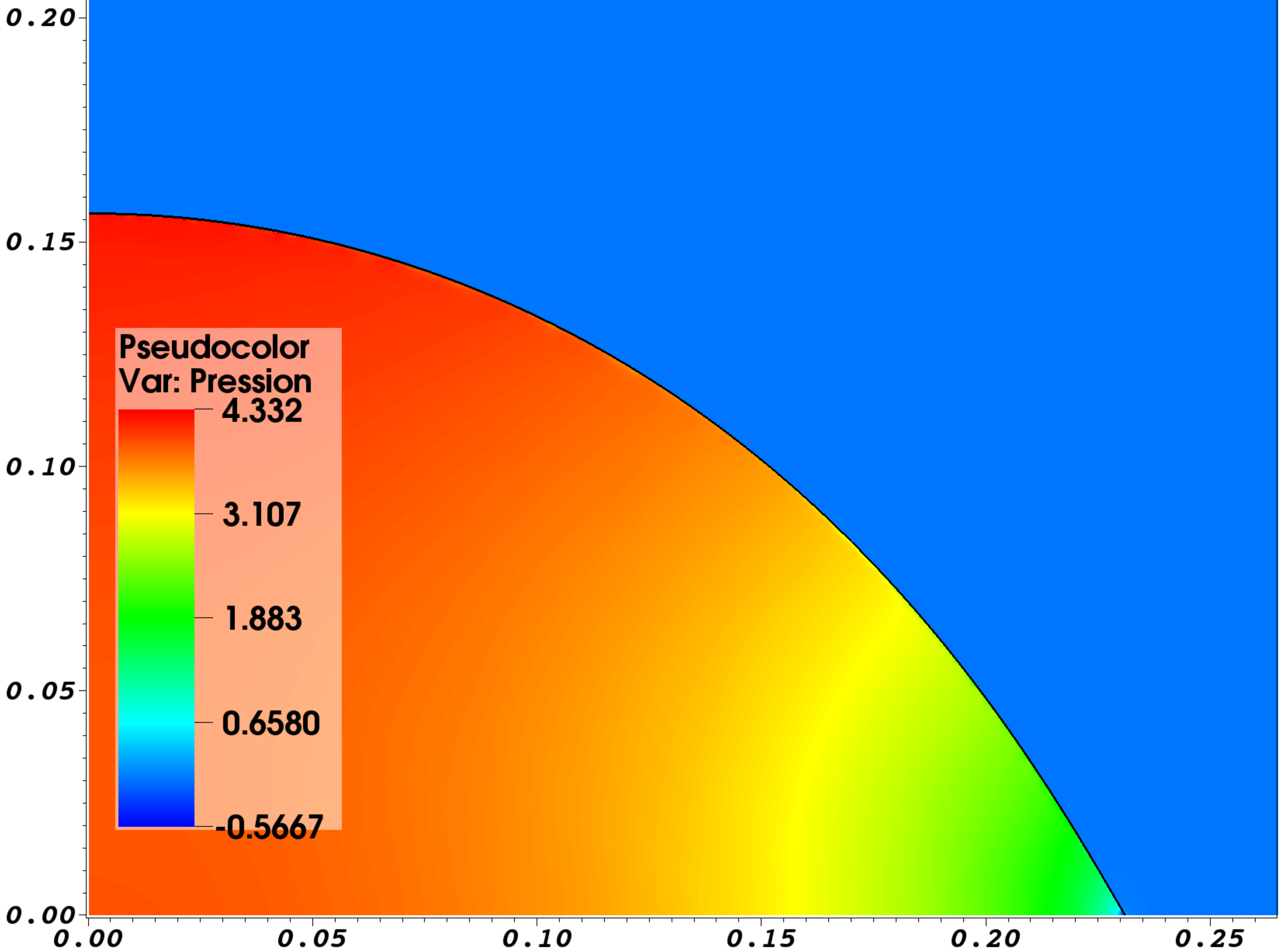}
      \label{sub:droplet_20} } \\
    \subfloat[$t = 6.0 \times 10^{-2}$]{
      \includegraphics[width=0.5\textwidth]{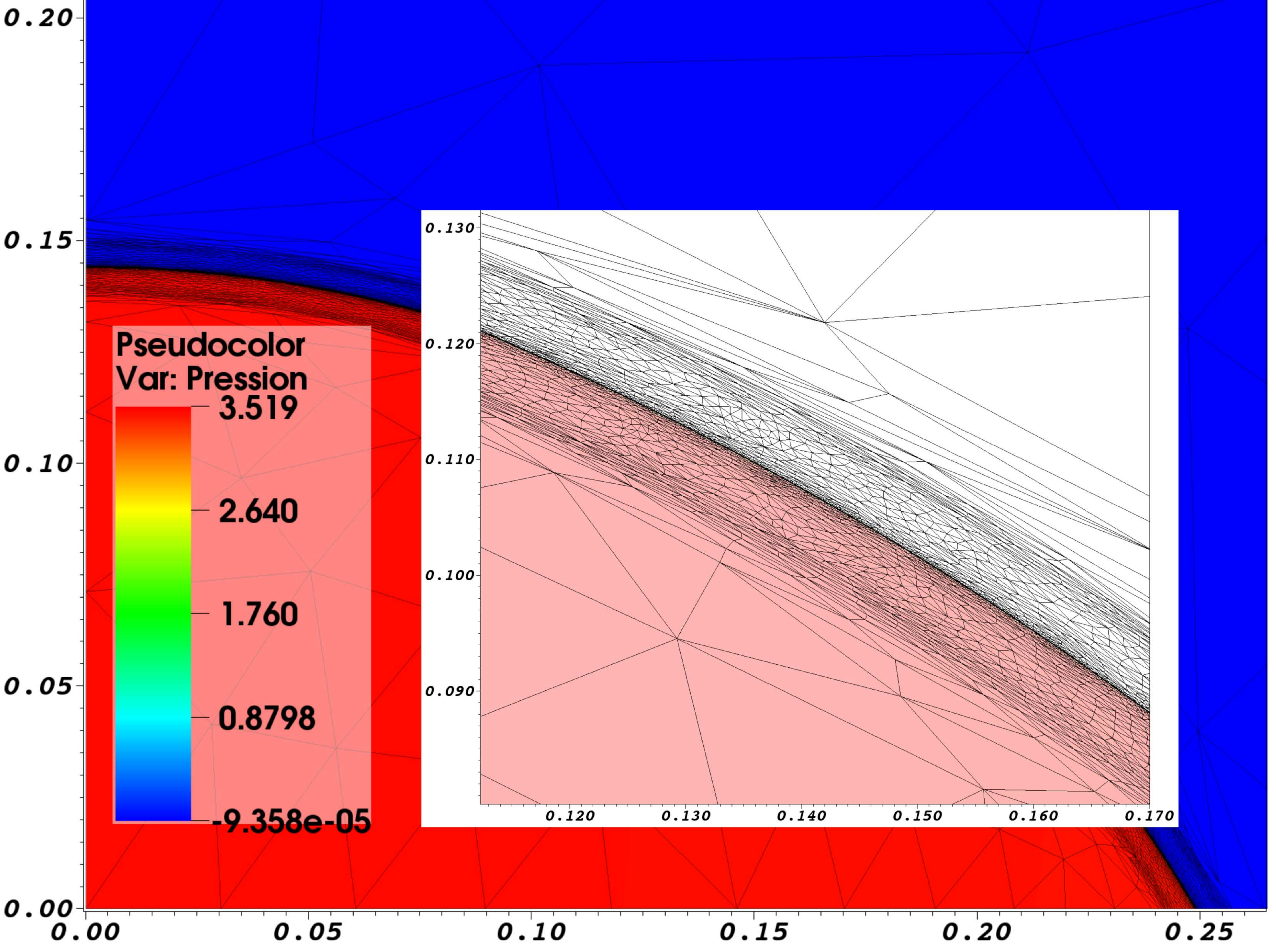}
      \label{sub:droplet_60} }
   \subfloat[$t = 1.0 \times 10^{-2}$]{
      \includegraphics[width=0.5\textwidth]{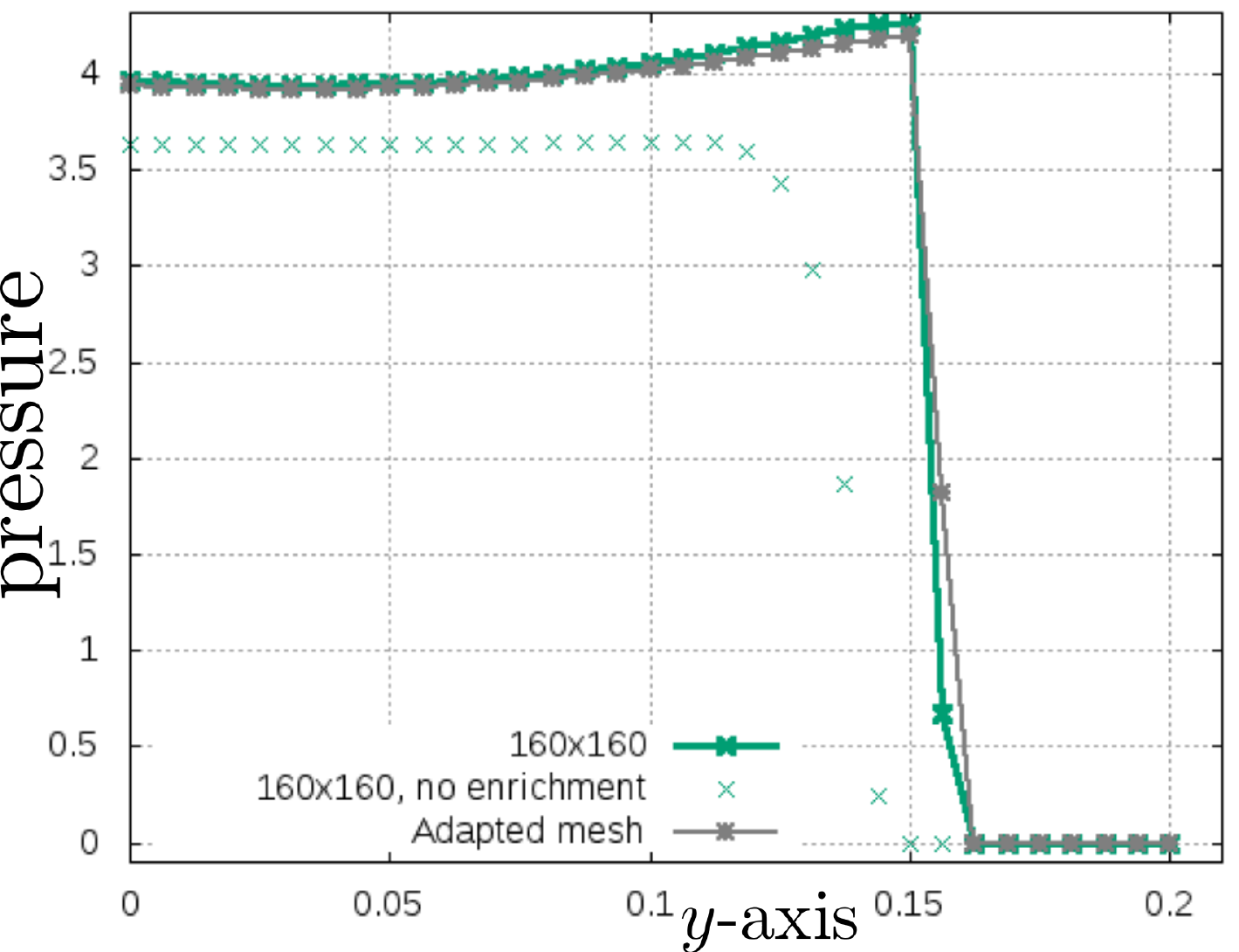}
      \label{sub:pressure} }

      \caption{2D-simulation of droplet spreading: (a)-(c) droplet profile (bold line) and pressure field at initial, intermediate and static equilibrium states, with a zoom-in on the adapted mesh in the vicinity of the interface in (c); (d) pressure plotted at $t=1.0 \times 10^{-2}$ along the $y$-axis.}
    \label{fig:droplet_maillage}
  \end{center}
\end{figure}

\begin{figure}[!hb]
  \begin{center}
    \subfloat[t=0]{
      \includegraphics[width=0.5\textwidth]{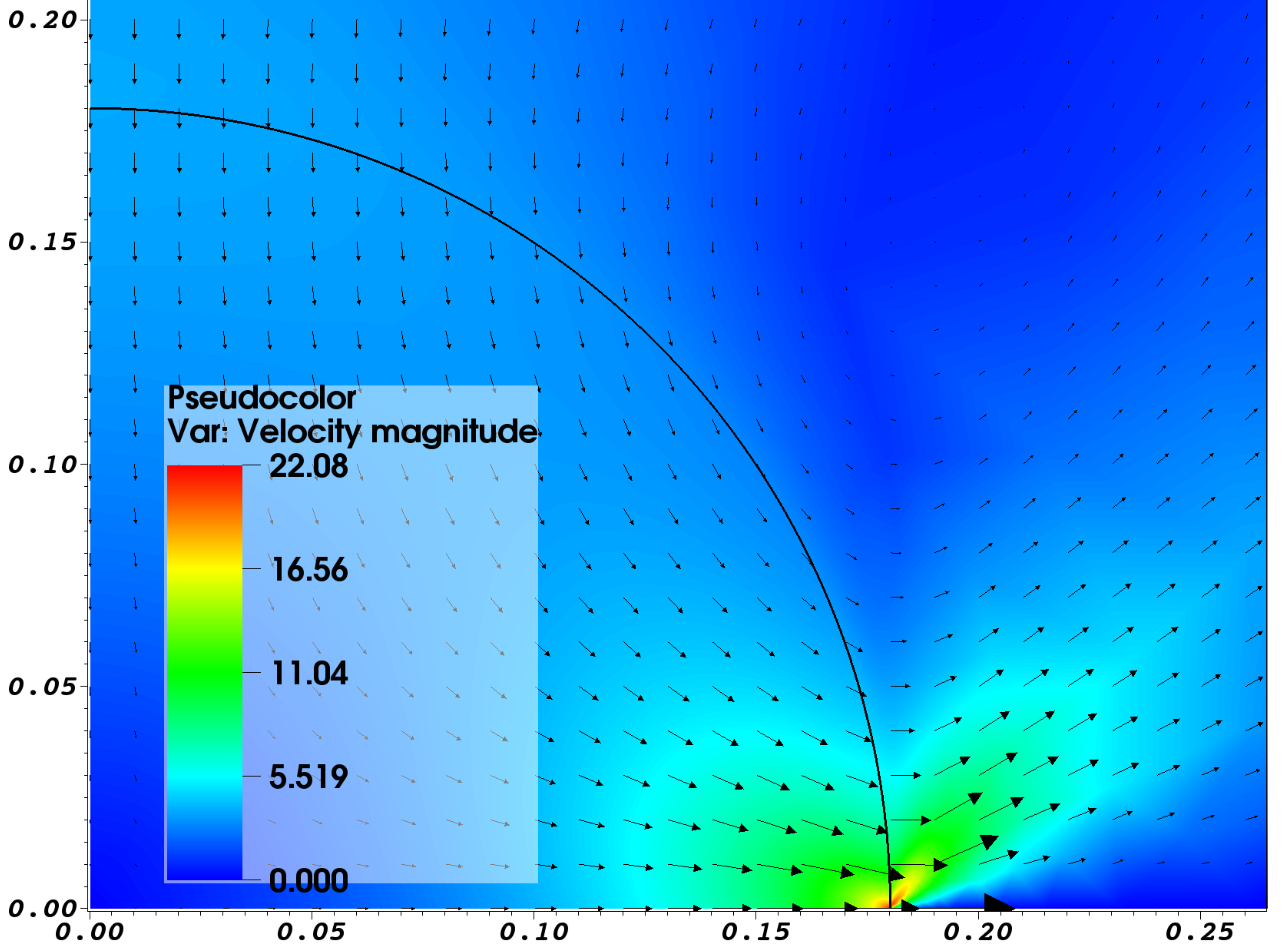}
      \label{sub:velocity_0} }
    \subfloat[$t = 1.0 \times 10^{-2}$]{
      \includegraphics[width=0.5\textwidth]{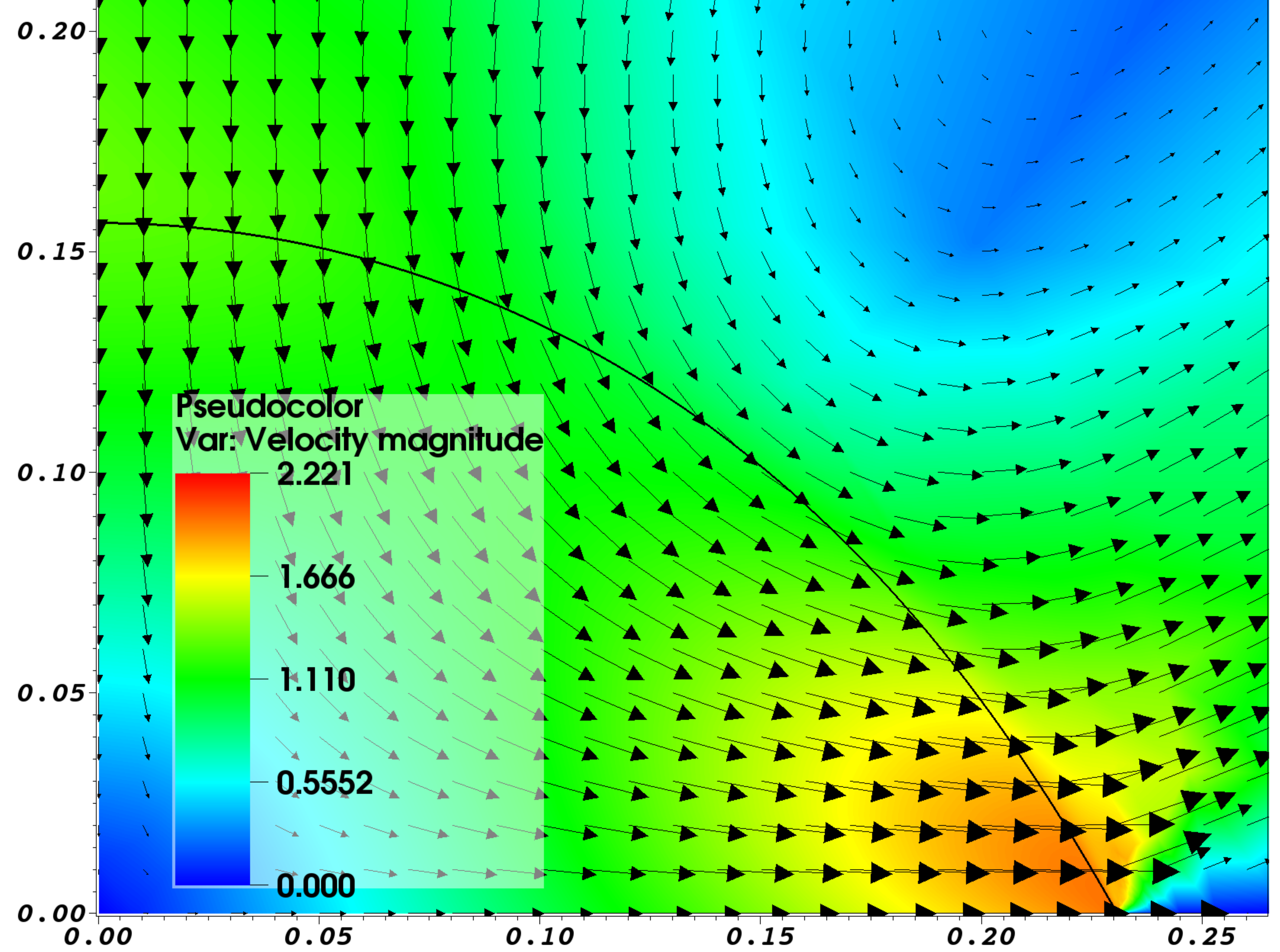}
      \label{sub:velocity_10} } \\
    \subfloat[$t = 6.0 \times 10^{-2}$]{
      \includegraphics[width=0.5\textwidth]{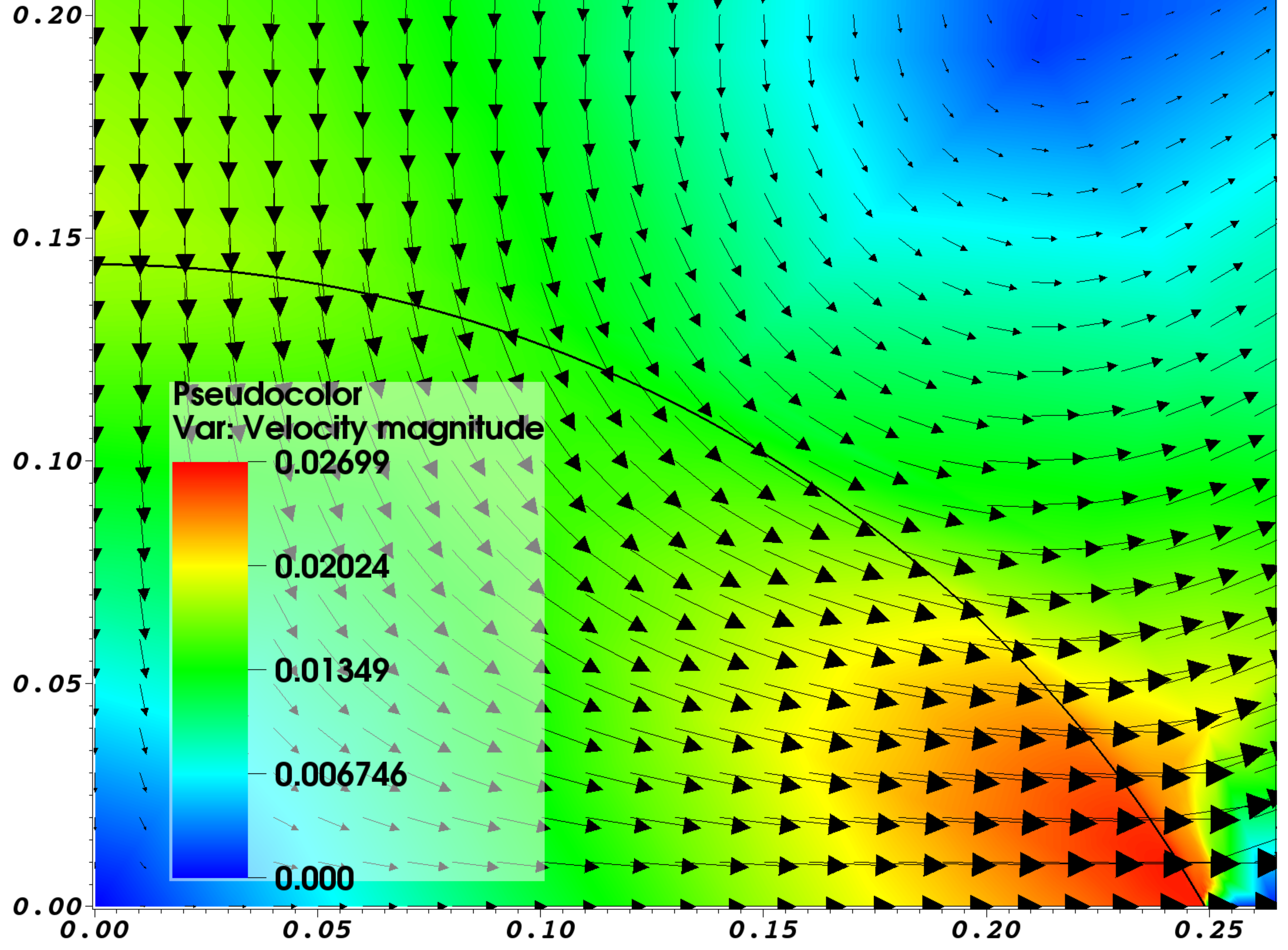}
      \label{sub:velocity_60} }
        \subfloat[Stationary state]{
      \includegraphics[width=0.5\textwidth]{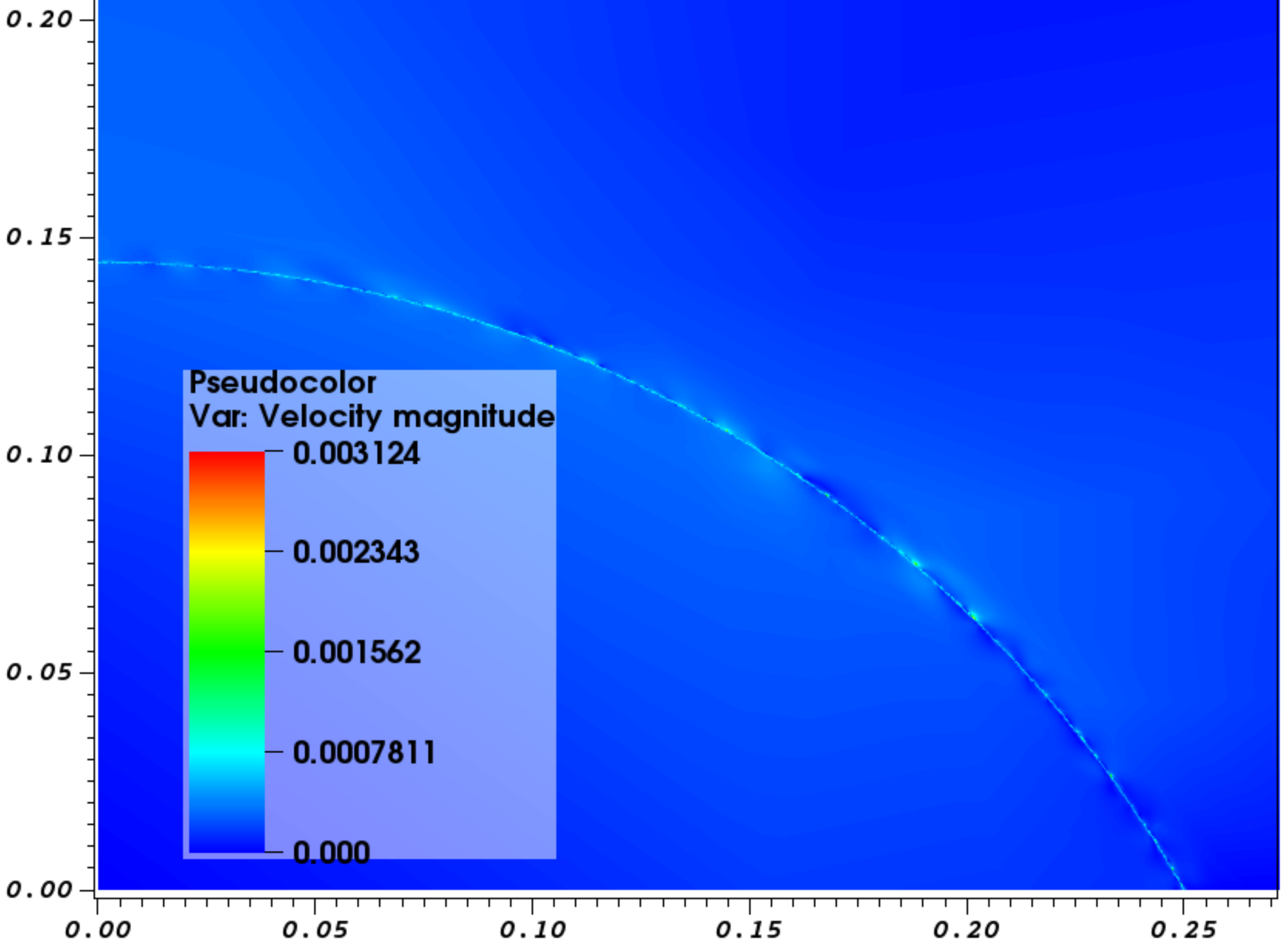}
      \label{sub:velocity_eq} }

      \caption{2D-simulation of droplet spreading: velocity magnitude field and velocity vectors (normalised and uniformly plotted through the mesh).}
    \label{fig:velocty}
  \end{center}
\end{figure}

The computational domain is the unit square, in which a liquid droplet, of initial radius 0.18, is spreading along the x-axis, as described in Figure~\ref{fig:modele}. The boundary conditions are given in this figure, the body forces are neglected ($\vb = \bfm{0}$), while the other parameters required for the simulations are provided in Table~\ref{tab:pamaters} with a time step $\Delta t = 10^{-4}$. The initial contact-angle is equal to 90\textdegree, $\gamma_{23} = 1$, and a contact-angle of 60\textdegree \, is consequently expected at equilibrium. A Navier boundary is considered with $f = 0.1$, while the coefficient $\xi$ is equal to zero. The simulations are carried out by using two structured meshes of, respectively, 80 and 160 nodes per side, as well as an unstructured mesh. To ameliorate the precision with the unstructured mesh, the adaptation step is performed every 3 time increments (with $\varepsilon = 6.0 \times 10^{-3}$ in Equation~\eqref{eq:delta}), providing a mesh of approximately 10,000 triangles and 5,000 nodes, with a mesh size ranging from $h_{min} \approx 5 \times 10^{-4}$ in the vicinity of the interface, to $h_{max} \approx 0.12$ in the surrounding medium, as shown in Figure~\ref{sub:droplet_60}. \\

The droplet profile, described through the zero-isovalue of the level-set function $\alpha_h$, is represented at initial, intermediate and static equilibrium states in Figures~\ref{sub:droplet_0}-\ref{sub:droplet_60}, with the associated pressure field $p_h$ as well. The corresponding velocity field is given in Figure~\ref{fig:velocty} both in magnitude and normalised vector form. These pictures correspond to a case carried out by combining mesh adaptation and pressure enrichment. Due to the fact that triple junction is far from equilibrium at $t=0$, a spike of pressure is observed in its vicinity during the first time-increments, as shown in Figure~\ref{sub:droplet_0}. This spike induces the velocity field of Figure~\ref{sub:velocity_0} which brings the system back to the geometrical equilibrium. Finally, at equilibrium  steady-state, droplet pressure is quasi-uniform (Figure~\ref{sub:droplet_60}), while velocity is not vanishing only because of parasitic currents located at the interface (Figure~\ref{sub:velocity_eq}). This crucial point will be further discussed.

In Figure~\ref{sub:pressure}, the pressure computed with the 160$\times$160 mesh, is plotted along the y-axis ($x=0$), from $y=0$ to $y = 0.2$, evaluated at nodes of the $160 \times 160$ meshes, with and without the enrichment. These results are superimposed on to the pressure obtained with mesh adaptation and thus evaluated at the same points. The pressure enrichment is clearly shown to improve the description of the pressure, capturing the transition between liquid and air pressures. Note that, due to the dramatically small mesh size, the pressure discontinuity is always well-captured with the mesh adaptation, with or without enrichment.

\begin{figure}[!hbt]
\begin{center}
 \subfloat[Relative change in surface area]{
      \includegraphics[width=0.5\textwidth]{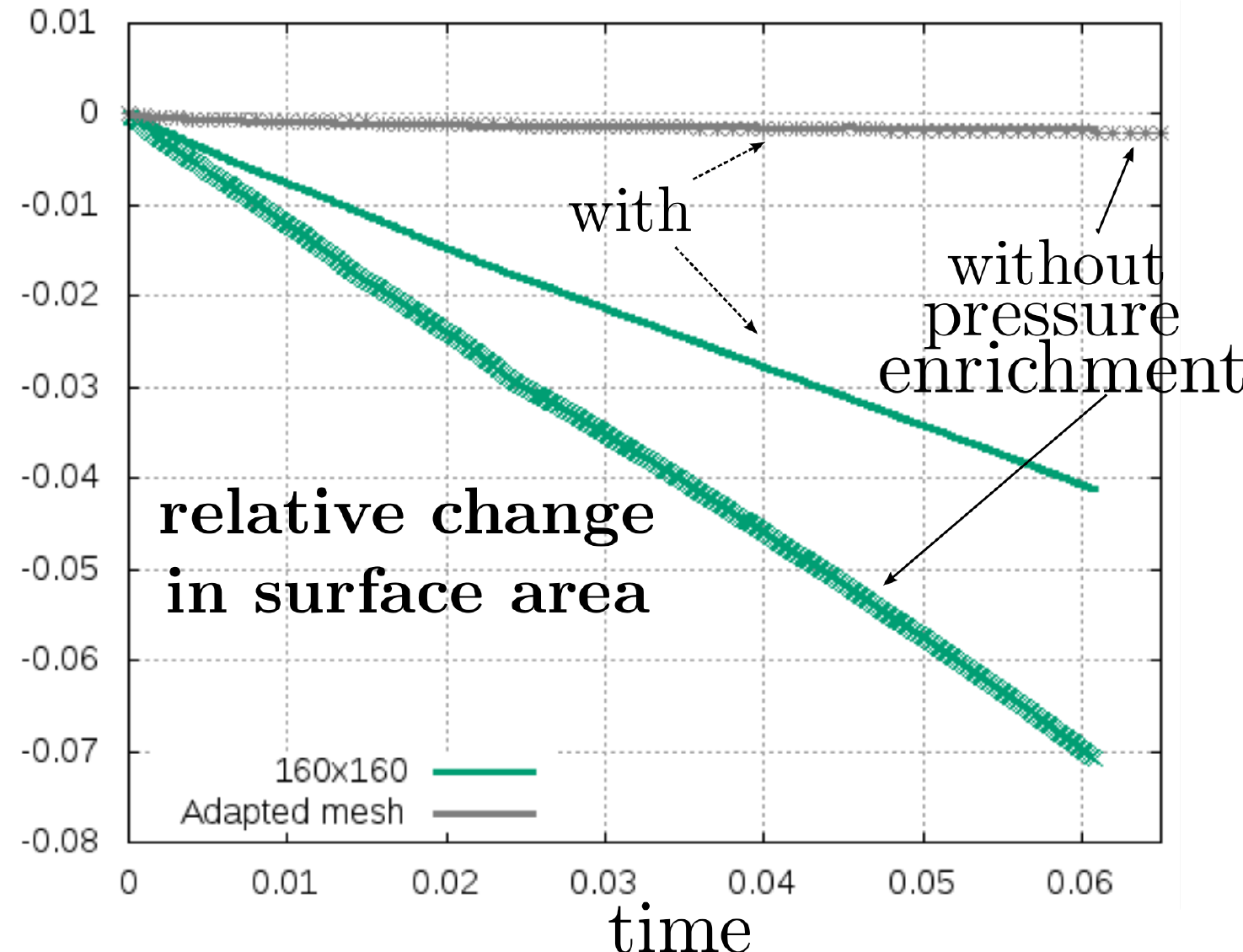}
      \label{sub:mass_loss} }
    \subfloat[Maximum of the velocity magnitude]{
      \includegraphics[width=0.5\textwidth]{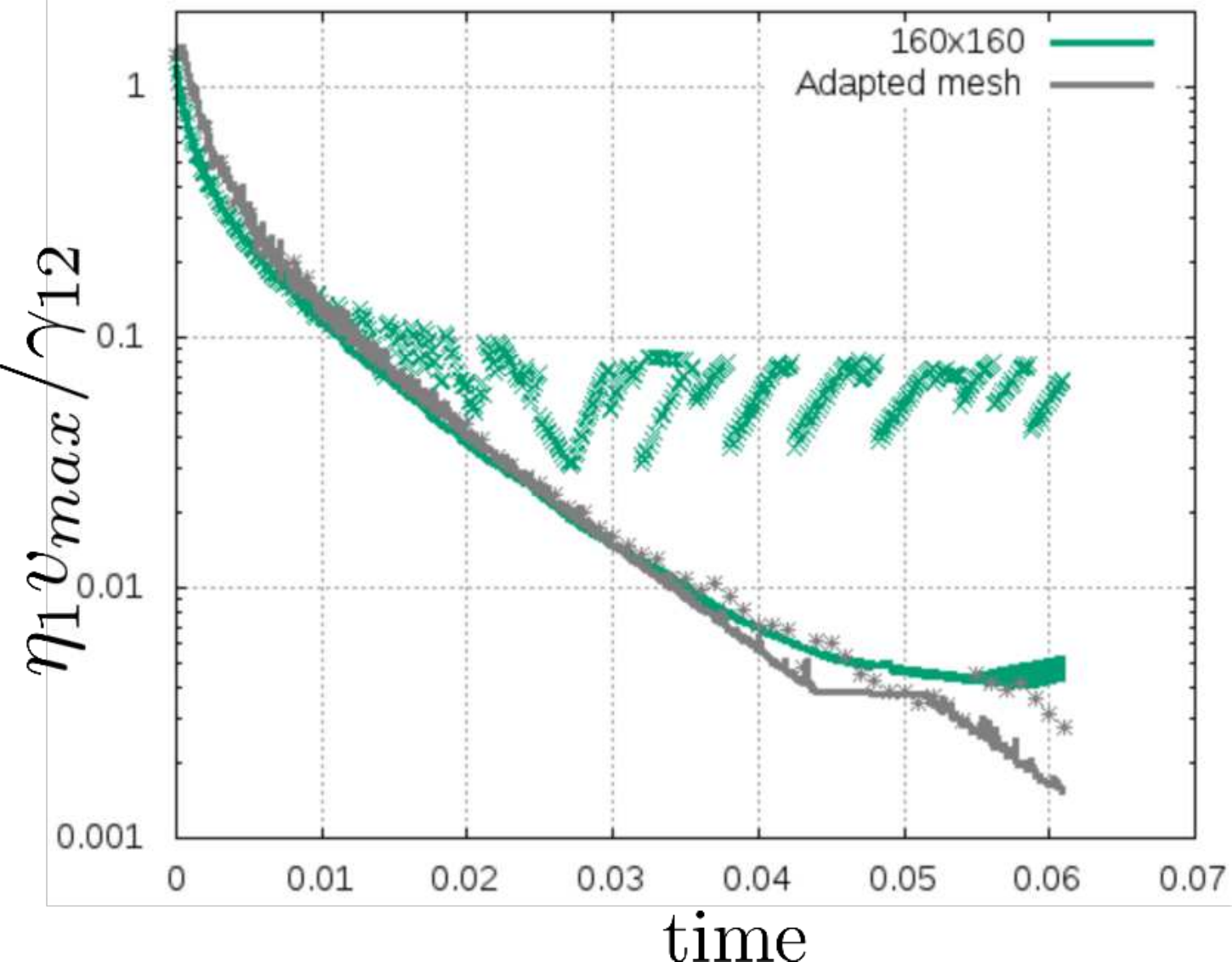}
      \label{sub:vmax} }
\end{center}
\caption{Effect of parasitic currents: relative change in droplet surface area (a) and maximum of velocity magnitude (b), computed  with different meshes.}
\label{fig:currents}
\end{figure}

The consequence of this pressure description improvement, especially without mesh adaptation, appears clearly when analysing the parasitic current intensity and the subsequent mass loss. The relative change in droplet surface area is plotted in Figure~\ref{sub:mass_loss} for the 160$\times$160 and adapted meshes. Since liquid droplet is incompressible, the variation in surface area (which, in the 2D-context of this study, plays the role of the droplet volume or mass) is only due to discretisation errors. The solid lines correspond to simulations carried out with pressure enrichment, while crosses denote those obtained without. It can be seen that, when adapting the mesh with or without pressure enrichment, the mass loss is unchanged: 0.17\% after 600 time increments. However, the benefit of the enrichment is evident on the structured case, since the mass loss passes from 7\% to 4\% after 600 increments. Additionally, having an adapted mesh refined in the vicinity of the interface, does not ensure necessarily a good mass conservation. For example, if the mesh adaptation is based on an error analysis performed exactly in the same conditions as before, but considering only the smooth delta function~\eqref{eq:delta}, and not the pressure field, a mass loss of more than 7\% is observed after 60 increments.
At equilibrium, the parasitic current intensity can be estimated according to the surface tension and dynamic viscosity of the drop, through the dimensionless number $C_p = \frac{v_{max} \eta_1}{\tensionsuperficielle_{12}}$, where $v_{max}$ is the maximum of the velocity magnitude~\cite{Strubelj2009}. This capillary number, characteristic of the quality of the surface tension force modelling, has, still at equilibrium, an optimal value equal to zero. $C_p$ is plotted over time in Figure~\ref{sub:vmax}. The maximum of velocity is first located at the triple point, at least while the parasitic currents are not preponderant. Afterwards, Figure~\ref{sub:vmax} manifestly shows spurious oscillations of the velocity when the pressure is not enriched with the structured mesh. These oscillations, responsible for a large part of the mass loss, are drastically reduced when using the enrichment in pressure. Furthermore, the pressure enrichment has only a slight effect when using mesh adaptation.

\begin{figure}[!hbt]
\begin{center}
\subfloat[Contact-angle]{
      \includegraphics[width=0.5\textwidth]{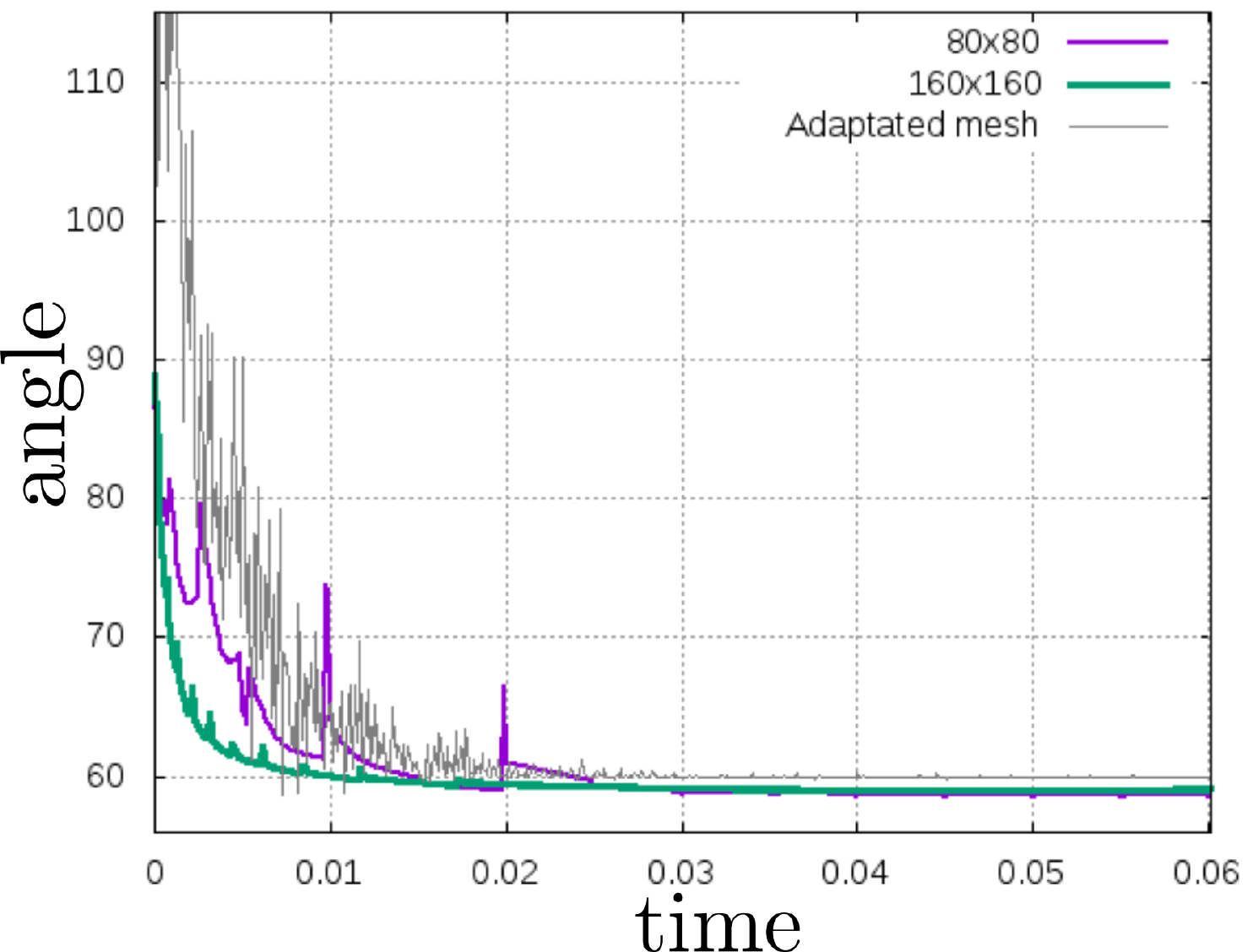}
      \label{sub:angles} }
\subfloat[Triple point position]{
      \includegraphics[width=0.5\textwidth]{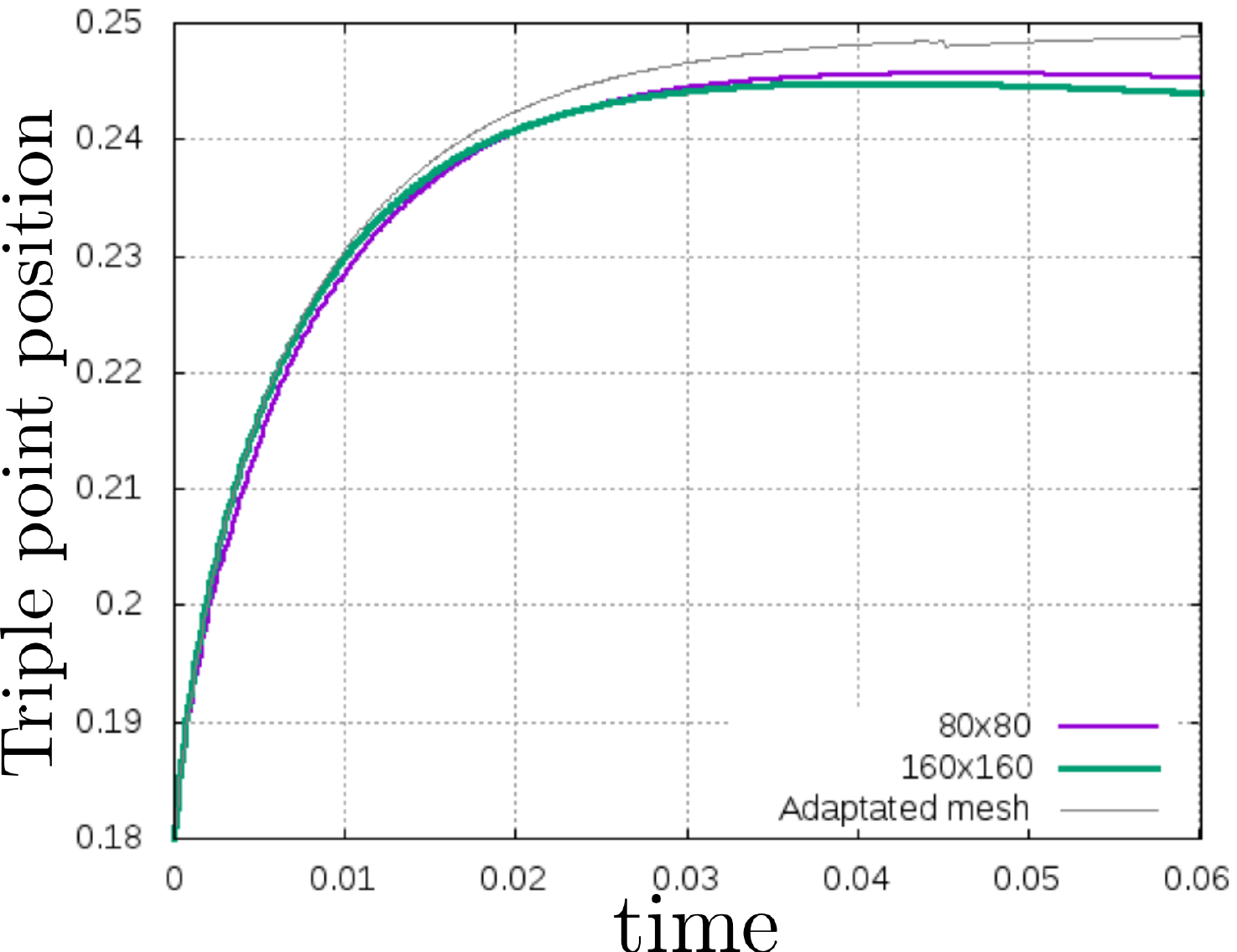}
      \label{sub:tp} }
      \end{center}
\caption{2D-simulation of droplet spreading: change in contact-angle and triple point position, for different meshes, with $f = 0.1$.}
\label{fig:2d_angles_pt}
\end{figure}

Concerning the contact-angle, convergence curves with respect to the mesh size, are presented in Figure~\ref{sub:angles}. This angle $\theta_h$ is computed in the element containing the triple point as:
\begin{equation}
\label{eq:angle_calcul}
\theta_h = \pi - \cos^{-1}\left(\frac{\grad \alpha_h}{\|\grad\alpha_h\|} \cdot \normale_h \right)
\end{equation}
where $\normale_h$ is the unit outward normal to the computational domain.

The angle is shown to be stabilised at 59\textdegree, 59.5\textdegree \, and 60\textdegree \, on the 80$\times$80, 160$\times$160 and adapted meshes, respectively. Hence, the expected equilibrium angle is reached with a precision that depends on the mesh size. Moreover, the convergence rate toward this equilibrium angle is shown to be independent of the considered mesh. Note that, when considering the mesh adaptation, due to the velocity profile shown in Figure~\ref{sub:velocity_0}, a kind of bump is first formed just above the triple junction, explaining the  values higher than 90\textdegree shown in Figure~\ref{sub:angles} in the first steps of the simulation. The oscillations observed, especially with the adapted mesh, correspond to the  passage of the interface from an element to another. This explains why these oscillations are more frequent on a fine mesh, since the passage from an element to another is more frequent, and why they disappear when reaching the steady state. Such oscillations are mentioned in reference~\cite{buscaglia_variational_2011}. They are mainly due to mesh-dependent numerical errors made during the transport of the level-set function. However, those errors do not depend on the velocity field computed in the mechanical problem. Indeed, the same phenomenon appears when a droplet, having a given initial contact-angle, is just translated by solving the transport equation~\eqref{eq:transport} with a constant velocity $\vv = (\pm1,0)$. To complete this study, the change in triple-point location over time is plotted in Figure~\ref{sub:tp}. It can be observed that the oscillations just mentioned do not affect the dynamic of the triple junction. The slight difference in the steady positions obtained for the adapted and structured cases, is only due to the parasitic currents which force the triple-point to move backward in the structured case.

\begin{figure}[!hbt]
\begin{center}
\subfloat[Contact-angle]{
      \includegraphics[width=0.5\textwidth]{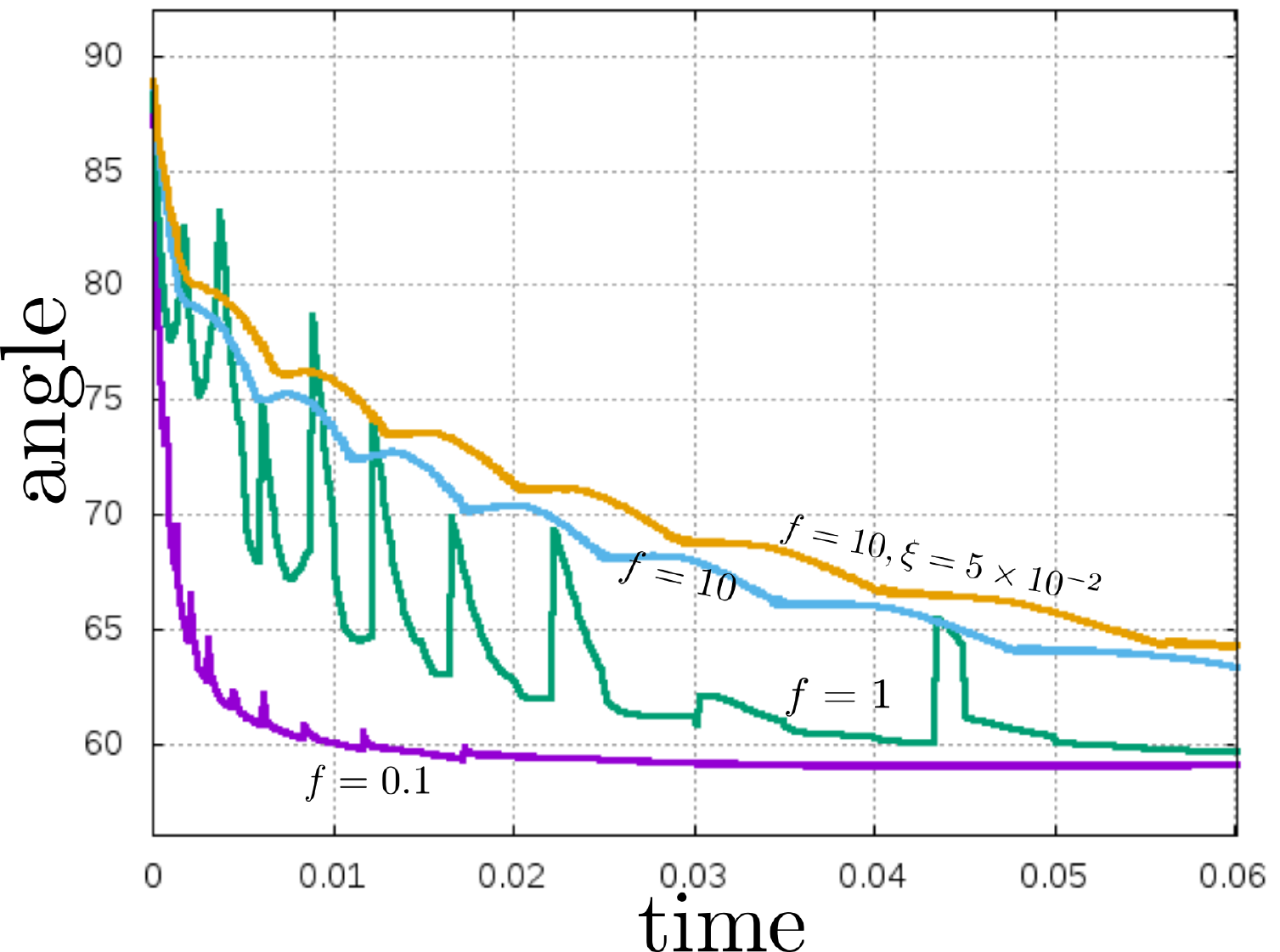}
      \label{sub:angles_f} }
\subfloat[Triple point position]{
      \includegraphics[width=0.5\textwidth]{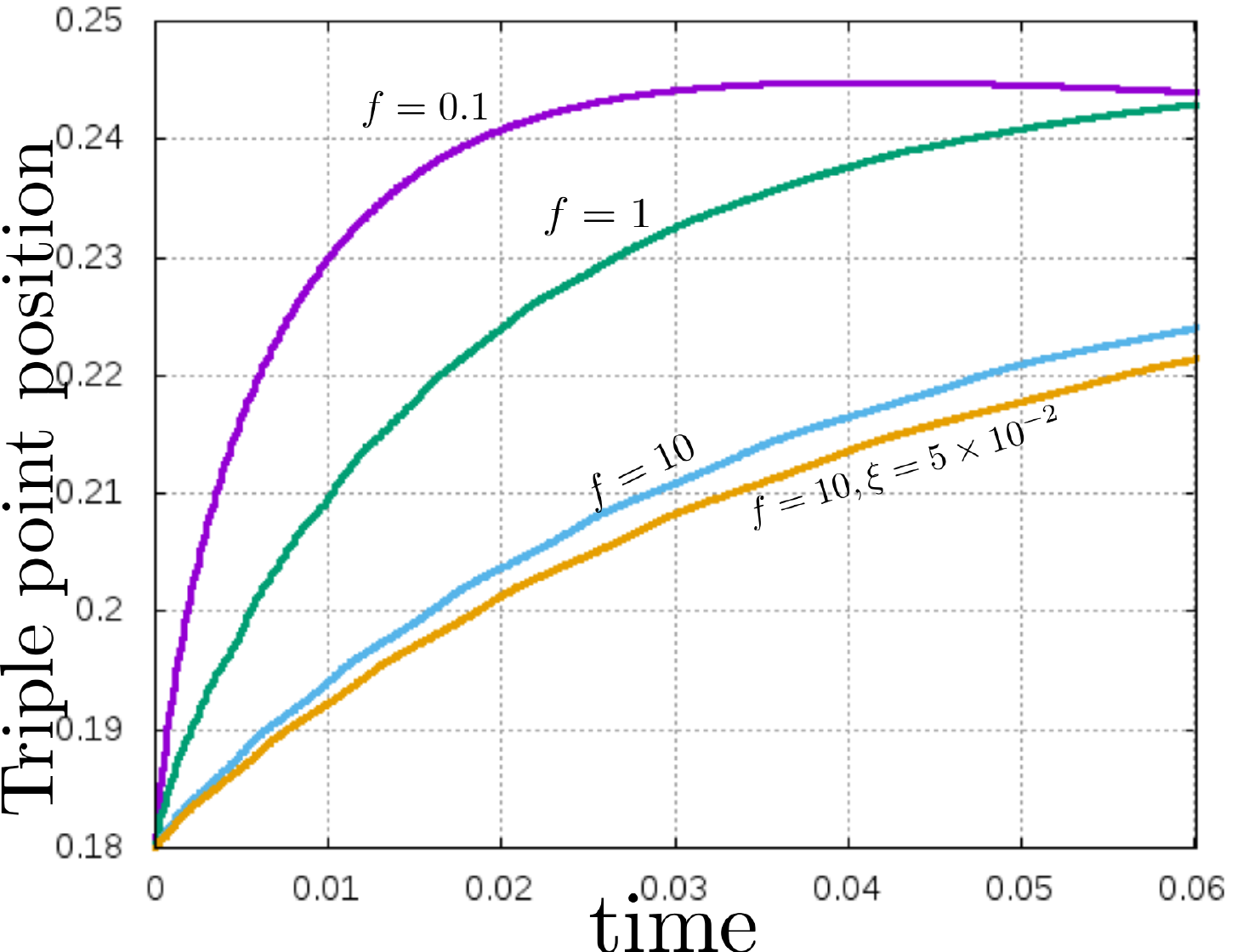}
      \label{sub:tp_f} }
      \end{center}
\caption{2D-simulation of droplet spreading: effect of the friction coefficient $f$ on the change in contact-angle and triple point dynamic (160$\times$160 mesh).}
\label{fig:2d_angles_pt_f}
\end{figure}

Heretofore, only $f = 0.1$ was considered. Figure~\ref{fig:2d_angles_pt_f} shows the effect of the friction coefficient on the dynamic of the contact angle, as well as on the one of the triple point. Note that in all cases $\xi = 0$, except for one simulation where $\xi = 5.0 \times 10^{-2}$, leading to an additional dissipation at the triple junction. Subsequently, the return to equilibrium is delayed. Hence, at $t = 0.06$, this steady state has already been reached only for $f = 0.1$ and $f = 1$.

\begin{figure}[!hbt]
\begin{center}
\includegraphics[width=0.55\textwidth]{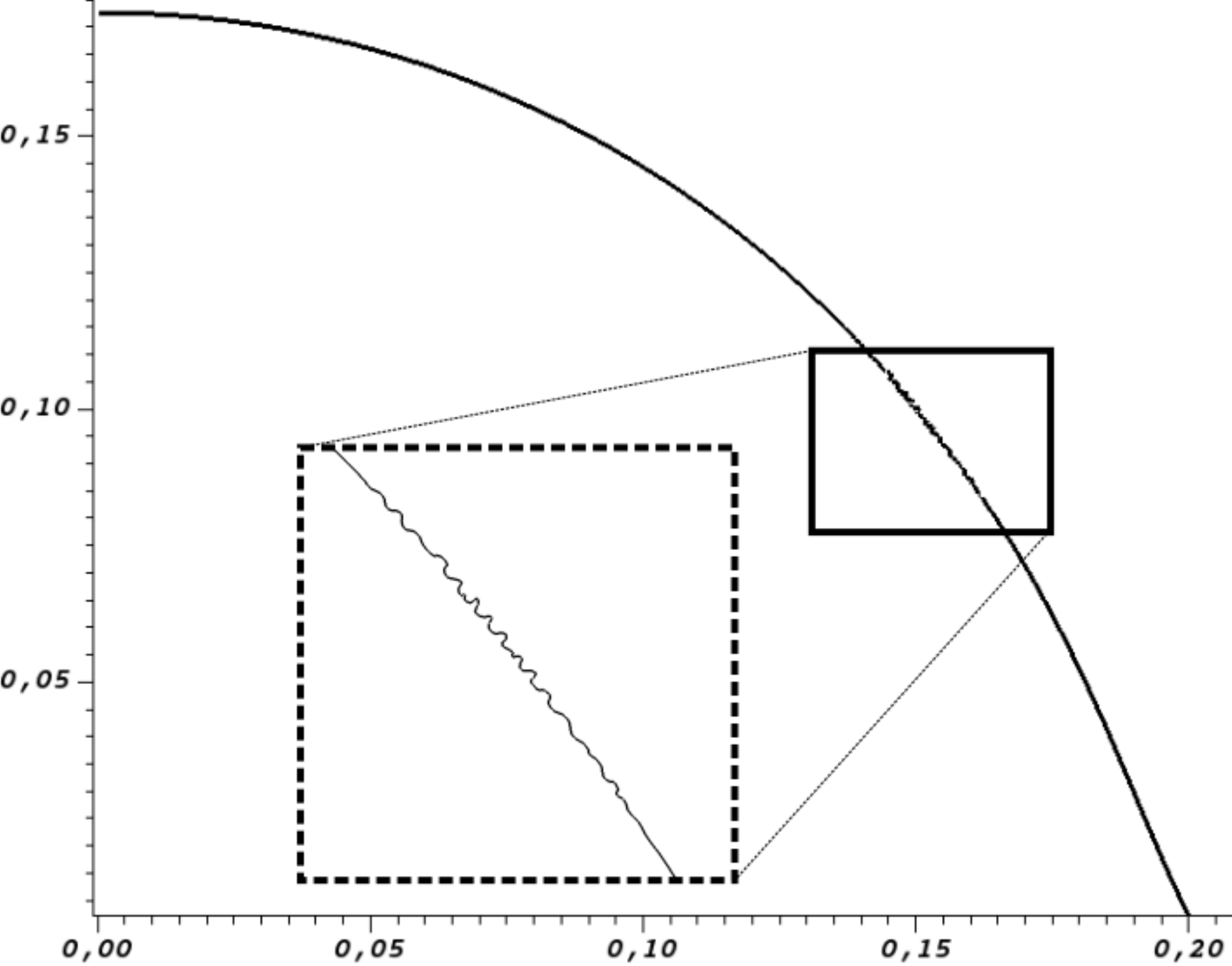}
\end{center}
\caption{Wiggle instabilities developed in the interface when not considering the semi-implicit treatment of surface tension term.}
\label{fig:oscillations}
\end{figure}

Some remarks on the choice of the time step should be done. First, the time step is highly related to the mesh size when not considering the semi-implicit formulation of the surface tension term, introduced in subsection~\ref{sec:semi_implicit}. Typically, the interface quickly develops the kind of wiggle instabilities shown in Figure~\ref{fig:oscillations} when $\Delta t = 10^{-3}$ on the $160\times 160$ mesh, and $\Delta t = 10^{-4}$ on the adapted mesh. Such instabilities, arising because of the discretisation of the level-set function, are well-known when simulating interface movements driven by the curvature or its spatial derivatives as in reference~\cite{bruchon_3d_2011}.
But, once the semi-implicit treatment of subsection~\ref{sec:semi_implicit} is considered, the oscillations completely disappear. The Stokes - Level-set coupling is shown to be stable for any time step ranging from $10^{-1}$ to $10^{-4}$, with all the meshes investigated in this section. Of course, the error committed on the transport of the level-set function depends on the time step, and is the real limiting factor, as in the 3D-simulation presented in Section~\ref{sec:capillary_rise}. The steady value of contact-angle is  reached for any time step considered, but with oscillations (Figure~\ref{sub:angles}) whose amplitude increases with the time step. \\

To conclude, this section has shown the ability of our numerical strategy, to simulate with adequate accuracy a wetting problem. Even if mesh adaptation is not indispensable, this technique, while limiting the number of nodes, both improves the description of the interface, and produces a mesh which captures the pressure discontinuity and limits the parasitic currents. Hence, the CPU-time of the previous simulations, is approximately 1900 seconds with the 160$\times$160 mesh (25,600 nodes) and drops to 750 seconds with the adapted mesh (5,000 nodes), on one core for 600 increments. Consequently, the 3D-simulations presented in next sections, are conducted with the combination of mesh adaptation and pressure enrichment.

\subsection{Cubic droplet spreading}
\label{sec:spreading}

\begin{figure}[!hb]
  \begin{center}
    \subfloat[$t=0$]{
      \includegraphics[width=0.5\textwidth]{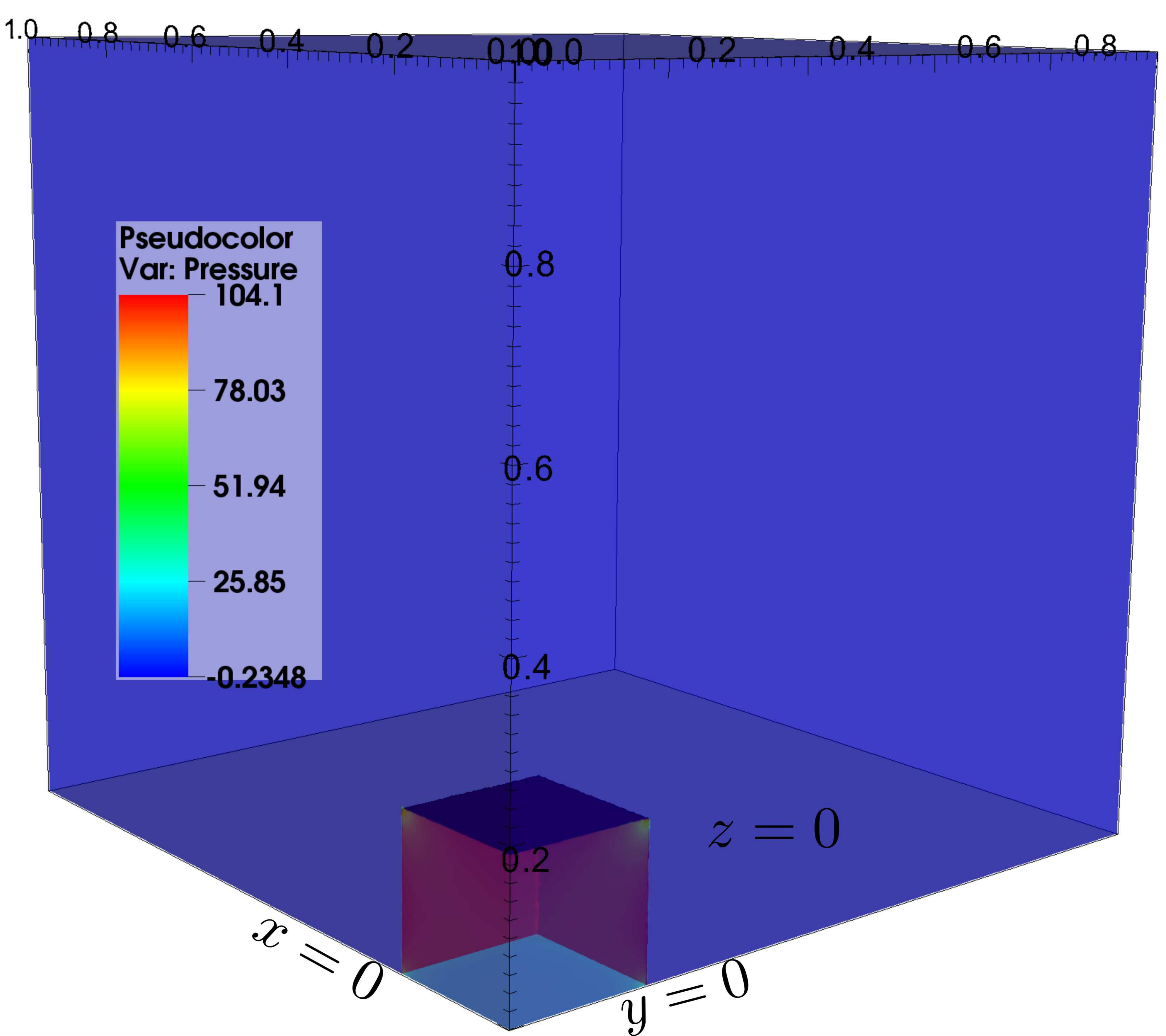}
      \label{sub:cube1_pressure_0}} \\
    \subfloat[$t = 0$]{
      \includegraphics[width=0.33\textwidth]{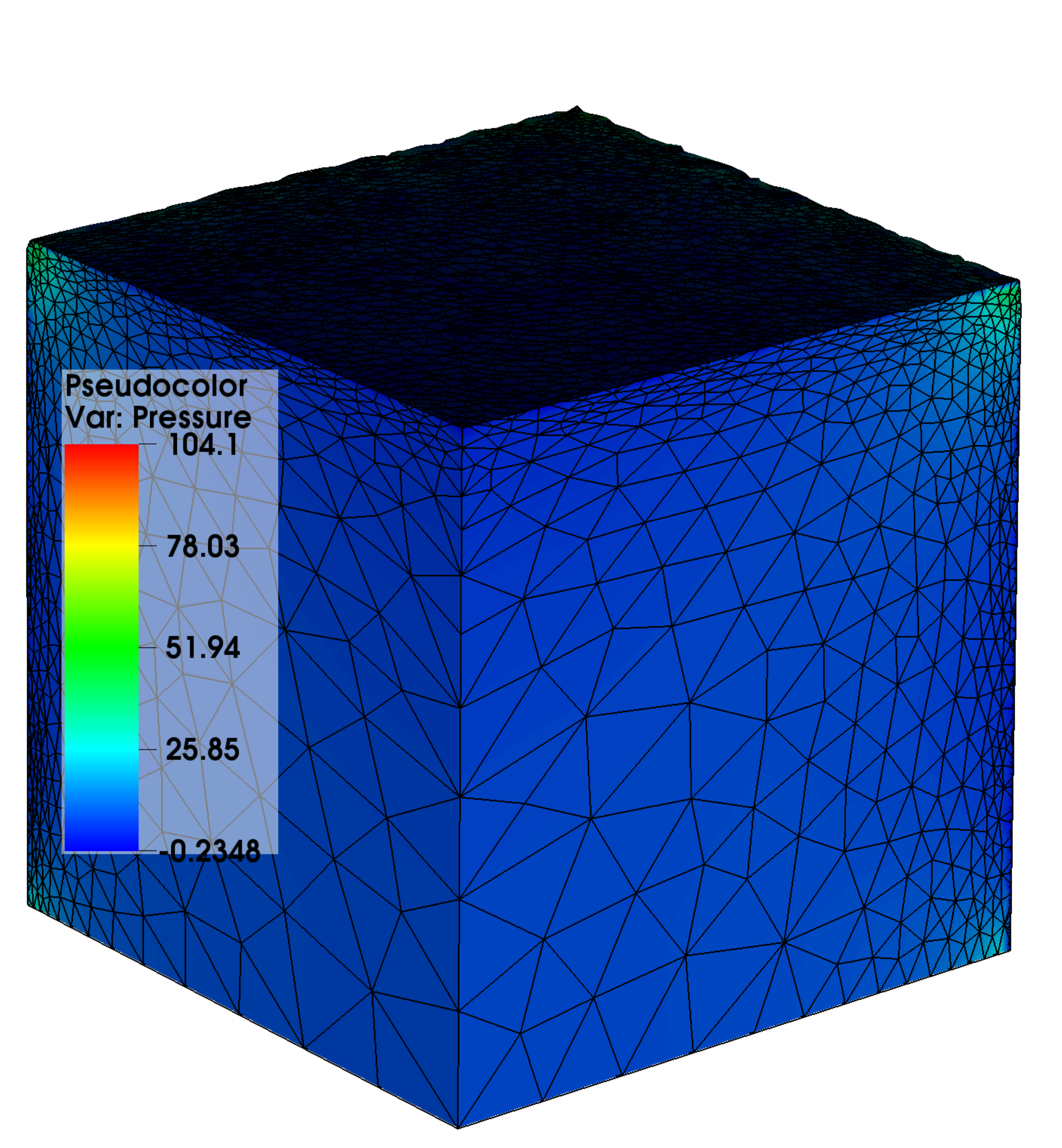}
      \label{sub:cube1_pressure_0b}}
    \subfloat[$t = 4.0 \times 10^{-3}$]{
      \includegraphics[width=0.33\textwidth]{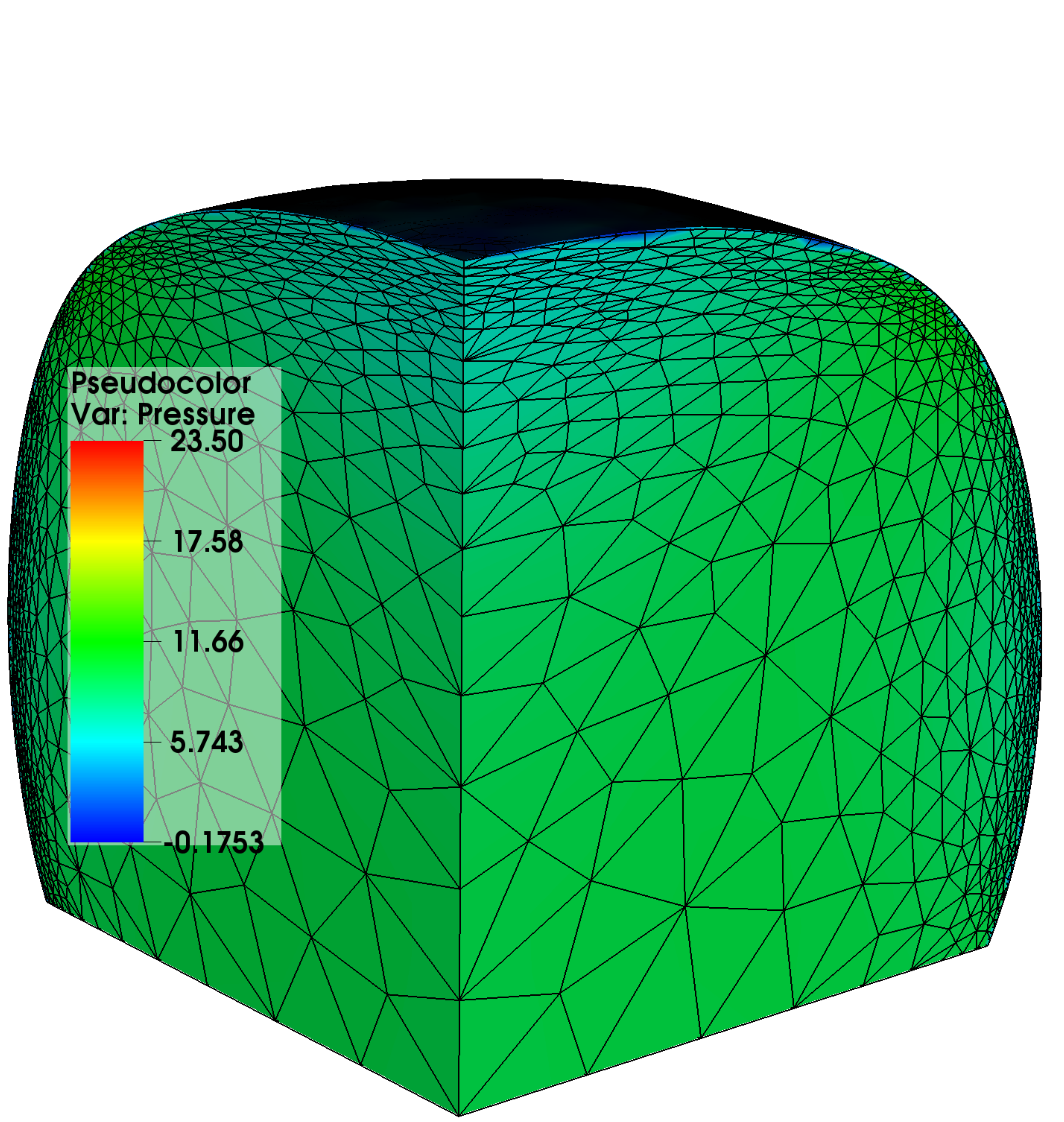}
      \label{sub:cube1_pressure_3}}
      \subfloat[$t = 3.0 \times 10^{-2}$]{
      \includegraphics[width=0.33\textwidth]{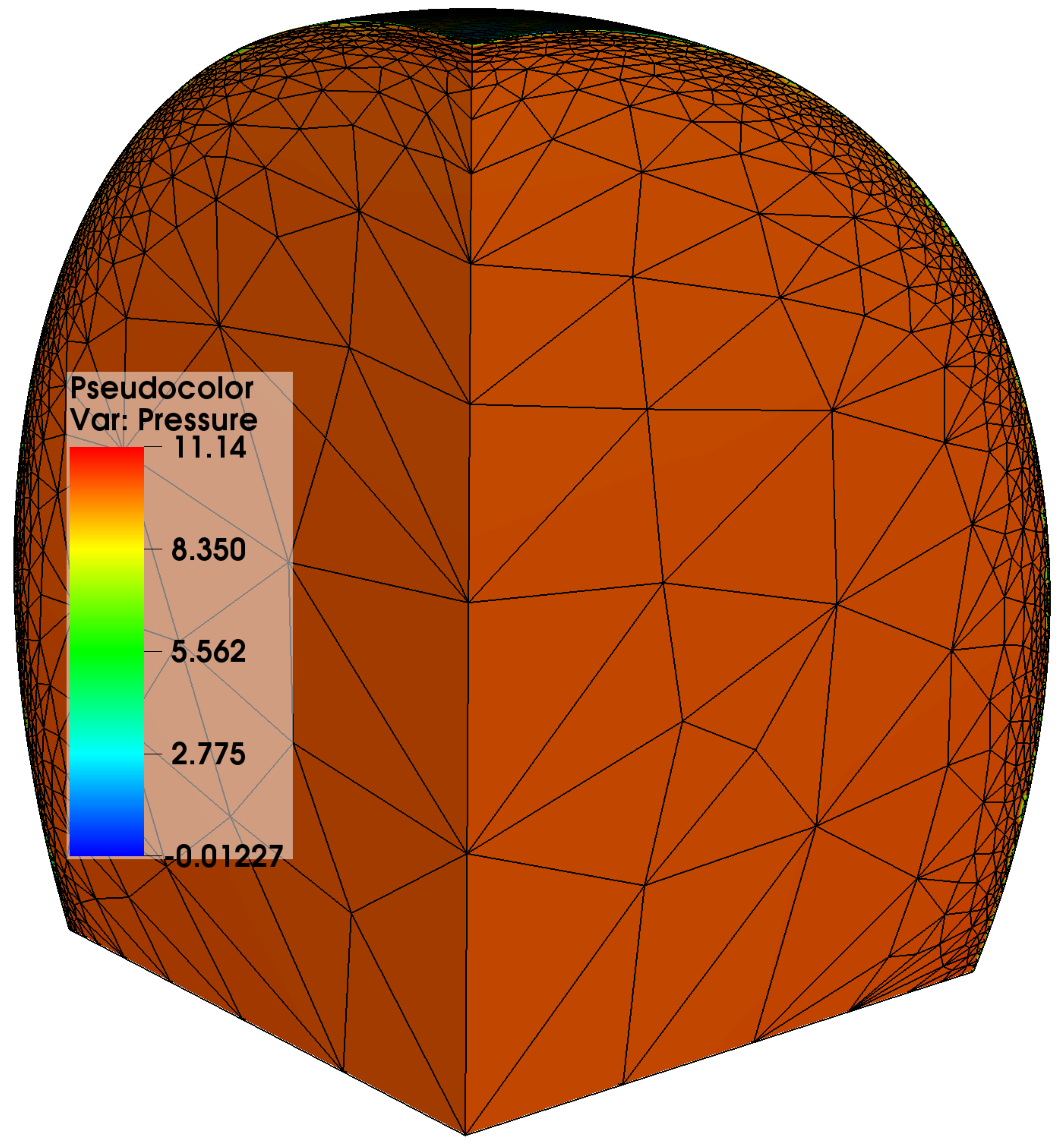}
      \label{sub:cube1_pressure_30}}
    \caption{Spreading of a droplet having an initial cubic shape. Parameters of the simulation are those given in Table~\ref{tab:pamaters} with $\xi = 0$ and $\gamma_{23} = 0$. (a) Whole computational domain and pressure field. (b)-(d) Pressure field and adapted mesh, both projected onto the droplet surface.}
    \label{fig:cube1_pressure}
  \end{center}
\end{figure}

The initial droplet is the $[0,0.18]^3$ cube immersed into the computational domain in Figure~\ref{sub:cube1_pressure_0}. Applying a cubic shape is only to demonstrate the ability of the approach to treat a case for which the strong form of the Laplace's law is ill-posed: the curvature of the liquid-air interface is not defined along the edges of the cube and vanishes over its faces. The computational domain is a unit cube, discretised with an adapted mesh, made up of approximately 278,000 tetrahedrons and 50,000 nodes. Mesh size, in the vicinity of the interface and perpendicular to it, is in the range $[2.0\times 10^{-3} : 5.0 \times 10^{-3}]$. The projections of both adapted mesh and pressure field onto the zero-isosurface of the level-set function, are shown in Figures~\ref{sub:cube1_pressure_0b}-\ref{sub:cube1_pressure_30}. The other parameters of the simulation are given in Table~\ref{tab:pamaters}, with a time step $\Delta t = 10^{-3}$. Moreover, as previously, the body forces are neglected: $\vb = \bfm{0}$.  Note that the simulated droplet represents in fact only one quarter of the entire droplet, since the normal velocity is vanishing over all the faces of the computational domain, except over the plane $\{z = 1\}$ which is a free boundary.

\begin{figure}[!hbt]
  \begin{center}
    \subfloat[$t=0$]{
      \includegraphics[width=0.32\textwidth]{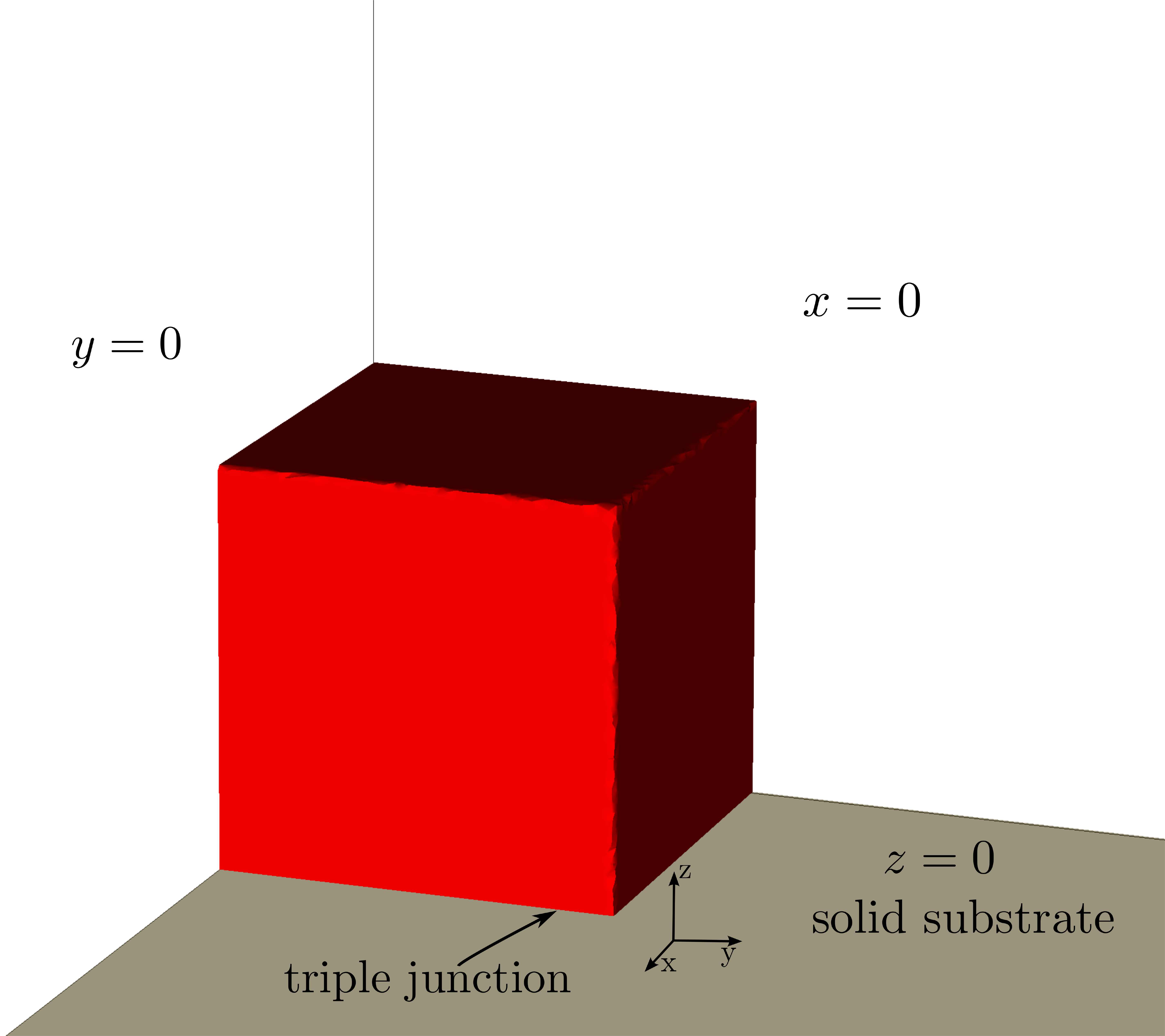}
      \label{sub:cube1_1}}
    \subfloat[$t = 3.0 \times 10^{-3}$]{
      \includegraphics[width=0.32\textwidth]{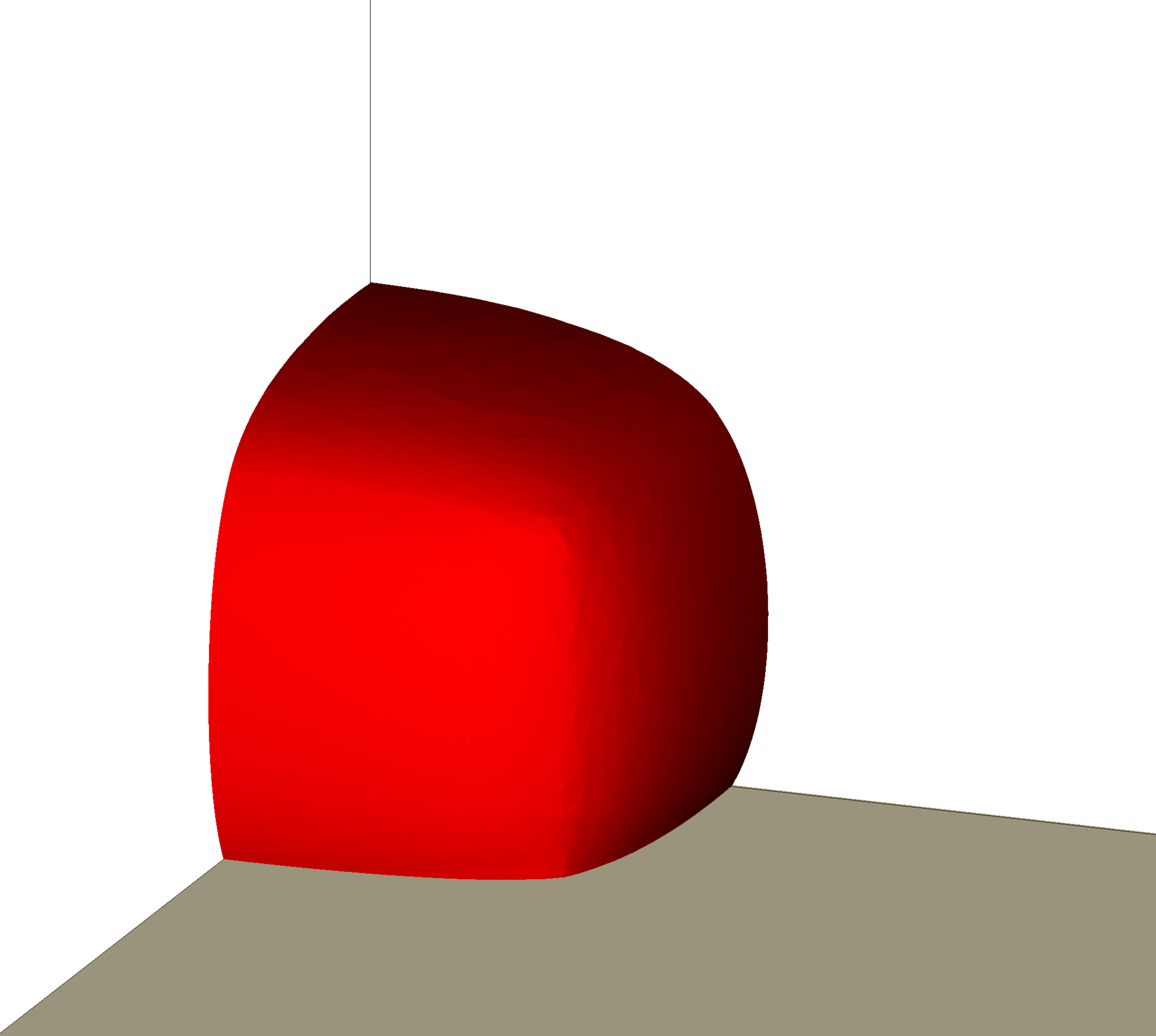}
      \label{sub:cube1_3}}
    \subfloat[$t = 3.0 \times 10^{-2}$]{
      \includegraphics[width=0.32\textwidth]{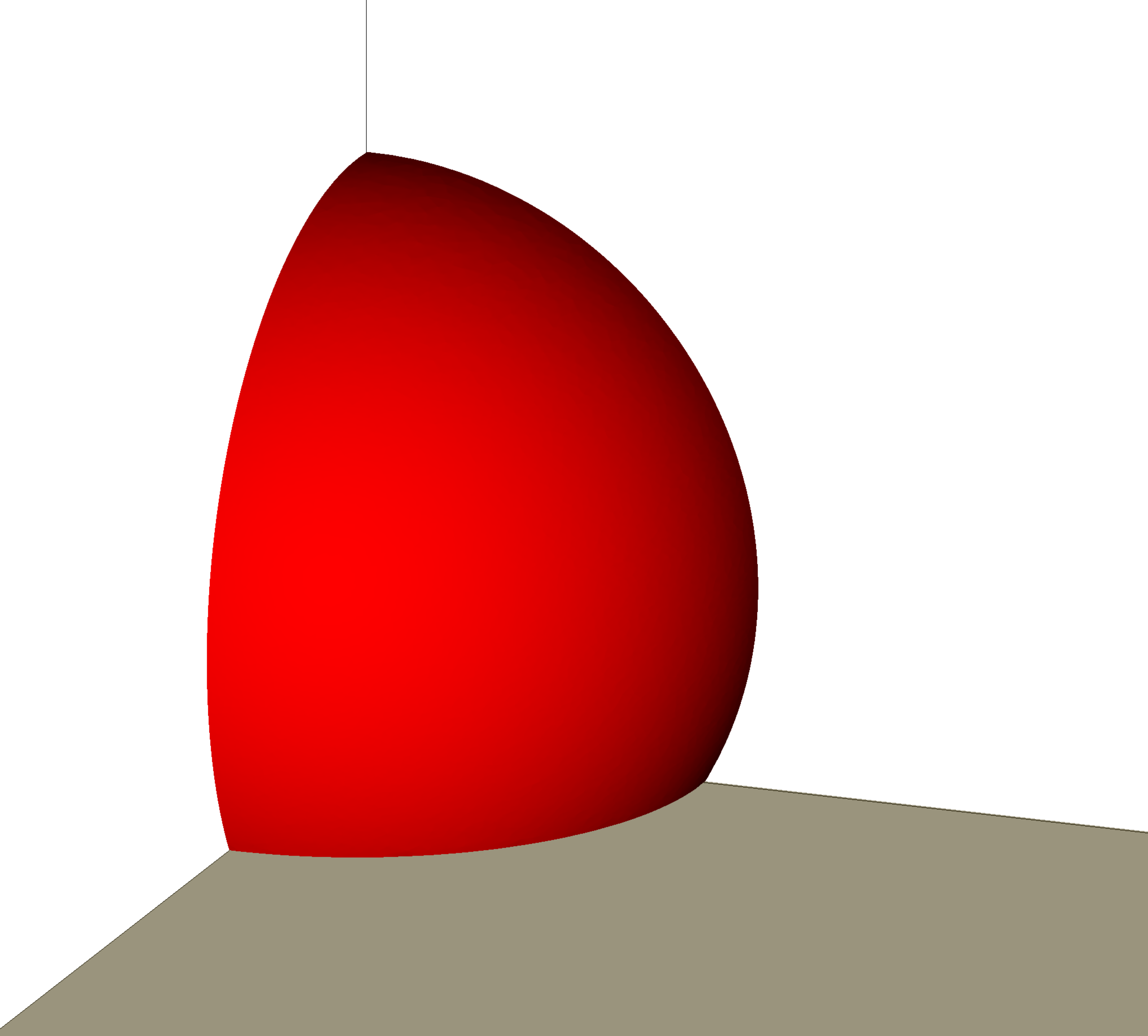}
      \label{sub:cube1_30}}
    \caption{Spreading of a droplet having an initial cubic shape. Parameters of the simulation are those given in Table~\ref{tab:pamaters} with $\gamma_{23} = 0$.}
    \label{fig:cube1}
  \end{center}
\end{figure}

\begin{figure}[!hbt]
  \begin{center}
    \subfloat[$t=0$]{
      \includegraphics[width=0.32\textwidth]{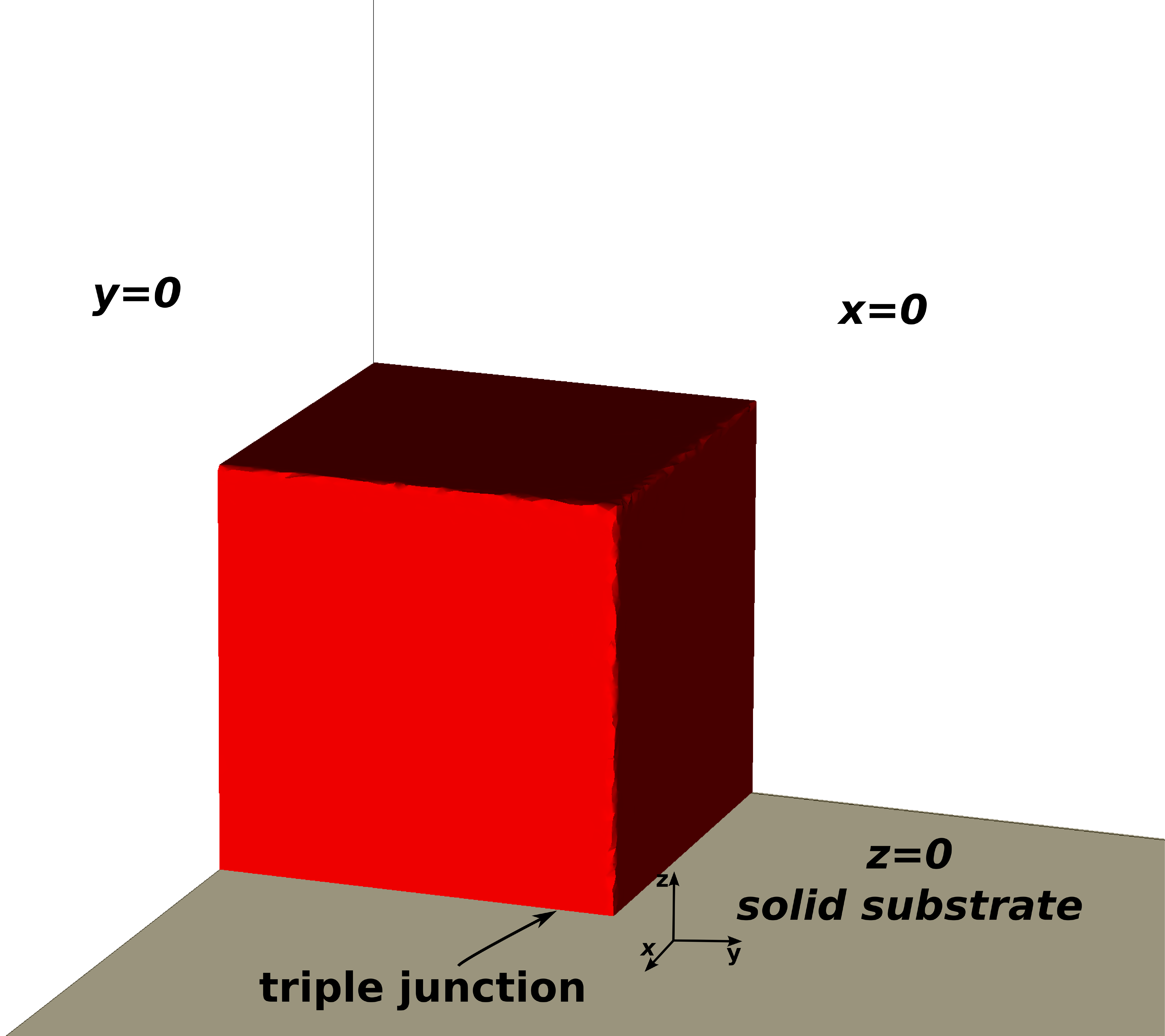}
      \label{sub:cube2_1}}
    \subfloat[$t = 3.0 \times 10^{-3}$]{
      \includegraphics[width=0.32\textwidth]{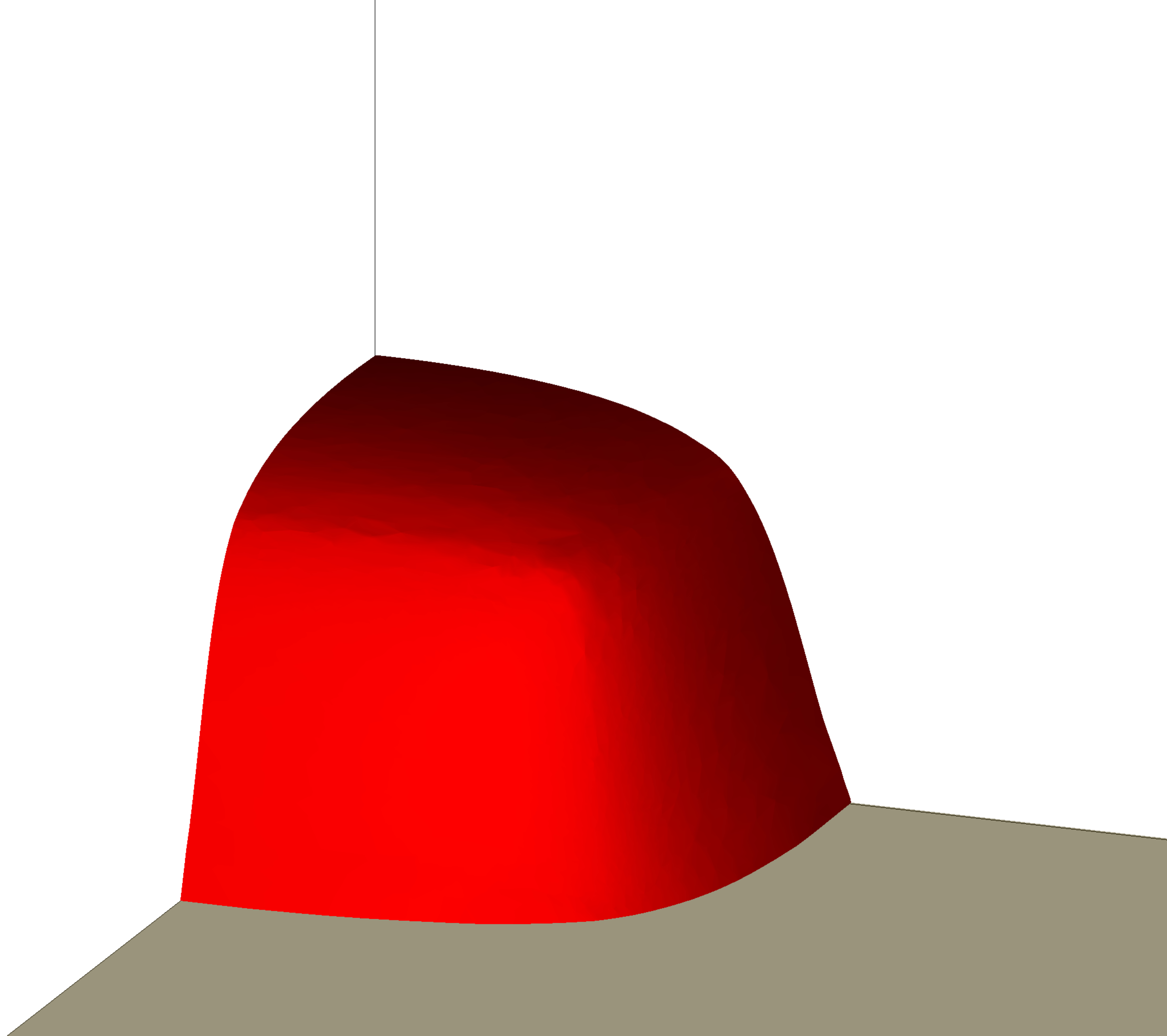}
      \label{sub:cube2_3}}
    \subfloat[$t = 3.0 \times 10^{-2}$]{
      \includegraphics[width=0.32\textwidth]{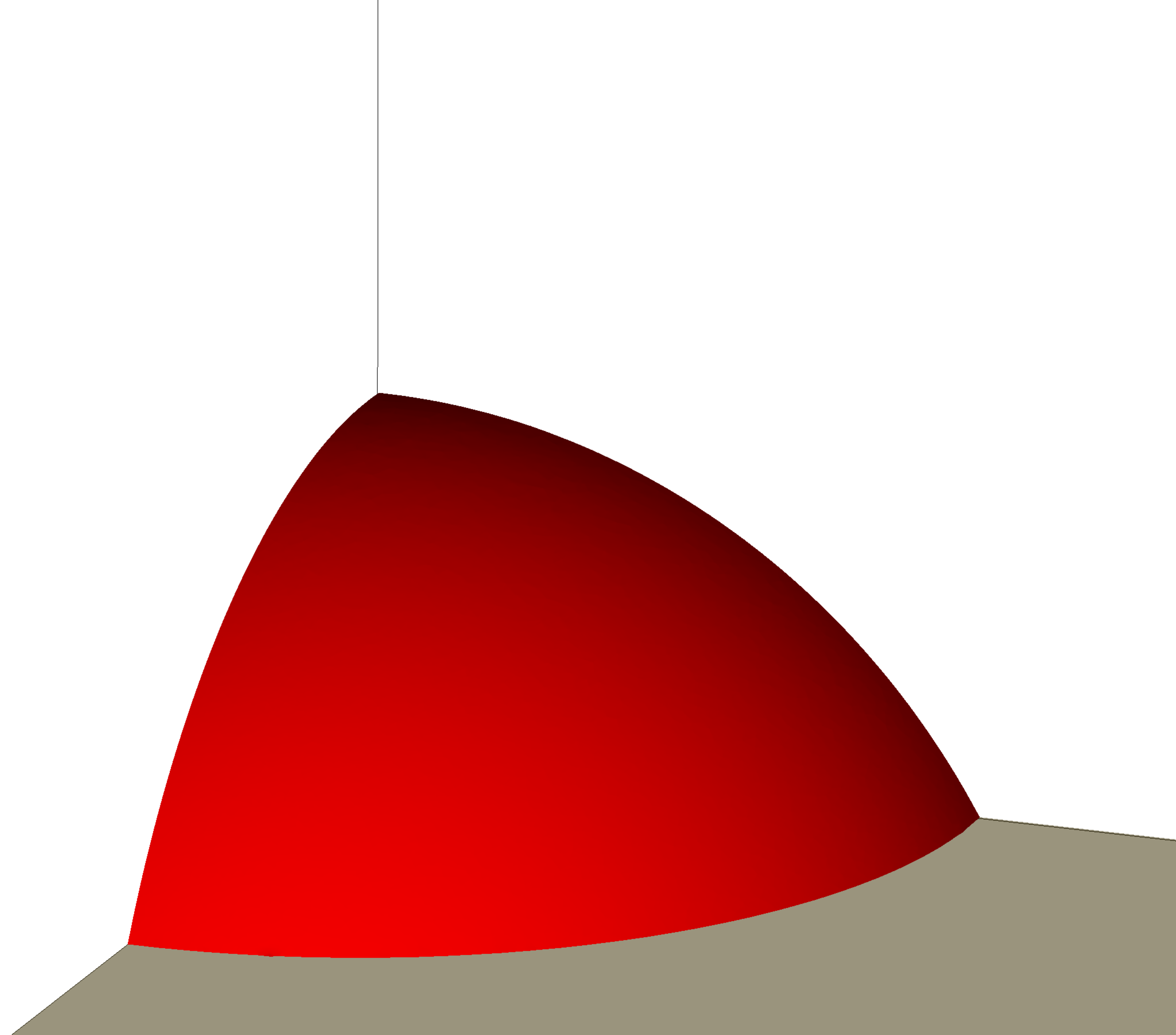}
      \label{sub:cube2_30}}
    \caption{Spreading of a droplet having an initial cubic shape. Parameters of the simulation are those given in Table~\ref{tab:pamaters} with $\gamma_{23} = 1$.}
    \label{fig:cube2}
  \end{center}
\end{figure}

\begin{figure}[!hbt]
\begin{center}
\subfloat[Contact-angle]{
      \includegraphics[width=0.5\textwidth]{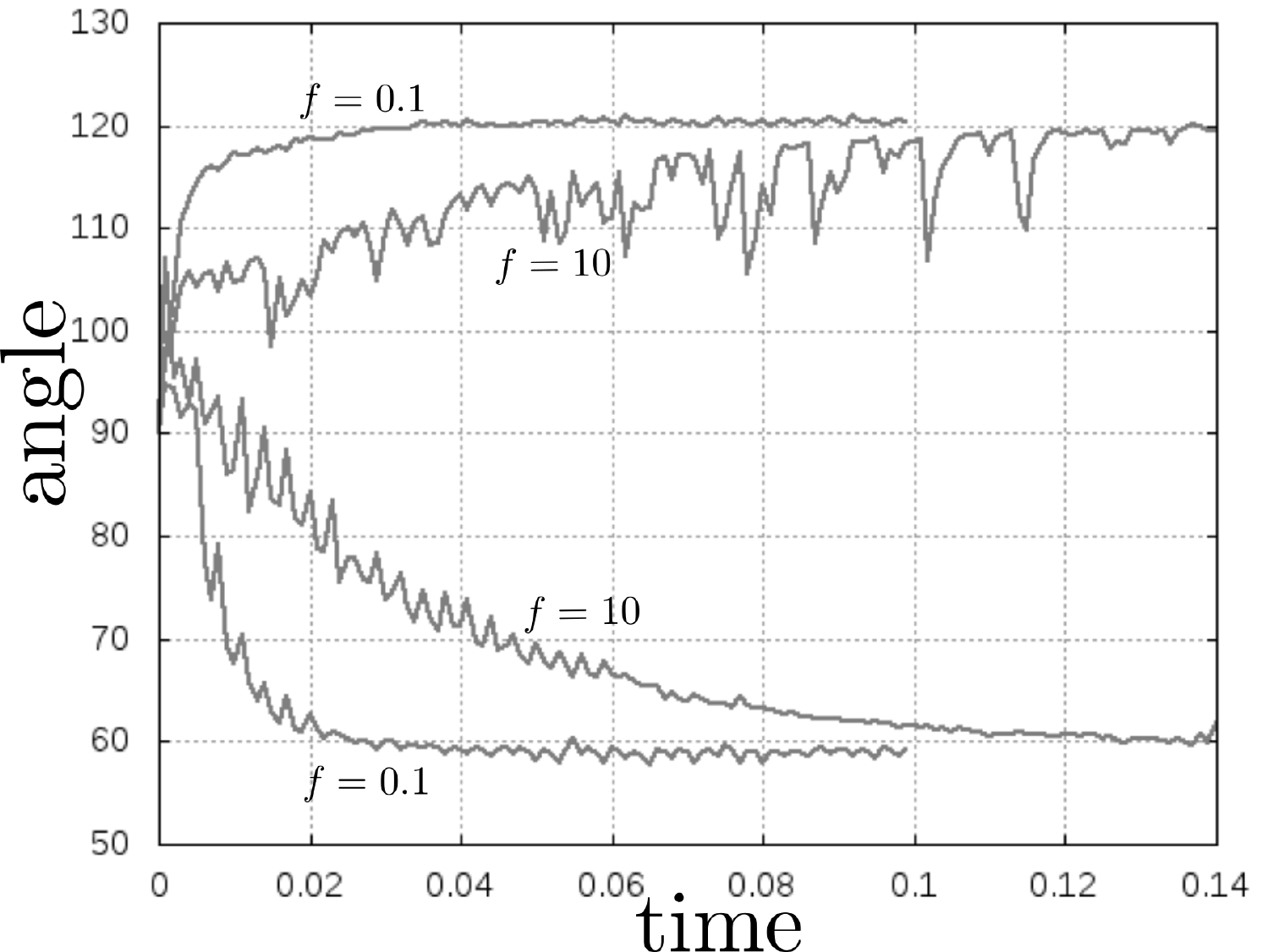}
      \label{sub:angles3D} }
\subfloat[Triple line position]{
      \includegraphics[width=0.5\textwidth]{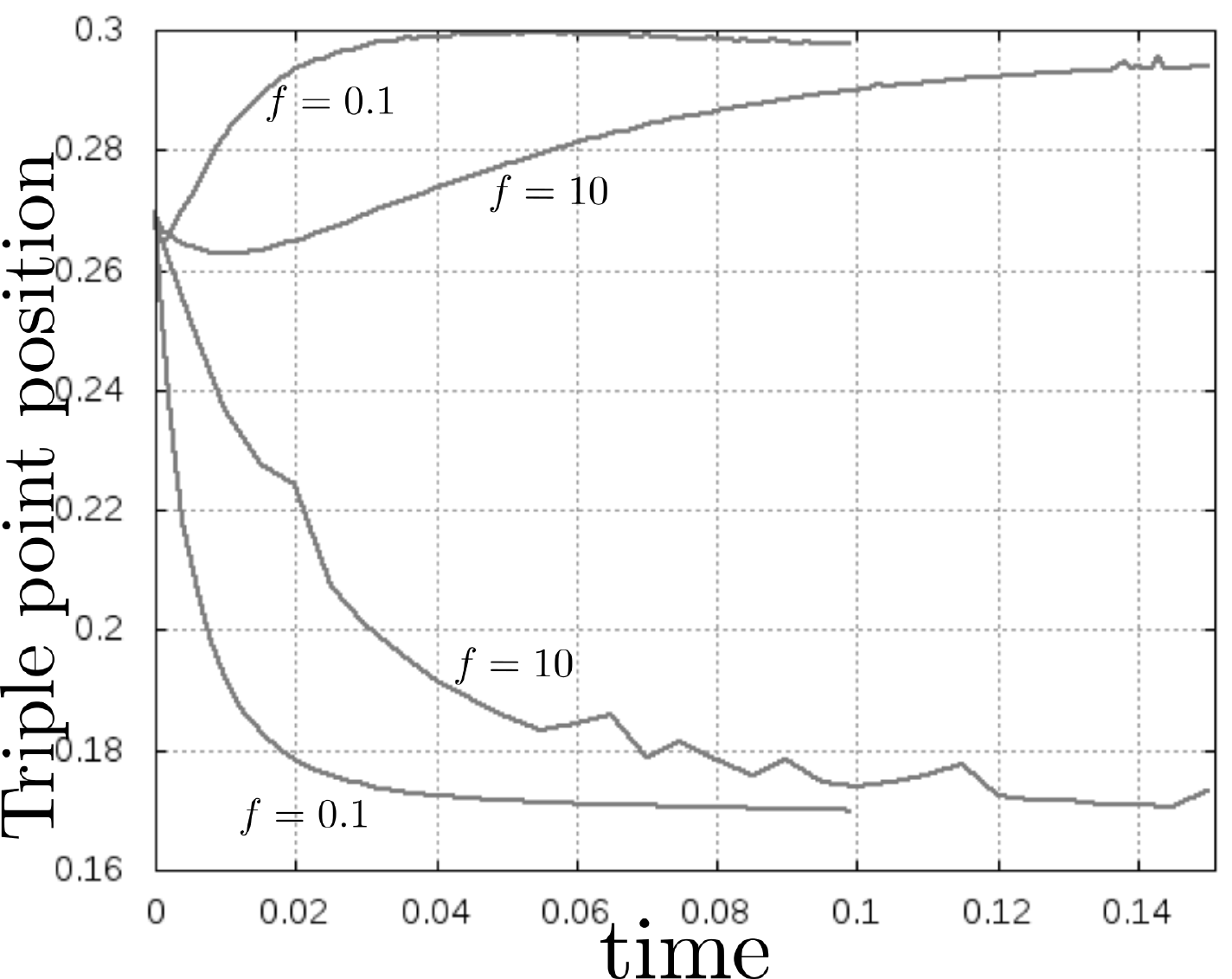}
      \label{sub:tp3D} }
      \end{center}
    \caption{Changes in contact-angle over time and associated contact line dynamic, for two values of the friction parameter ($f = 0.1$ and $f = 10$). Two cases are represented: $\gamma_{23} = 0$ and $\gamma_{23} = 1$, leading respectively to equilibrium angles of 120\textdegree and 60\textdegree.}
    \label{fig:angles3d}
\end{figure}

The droplet surface is presented in Figures~\ref{fig:cube1} ($\gamma_{23} = 0$) and~\ref{fig:cube2} ($\gamma_{23} = 1$) at initial, intermediate and equilibrium states. Because of Laplace's law, the droplet recovers quickly a spheroidal shape, minimising the surface energy. The corresponding pressure is visualised in Figure~\ref{fig:cube1_pressure}. While it reaches its maximal value near the corners of the initial cubic droplet, the pressure field becomes progressively uniform as the droplet tends to the steady-state. The change in contact-angle, averaged along the triple line, is plotted over time in Figure~\ref{sub:angles3D} for two values of the friction parameter, $f = 0.1$ and $f = 10$. The same remarks as in the 2D-case can be done. In particular, the oscillations, due to numerical reasons, do not affect the dynamics of the triple line, nor the final equilibrium state, as demonstrated by Figure~\ref{sub:tp3D}. This figure shows the changes over time of the distance to the origin, of the material point located initially at coordinates $(0.18,0.18,0)$ (a corner of the droplet).  Note that for this point, because of the initial droplet cubic shape, two types of trajectories can be observed, depending on either $\gamma_{12} = 0$ or $\gamma_{12} = 1$.

\begin{figure}[!hbt]
  \begin{center}
    \subfloat[$t=0$]{
      \includegraphics[width=0.19\textwidth]{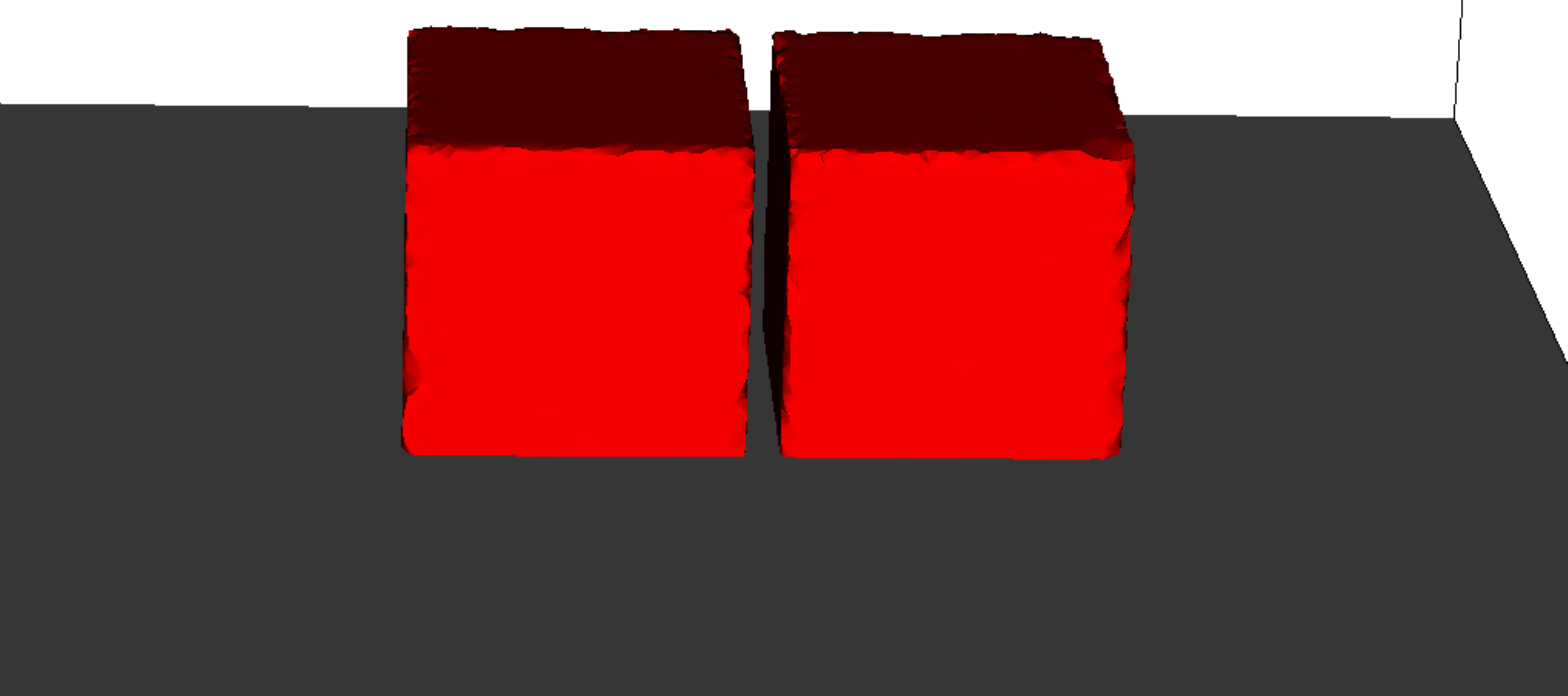}
      \label{sub:coal0}}
    \subfloat[$t = 5.0 \times 10^{-3}$]{
      \includegraphics[width=0.19\textwidth]{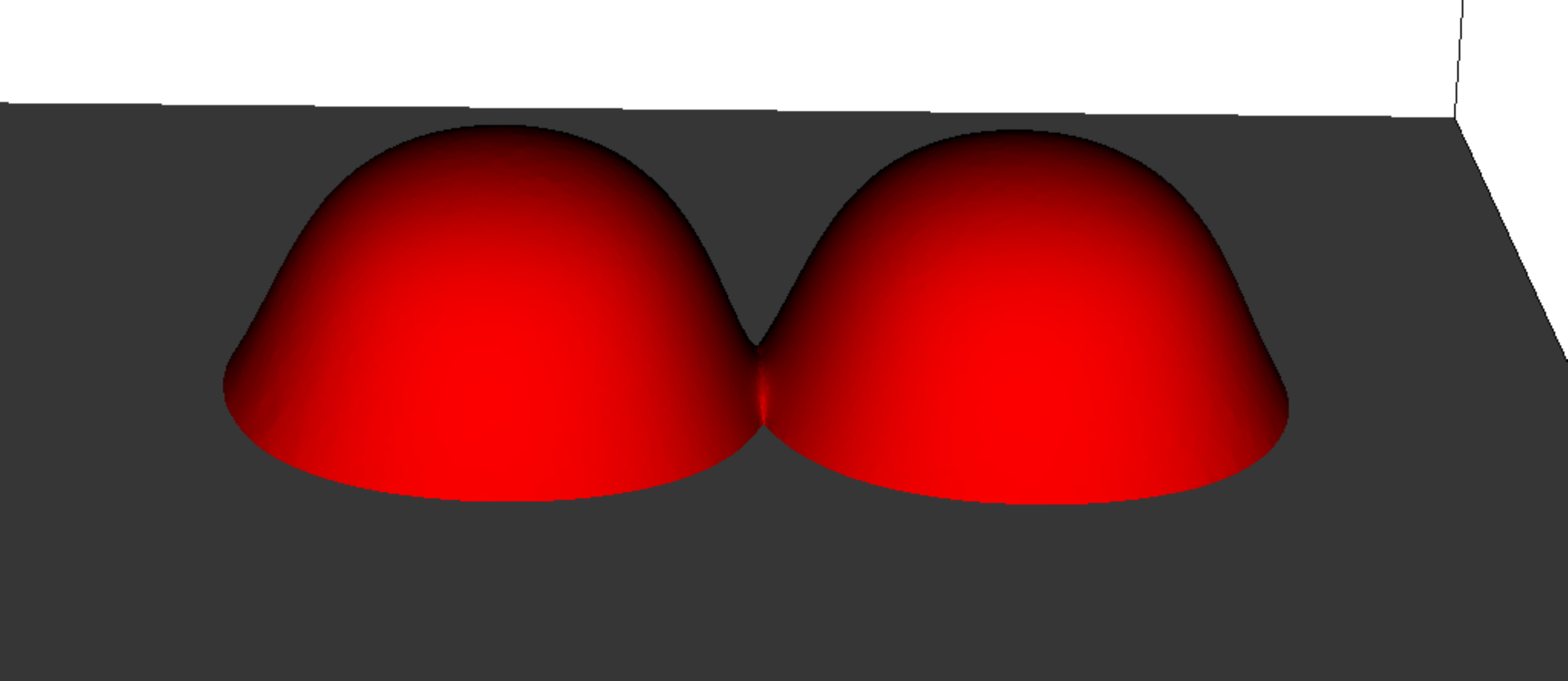}
      \label{sub:coal10}}
    \subfloat[$t = 1.0 \times 10^{-2}$]{
      \includegraphics[width=0.19\textwidth]{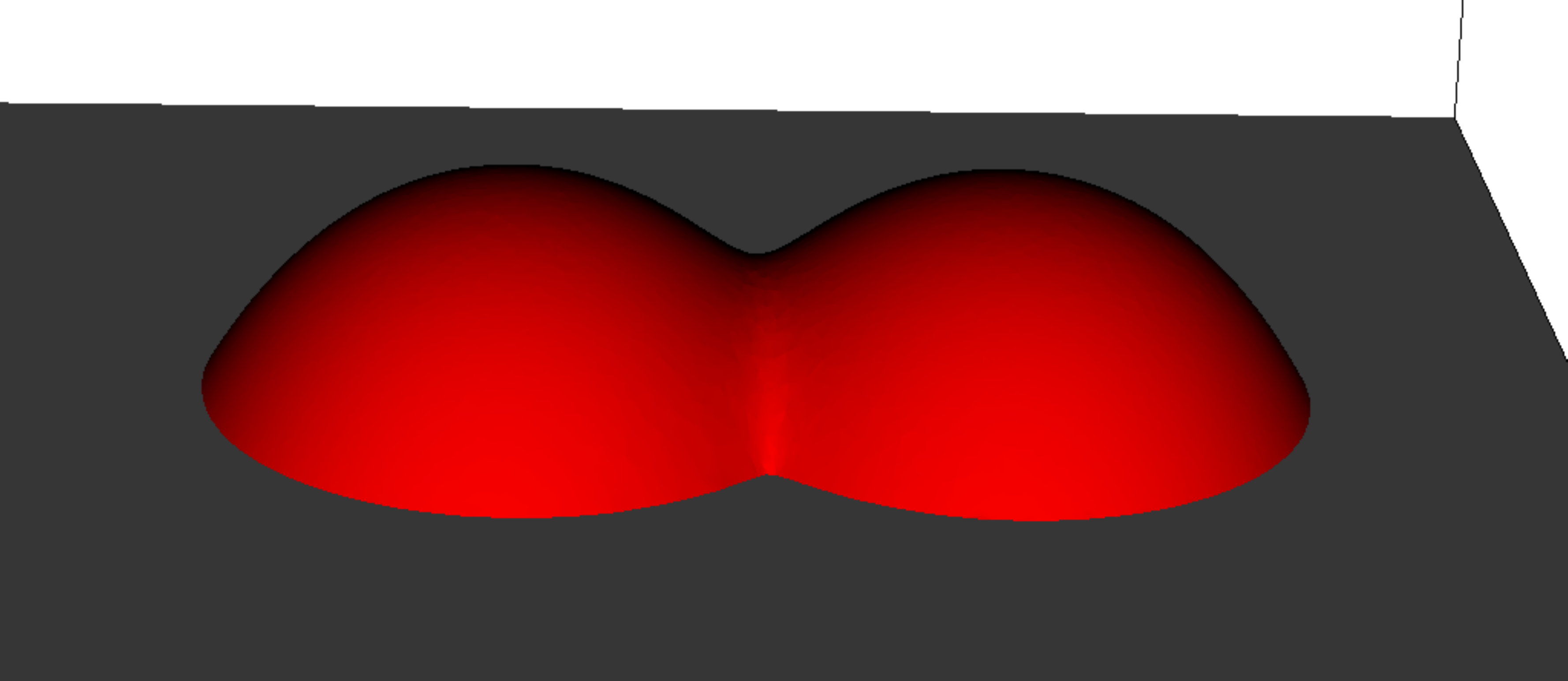}
      \label{sub:coal20}}
      \subfloat[$t = 1.5 \times 10^{-2}$]{
      \includegraphics[width=0.19\textwidth]{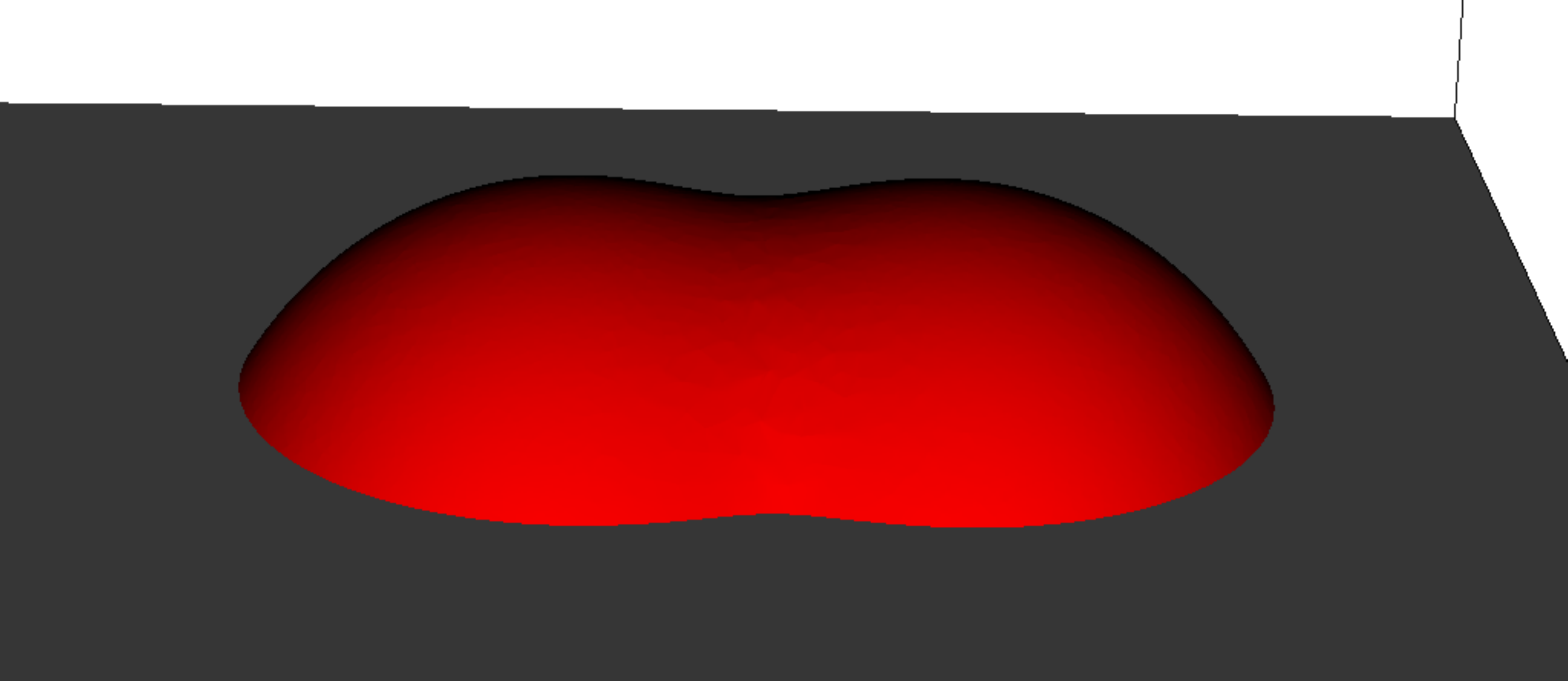}
      \label{sub:coal30}}
            \subfloat[$t = 2.5 \times 10^{-2}$]{
      \includegraphics[width=0.19\textwidth]{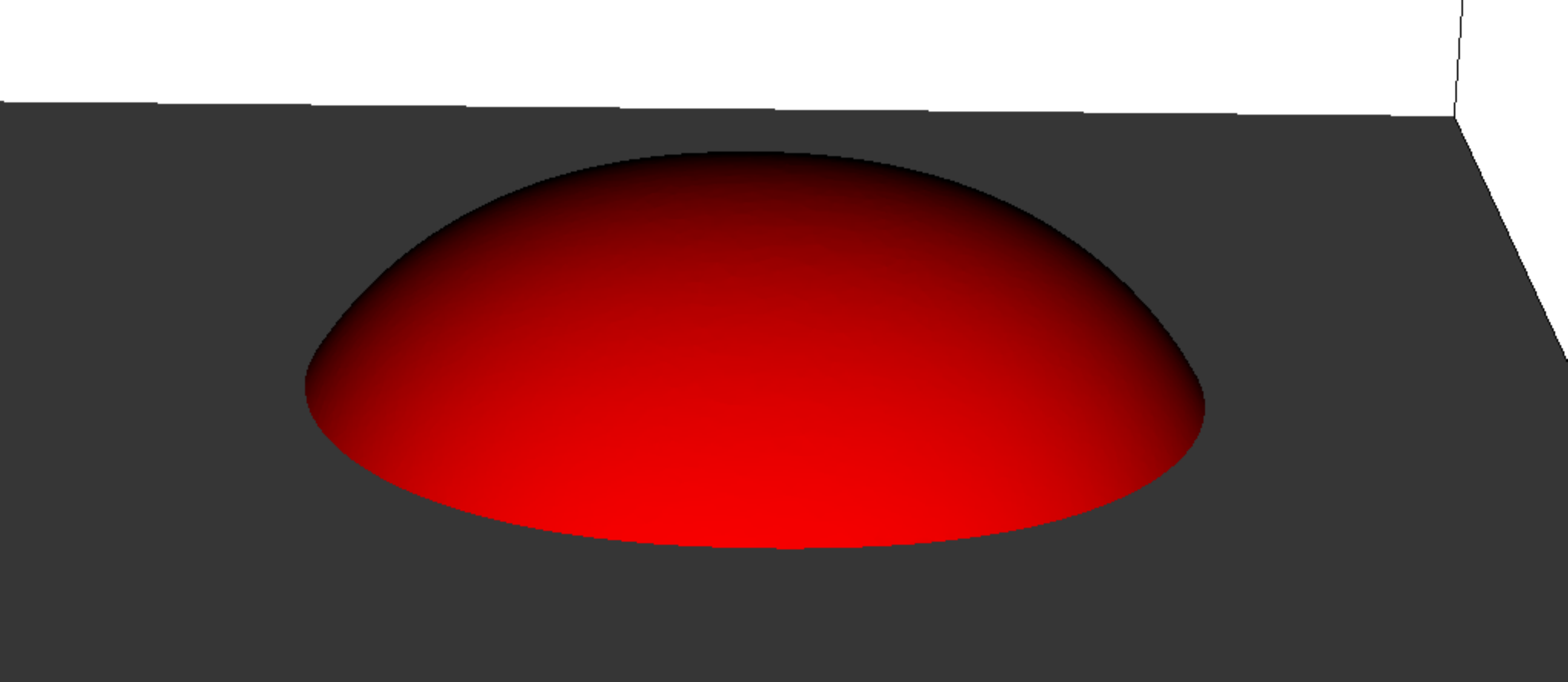}
      \label{sub:coal50}}
    \caption{Coalescence between two droplets ($\tensionsuperficielle_{23} = 1$).}
    \label{fig:coalescence}
  \end{center}
\end{figure}

Finally, Figure~\ref{fig:coalescence} shows a straightforward extension of previous simulations, the direct simulation of the coalescence between two identical droplets. The only differences with previous cases, are that the level-set function describes initially two droplets, and these droplets are totally represented, without symmetry assumption. The corresponding adapted mesh is made up of approximately 75,000 nodes and 425,000 tetrahedrons. All the other material and numerical parameters are identical to those used previously, and summarised in Table~\ref{tab:pamaters}, with a time step, here equal to $5.0 \times 10^{-4}$ in order to be able to visualise the early instants of the coalescence. Note that no special coalescence model has been used in this simulation. Consequently, the merging arises only when the distance between the surfaces of each droplet, is smaller than the mesh size.

\FloatBarrier
\subsection{Capillary rise}
\label{sec:capillary_rise}
\begin{figure}[!hbt]
  \begin{center}
    \subfloat[$t=0$]{
      \includegraphics[width=0.31\textwidth]{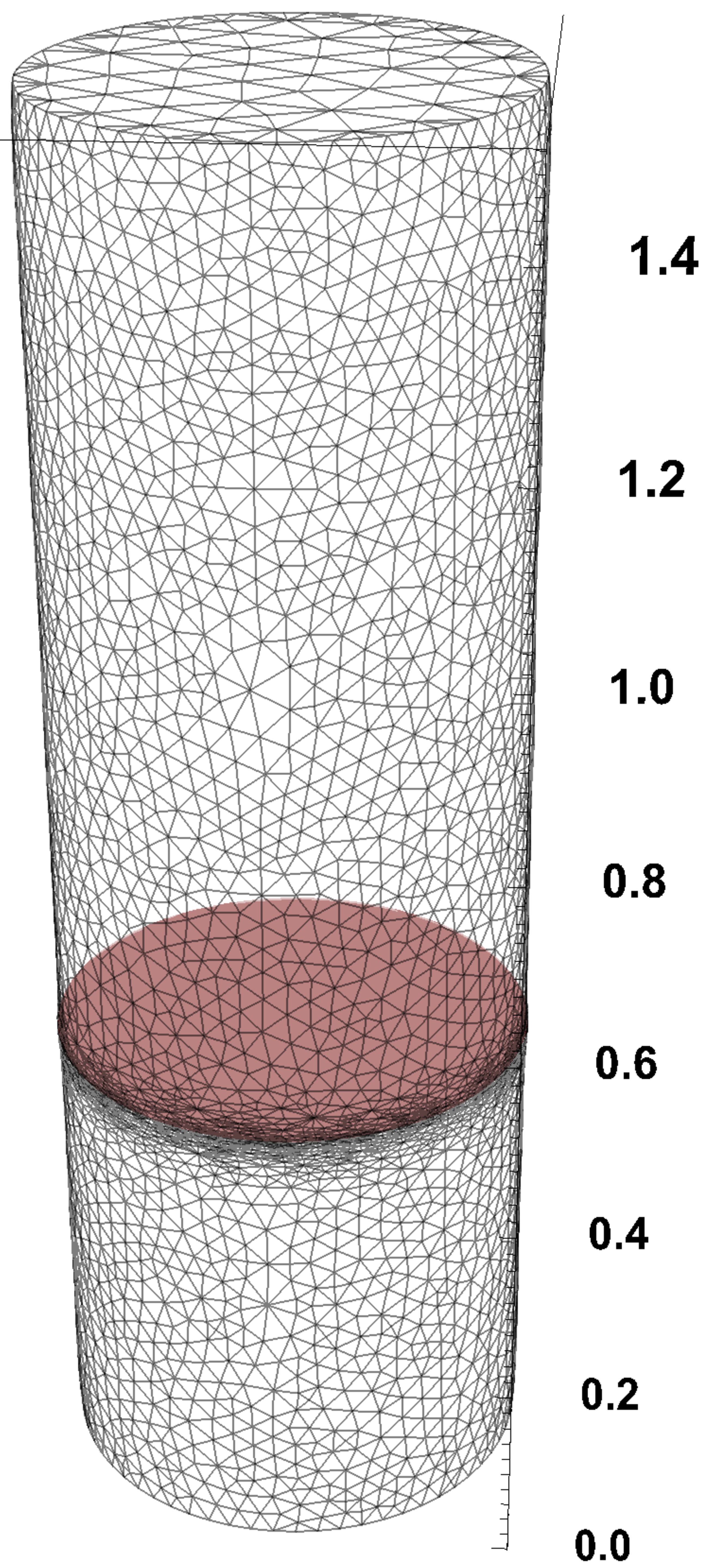}
      \label{sub:jurin_1_1}}
    \subfloat[$t = 1.5$]{
      \includegraphics[width=0.31\textwidth]{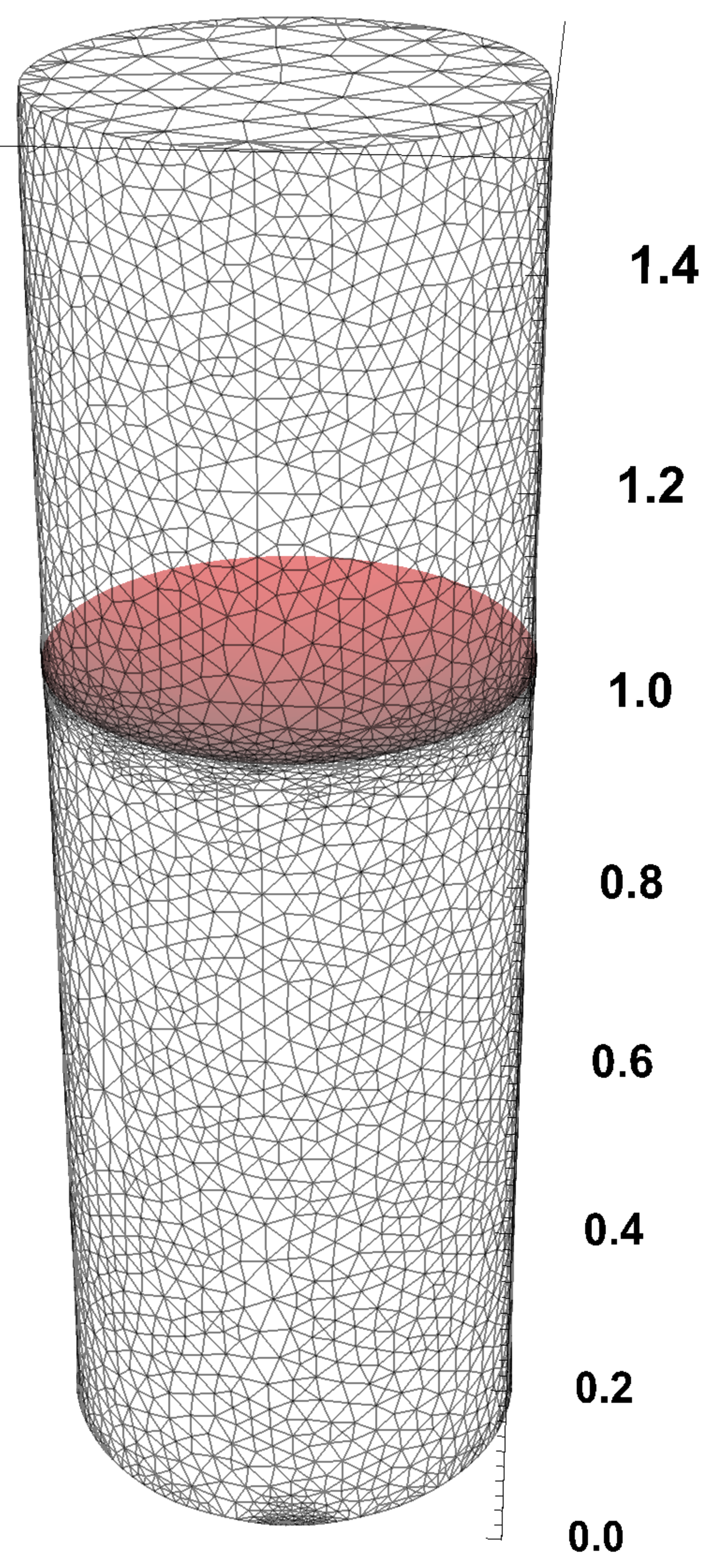}
      \label{sub:jurin_1_2}}
    \subfloat[Parallelepipedic tube]{
      \includegraphics[width=0.34\textwidth]{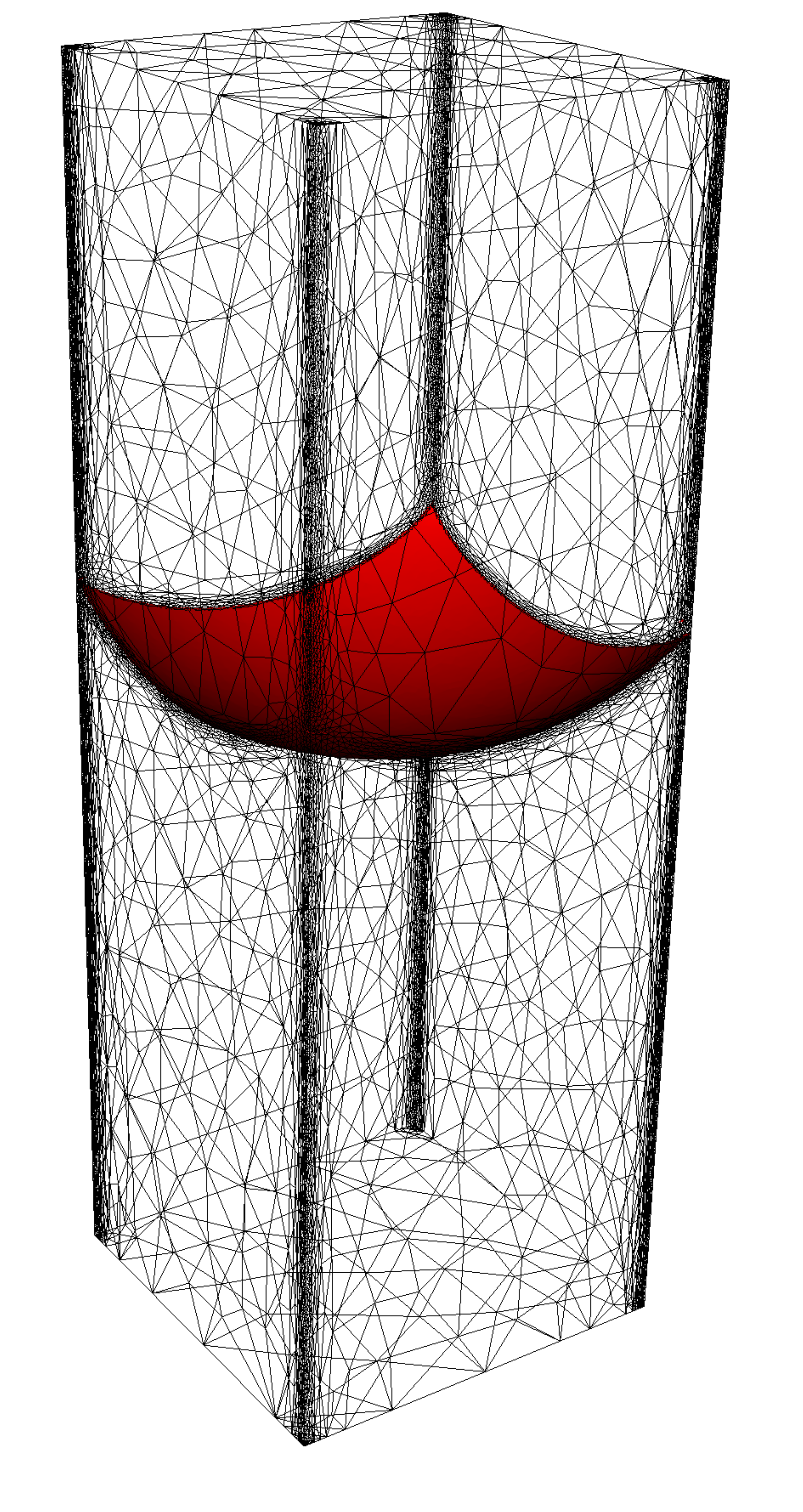}
      \label{sub:jurin_1_3}}
    \caption{Capillary rise with $\gamma_{23} = 1$. Initial- (a) and steady- (b) states in a cylindrical tube. (c) Steady state in a parallelepipedic tube. }
    \label{fig:jurin_1}
  \end{center}
\end{figure}

The simulation of the capillary rise is described in Figure~\ref{fig:jurin_1}: the capillary tube is a cylinder of radius $R_T = 0.25$ and height $h_T = 1.5$ (Figure~\ref{sub:jurin_1_1}). The tube is filled with two Newtonian fluids (liquid and gas), separated initially  by a flat interface.  The tube is discretised by a mesh made up of approximately 16,000 nodes and 73,000 tetrahedrons. In this way, the mesh size in the vicinity of the interface and perpendicular to it, is in the range $[3.0\times 10^{-3} : 8.0 \times 10^{-3}]$. The normal velocity vanishes at the lateral faces, while the top $\{z = 1\}$ and bottom $\{z = 0\}$ are considered as free boundaries. Furthermore body forces, typically the gravity forces, are considered acting only in the liquid domain (that is, the gas density is neglected). Hence, the body force vector $\vb$ is chosen equal to (0,0,-5) in the liquid, while $\vb = \bfm{0}$ in the gas. The value in the liquid is  arbitrarily, \ie chosen without respect to any realistic density, but only in order to have the interface in the computational domain when the steady-state is reached. The other parameters of the simulation are again those given in Table~\ref{tab:pamaters}, with a time step $\Delta t = 5.0 \times 10^{-4}$.

\begin{figure}[!hbt]
\begin{center}
\includegraphics[width=0.55\textwidth]{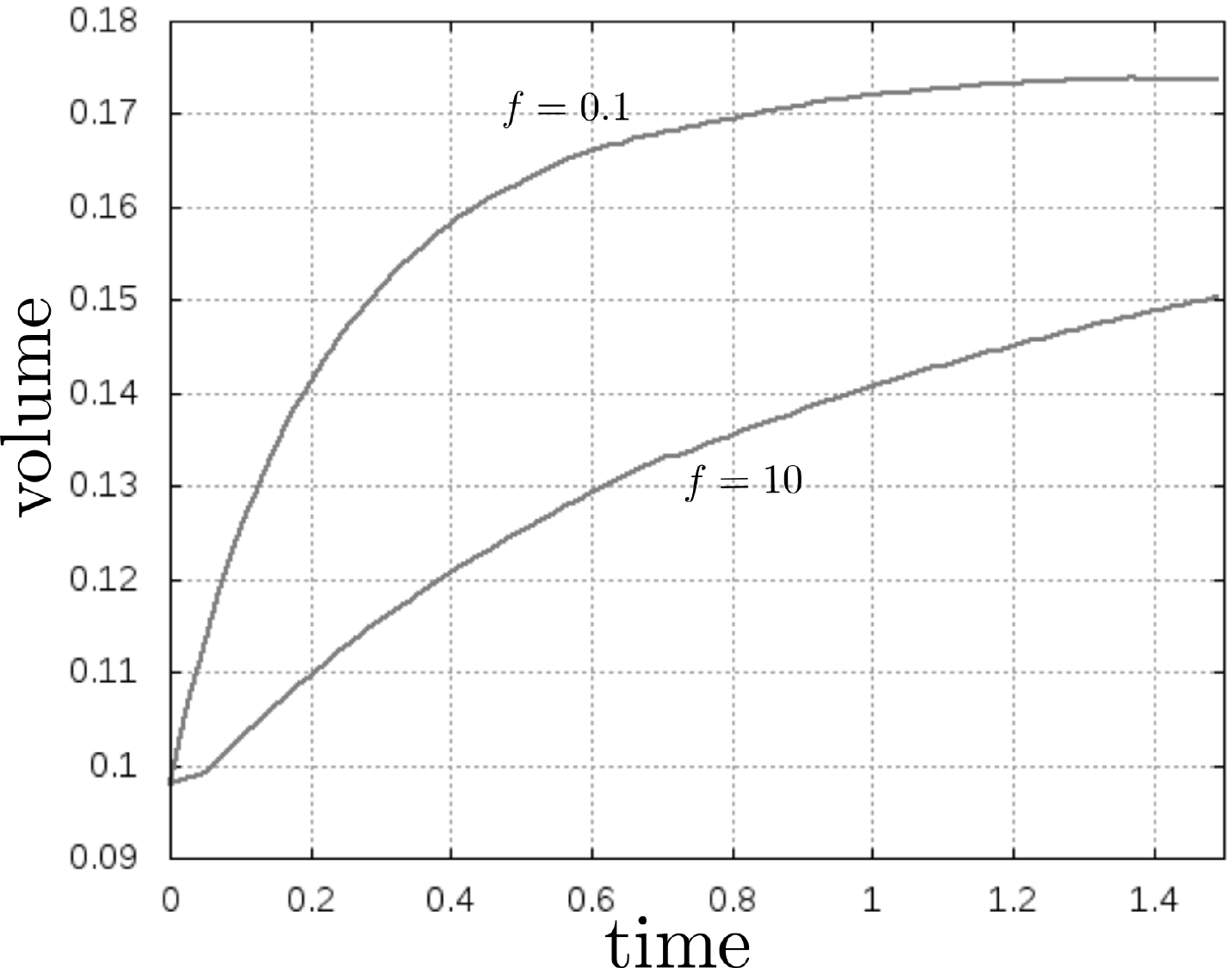}
\end{center}
\caption{Capillary rise: liquid column volume vs time.}
\label{fig:montee}
\end{figure}

Due to the capillary forces, the liquid-air interface adopts quickly a curved shape, to form an angle of 60\textdegree\, with the lateral faces of the tube. Moreover, since the liquid is free to flow into the tube through the face $\{z = 0\}$, the height of the liquid column increases, until reaching the equilibrium position of Figure~\ref{sub:jurin_1_2}. The volume of the liquid column is plotted over time in Figure~\ref{fig:montee} for two values of the friction parameter, $f = 0.1$ and $f = 10$. With this last value, the steady-state is not reached at $t = 1.5$ yet. However, in  both cases, equilibrium is of course the same, characterised by the balance between capillary and volume forces. Denoting $|\Omega_1|$ the volume of the liquid domain, $b$ the magnitude of the volume forces, $|\Gamma_{12}|$ the interface surface area, this force balance can be expressed as
\begin{equation}
\label{eq:force_balance}
b|\Omega_1| = \tensionsuperficielle_{12} \kappa |\Gamma_{12}|
\end{equation}
where $\kappa$ is the mean curvature of the liquid-air interface. This curvature is given by $\kappa = 2\cos\theta/R_T$, here $\kappa = 4$. Note that this derivation, neglecting the variations of $\kappa$, is valid only when the tube diameter is very small compared to the capillary length $\sqrt{\gamma_{12} b} \approx 1.44$ (see de Gennes~\cite{deGennes_book}, Section 2.4). Even if this condition is not fully satisfied here, a post-treatment using the Visit visualisation software, provides an average mean curvature approximately equal to 4. With this same software, the value of the interface area is found to be $|\Gamma_{12}| = 0.21$, leading to a corresponding liquid volume, given by relation~\eqref{eq:force_balance}, of $|\Omega_1| = 0.168$. The volume of the liquid column obtained by simulation (Figure~\ref{fig:montee}) is equal, at the equilibrium, to 0.173, showing a good description of the steady state. Still neglecting the variations of $\kappa$, the equilibrium height of the liquid column (referred to as Jurin's height) is $H = 2\gamma_{12} \cos\theta / bR_T$, here $H = 0.8$. Again, the simulation is in agreement with this value, since, at equilibrium, the point of the interface with a minimal $z$-coordinate is located at $z = 0.86$. Finally, note that with the presented numerical methodology, such simulations can be carried out using capillary tubes having a lateral surface not as smooth as the cylinder. For example, Figure~\ref{sub:jurin_1_3} shows a parallelepipedic tube. The curvature is then equal to zero on the lateral faces, and is no longer defined along the edges of the parallelepiped. However,  the normal vectors remain well- and uniquely- defined over each face. Thus, the weak formulation of the Laplace's law remains relevant in this case.

\FloatBarrier
\section{Conclusion}

A numerical framework for simulating, at a macroscopic scale, wetting problems and associated capillary effects, has been suggested. These developments rely on two key ingredients: a variational formulation of the mechanical problem, and a FE setting for discretising the resulting Stokes' equations. First, by expressing Laplace's law in a weak form, at liquid-air, liquid-solid and air-solid interfaces, the triple junction condition is naturally included in the weak formulation of the mechanical problem, without enforcing explicitly the numerical value of the contact-angle itself.  An additional dissipation term can be defined along the wetting line, ensuring an implicit coupling between geometry (dynamic contact-angle) and motion of this line. The discretisation of Stokes' equations is achieved by using FEs, linear both in velocity and pressure, stabilised by the ASGS method. In the elements cut by the moving interface, the  discrete pressure field is enriched, in order to capture the pressure discontinuity and consequently limit the effects of parasitic currents. The moving interface is described by using a Level-set method, combined with an anisotropic mesh adaptation technique with respect to both Level-Set and pressure fields.

This FE setting was first evaluated through 2D-simulations of droplet spreading. Even if the mesh adaptation is not absolutely necessary to such simulations, it was shown that the adaptation with respect to the pressure allows an accurate description of the pressure discontinuity, and therefore reduces the parasitic currents and improves mass conservation. The benefits of the pressure enrichment were also highlighted, particularly on meshes which are not refined. 3D-simulations of standard wetting problems were also investigated: spreading of droplets with possibly coalescence, and capillary rise into a tube. In all these simulations, the expected steady equilibrium state was reached. This state was described in term of contact-angle, and additionally for the capillary rise, volume of liquid column. This methodology simulates capillary-driven flows in 2D and 3D with an adequate precision. Furthermore, the investigated situations outline the ability of the exposed method to deal with ill-defined curvatures, at the liquid-gas interface, or along the rigid substrate.

\section*{Acknowledgement}

The authors wish to thank Chris Yukna for his help in proofreading.

\bibliographystyle{elsarticle-num}
\bibliography{bibliography}

\appendix
\section{ }
\label{appendix:A}
The objective of this appendix is to justify the semi-implicit expression of surface tension~\eqref{eq:implicite}, using the tensor analysis framework developed in this paper. First, one additional mathematical object, the shift tensor, has to be introduced. With the notations of Section~\ref{sec:math}, let us consider a point lying on the surface $\surface$. The position vector $\position$ of this point can be expressed with respect to ambient coordinates, generically denoted by $X = (X^1,X^2,X^3)$, or with respect to surface coordinates $S = (S^1,S^2)$. The connection between these two descriptions comes from the identity $\position(S) = \position(X(S))$. Differentiating this with respect to $S^\alpha$ leads to
\begin{equation}
\label{eq:surface_ambient}
\vS_\alpha = X^i_\alpha \vx_i
\end{equation}
where $\vx_i$ is the $i^{th}$ vector of the standard basis of $\R^3$, and $X^i_\alpha = \frac{\partial X^i}{\partial S^\alpha}$ is known as the shift tensor. The entries of this tensor are the components of the surface covariant basis $\vS_\alpha$ with respect to the ambient basis $\vx_i$.

Next, let us consider a vector $\vV$ of $\R^3$. The normal projection of $\vV$ is the vector $\vP$ defined by $\vP = (\vV \cdot \normale) \normale$, with $\normale$ the unit normal to $\surface$. In component form, this can be rewritten as $P^i = (n^i n_j) V^j$, and the operator $n^i n_j$, or in dyadic notation $\normale \otimes \normale$, is therefore the projection operator onto the normal $\normale$. The orthogonal projection of $\vV$ can also be considered as the tangent plane vector $\vT$ defined by $\vT = (\vV \cdot \vS^\alpha) \vS_\alpha$. In component form, this relation turns into $T^i = X^i_\alpha X^\alpha_j V^j$, and the tensor $X^i_\alpha X^\alpha_j$ appears to be the operator projection onto the surface $\surface$.

Finally, considering that any vector $\vV$ is equal to the sum of its normal and orthogonal projections, $\vV = \vP + \vT$, so the sum of normal and orthogonal projection operators equals the identity:
\begin{equation}
\label{eq:projection}
n^i n_j + X^i_\alpha X^\alpha_j = \delta^i_j
\end{equation}

This relation is all that we need to conclude. First, the equality $\vS^\alpha \cdot \grad_\alpha \vw = (\identity - \normale \otimes \normale) : \grad \vw$, used when passing from Equation~\eqref{eq:momentum_weak8} to~\eqref{eq:weak_form2} can easily be derived. On the left hand side, just use expansion~\eqref{eq:surface_ambient}, understand that by the chain rule, $\grad_\alpha \vw = Z^i_\alpha \frac{\partial \vw}{\partial X^i}$, and apply relation~\eqref{eq:projection}. Second, the semi-implicit treatment of surface tension term consists, in Equation~\eqref{eq:momentum_weak8}, in computing $\vS^\alpha \cdot \grad_\alpha \vw$ at time $t_n$, by using a predictor of $\vS^\alpha$ at $t^{n+1}$. Let $\vS_\alpha^{n+1/2}$ be this predictor, defined by:
\begin{equation}
\label{eq:prediction}
\vS_\alpha^{n+1/2}(\vx) = \frac{\partial}{\partial S^\alpha} \left( \position(\vx,t_n) + \Delta t \vv (\vx,t_n) \right)
\end{equation}
for any point $\vx  \in \Gamma_{12}(t_n)$, with $\vv$ the velocity field and $\Delta t$ the time step. The semi-implicit expression of surface tension term becomes
\begin{equation}
\label{eq:prediction2}
\int_{\Gamma_{12}} \gamma \vS_\alpha^{n+1/2} \cdot \grad^\alpha \vw \, \dsurface = \int_{\Gamma_{12}} \gamma (\identity - \normale \otimes \normale) :  \grad \vw \, \dsurface + \Delta t \int_{\Gamma_{12}} \gamma (\grad_\alpha \vv) \cdot (\grad^\alpha \vw) \, \dsurface
\end{equation}

The second term on the right-hand side of expression~\eqref{eq:prediction2} is the term announced in~\eqref{eq:implicite}. To show the equivalence, let us expand vectors $\vv$ and $\vw$ on the ambient basis, $\vv = v^i \vx_i$ and $\vw = w_j \vx^j$. As the ambient coordinate system has been chosen to be Cartesian, $\grad_\alpha \vv \cdot \grad^\alpha \vw = (\grad_\alpha v^i) (\grad^\alpha w_i)$. By the chain rule, we obtain $\grad_\alpha \vv \cdot \grad^\alpha \vw = (\grad_j v^i X^j_\alpha) (\grad^k w_i X^\alpha_k)$. Then, relation~\eqref{eq:projection} allows us to conclude:
\begin{equation}
\label{eq:prediction3}
\begin{array}{lcl}
\displaystyle \Delta t \int_{\Gamma_{12}} \gamma (\grad_\alpha \vv) \cdot (\grad^\alpha \vw) \, \dsurface &=& \displaystyle \Delta t \int_{\Gamma_{12}} \gamma (\grad_j v^i) (\delta^j_k - n^j n_k) (\grad^k w_i) \, \dsurface \\ &=& \Delta t \int_{\Gamma_{12}} \gamma
 \left( (\grad \vv) \cdot (\identity - \normale \otimes \normale) \right) : \grad \vw \, \dsurface
\end{array}
\end{equation}
which is exactly the term given in Equation~\eqref{eq:implicite}.

\end{document}